\documentclass[10pt]{iopart}
\usepackage{epsfig}
\eqnobysec

\newcommand{\be}{\begin{equation}}
\newcommand{\ee}{\end{equation}}
\newcommand{\bea}{\begin{eqnarray}}
\newcommand{\eea}{\end{eqnarray}}
\newcommand{\beastar}{\begin{eqnarray*}}
\newcommand{\eeastar}{\end{eqnarray*}}
\newcommand{\nn}{\nonumber\\}
\newcommand{\lav}{\left\langle}
\newcommand{\rav}{\right\rangle}
\newcommand{\ts}{\textstyle}

\newcommand{\half}{\frac{1}{2}}

\newcommand{\eq}[1]{~(\ref{#1})}
\newcommand{\eqq}[2]{~(\ref{#1},\ref{#2})}

\newcommand{\order}{{{\mathcal O}}}

\newcommand{\ie}{{\it i.e.}}
\newcommand{\eg}{{\it e.g.}}

\newcommand{\tw}{t_{\rm w}}
\newcommand{\yw}{y_{\rm w}}
\newcommand{\dt}{\Delta t}

\newcommand{\Tc}{T_{\rm c}}
\newcommand{\zv}{\mathbf{0}}
\newcommand{\qv}{\mathbf{q}}
\newcommand{\rv}{\mathbf{r}}
\newcommand{\dr}{\Delta\mathbf{r}}
\newcommand{\dq}{\int(dq)\,}
\newcommand{\n}{\kappa}
\renewcommand{\sc}[1]{{\mathcal{F}}_{#1}}
\newcommand{\dz}{\Delta z}
\newcommand{\X}{s}
\newcommand{\Y}{r}
\newcommand{\rn}{N^{-1/2}}
\newcommand{\del}{\Delta}
\newcommand{\p}{K}
\newcommand{\Lone}{L^{(1)}}
\newcommand{\Ltwo}{L^{(2)}}
\renewcommand{\eql}{_{\rm eq}}
\newcommand{\eqlup}{^{\rm eq}}
\newcommand{\Eql}{_{E}^{\rm eq}}
\newcommand{\Om}{\Omega}

\newcommand{\COC}{C\Om C}
\newcommand{\bond}{_{\rm bond}}
\newcommand{\product}{_{\rm prod}}
\newcommand{\bl}{^{\rm \blword}}
\newcommand{\loc}{^{\rm loc}}
\newcommand{\blword}{block}
\newcommand{\ROC}{R\Om C}
\newcommand{\CC}{CC}
\newcommand{\CCt}{\Cc\Cc}
\newcommand{\Ld}{\lambda_d}
\newcommand{\CCd}{\gamma_d}
\newcommand{\G}{{\mathcal{G}}}
\newcommand{\Gt}{\tilde{\mathcal{G}}}
\newcommand{\omt}{w}
\newcommand{\Gd}{g_d}
\newcommand{\Gtd}{\tilde{g}_d}
\newcommand{\Ad}{\alpha_d}
\newcommand{\Atd}{\tilde{\alpha}_d}
\newcommand{\Bd}{\beta_d}
\newcommand{\nnn}{n}
\newcommand{\Cc}{\tilde{C}}
\newcommand{\Cz}{\Cc_\zv}
\newcommand{\Kc}{\tilde{K}}
\newcommand{\Lc}{\tilde{L}}
\newcommand{\dL}{L_\zv}
\newcommand{\dLtwo}{\Ltwo_\zv}
\newcommand{\Lctwo}{\tilde{L}^{(2)}}
\newcommand{\Dc}{C_{\rm m}}
\newcommand{\Q}{R_{\rm m}}
\newcommand{\CCtd}{\tilde\gamma_d}

\begin{document}

\title[Fluctuation-dissipation relations in the non-equilibrium
spherical ferromagnet]{Spin, bond and global fluctuation-dissipation relations
in the non-equilibrium spherical ferromagnet}

\author{Alessia Annibale\footnote[1]{Email
alessia.annibale@kcl.ac.uk}, Peter Sollich\footnote[2]{Email
peter.sollich@kcl.ac.uk}}

\address{King's College London, Department of Mathematics, London WC2R
2LS, UK}

\begin{abstract} We study the out-of-equilibrium dynamics of the
spherical ferromagnet after a quench to its critical
temperature. We calculate correlation and response functions
for spin observables which probe lengthscales much larger than the lattice
spacing but smaller than the system size, and find that the
asymptotic fluctuation-dissipation ratio (FDR) $X^\infty$ is the same
as for local observables. This is consistent with our earlier results
for the Ising model in dimension $d=1$ and $d=2$. We also check that
bond observables, both local and long-range, give the same asymptotic
FDR. In the second part of the paper the analysis is extended to global
observables, which probe correlations among all $N$ spins.
Here non-Gaussian fluctuations arising from the spherical
constraint need to be accounted for, and we develop a systematic
expansion in $1/\sqrt{N}$ to do this. Applying this to the global bond
observable, i.e.\ the energy, we find that non-Gaussian corrections
change its FDR to a nontrivial value which we calculate exactly for
all dimensions $d>2$. Finally, we consider quenches from magnetized initial
states. Here even the FDR for the global {\em spin} observable, i.e.\
the magnetization, is nontrivial. It differs from the one for
unmagnetized states even in $d>4$, signalling the appearance of a
distinct dynamical universality class of magnetized critical
coarsening. For lower $d$ the FDR is 
irrational even to first order in $4-d$ and $d-2$, the latter in contrast to
recent results for the $n$-vector model.
\end{abstract}

\section{Introduction}

A key insight of statistical mechanics is that {\em
equilibrium} states can be accurately described in terms of only a
small number of thermodynamic variables, such as temperature and
pressure. For non-equilibrium systems such as glasses no similar
simplification exists {\em a 
priori}; the whole past history of a sample is in principle required
to specify its state at a given time. This complexity makes
theoretical analysis awkward, and one is led
instead to look for a description of non-equilibrium states in
terms of a few effective thermodynamic parameters. Much work in recent
years has focussed on one such parameter, the {\em effective
temperature}. This can be defined on the basis of fluctuation-dissipation
(FD) relations between correlation and response functions and has
proved to be very fruitful in mean field
systems~\cite{CugKurPel97,CriRit03}.

The use of FD relations to quantify the out-of-equilibrium dynamics in
glassy systems is motivated by the occurrence of {\em
aging}~\cite{BouCugKurMez98}: the time scale of response to an
external perturbation increases with the age (time since preparation)
$\tw$ of the system. As a consequence, time translational invariance
and the equilibrium fluctuation-dissipation
theorem~\cite{Reichl80} (FDT) relating correlation and response
functions break down. To quantify this, one considers the correlation
function of two generic observable $A$ and $B$ of a system, defined as
\be
C(t,\tw)=\lav A(t)B(\tw)\rav -\lav A(t) \rav \lav B(\tw) \rav
\label{corr}
\ee
The associated (impulse) response function can be defined as
\[
R(t,\tw)=\left.\frac{\delta \lav A(t)\rav}{\delta h_B(\tw)}\right|_{h_B=0}
\]
and gives the linear response of $A$ at time $t$ to a small impulse in
the field $h_B$ conjugate to $B$ at the earlier time $\tw$. (The
latter is normally thought of as a ``waiting time'' since preparation
of the system at time 0.) Equivalently one can work with the
susceptibility
\be
\chi(t,\tw)=\int_{\tw}^t\! dt'\, R(t,t')
\label{switch_on}
\ee
which encodes the response of $A(t)$ to a small step
$h_B(t)=h_B\Theta(t-\tw)$ in the field starting at $\tw$. In
equilibrium, FDT implies that 
$-\partial_{\tw} \chi(t,\tw) = R(t,\tw) =
{T}^{-1}\partial_{\tw}C(t,\tw)$. Out of equilibrium, the
violation of FDT can be measured by an FD ratio (FDR), $X$, defined
through~\cite{CugKur93,CugKur94}
\be
\label{eqn:non_eq_fdt}
-\partial_{\tw} \chi(t,\tw) = R(t,\tw) =
\frac{X(t,\tw)}{T}\partial_{\tw}C(t,\tw)
\ee
This implies that $X$ can be read off from the slope $-X/T$ of a
parametric FD plot showing $\chi$ vs $C$, at fixed $t$ and with $\tw$
as the curve parameter. This remains the case if both axes are
normalized by the equal-time variance of $A$, $C(t,t)$, a procedure
which is helpful
in fixing the scale of the plot in situations where $C(t,t)$ varies
significantly with time~\cite{FieSol02,SolFieMay02}. In equilibrium, the FD
plot is a straight line with slope $-1/T$.

In mean-field spin glasses~\cite{CugKurPel97,CugKur93,CugKur94} one
finds that FD plots of autocorrelation and response of local spins and
similar observables approach a limiting shape for large $t$. This is
typically composed of 
two straight line segments. In the first of these one finds $X=1$,
corresponding to 
quasi-equilibrium dynamics for time differences $t-\tw$ that do not
grow with the age of the system. The second line segment has $X<1$ and
reflects the dynamics on aging timescales, i.e.\ time differences
growing (in the simplest case linearly) with $\tw$. One can use this
to define a non-equilibrium {\em effective temperature} $T_{\rm
eff}=T/X$, which has been shown to have many of the properties of a
thermodynamic temperature~\cite{CugKurPel97,CugKur93,CugKur94}.

How well this physically very attractive mean-field scenario transfers
to more realistic non-equilibrium systems with short-range
interactions has been a matter of intense research
recently~\cite{CriRit03}. A useful class of systems for studying this
question in detail is provided by ferromagnets quenched from high
temperature to the critical temperature (see e.g.\
Refs.~\cite{GodLuc00b,MayBerGarSol03} and the recent
review~\cite{CalGam05}) or below. The system then coarsens -- by the
growth of domains with the equilibrium magnetization, for $T<T_c$ --
and exhibits aging; in an infinite system equilibrium is never
reached. The aging is clearly related to the growth of a
lengthscale~\cite{Bray94} (domain size for $T<\Tc$, or correlation
length for $T=\Tc$), and this makes ferromagnets attractive
``laboratory'' systems for understanding general properties of
non-equilibrium dynamics. They are of course not completely generic;
compared to e.g.\ glasses they lack features such as thermal
activation over energetic or entropic barriers.

We focus in this paper mostly on ferromagnets quenched to $\Tc$, i.e.\
on critical coarsening dynamics. Some care is needed in this case with
the interpretation of limiting FD plots: while in the mean-field
situation $X$ becomes at long times a function of $C$ only, as implied
by the existence of a nontrivial limit plot, in critical coarsening
$X$ approaches a function of $t/\tw$~\cite{GodLuc00b}. In the
interesting regime where $t/\tw$ is finite but $>1$, time differences
$t-\tw=\order(\tw)$ are then large, and e.g.\ spin autocorrelation
functions have decayed to a small part of their initial value. In the
limit $t,\tw\to\infty$ the FD plot then assumes a pseudo-equilibrium
shape, with all nontrivial detail compressed into a vanishing region
around $C=0$.

The fact that the FDR is a smooth function of $t/\tw$ makes the
interpretation of $T/X$ as an effective temperature less obvious than
in mean-field spin glasses, where $T/X$ is constant within each time
sector ($t-\tw=\order(1)$ vs $t-\tw$ growing with $\tw$). To eliminate
the time-dependence one can consider the limit of times that are both
large and well-separated. This defines an {\em asymptotic FDR}
\be
X^\infty=\lim_{\tw\to\infty}\lim_{t\to\infty}X(t,\tw)
\ee
An important property of this quantity is that it should be {\em
universal}~\cite{GodLuc00b,CalGam05} in the sense that its value is
the same for different systems falling into the same universality
class of critical non-equilibrium dynamics. This makes a study of
$X^\infty$ interesting in its own right, even without an
interpretation in terms of effective temperatures.

If one nevertheless wants to pursue such an interpretation, the
resulting value of 
the effective temperature $T/X^\infty$ should be the same for all (or
at least a large class of) observables $A$. The observable-dependence
of $X^\infty$ therefore becomes a key
question~\cite{FieSol02,SolFieMay02,MayBerGarSol03,CalGam04}.
Conventionally, much work on non-equilibrium ferromagnets has focussed
on the local spin autocorrelation function and associated response. An
obvious alternative is the long-wavelength analogue, i.e.\ the
correlation function of the fluctuating magnetization. Exact
calculations for the Ising chain~\cite{MayBerGarSol03,MaySol04} as
well as numerical simulations~\cite{MayBerGarSol03,MayBerGarSol04} in
dimension $d=2$ show that the resulting global $X^\infty$ is always
identical to the local version. This local--global correspondence,
which can also be obtained by field-theoretic
arguments~\cite{CalGam05,CalGam02c,CalGam02}, arises physically
because the long wavelength Fourier components of the spins are
slowest to relax and dominate the long-time behaviour of both local
and global quantities.

The local--global correspondence does of course not address the full
range of observable-dependence of the asymptotic FDR; one might ask
about other observables which are linear combinations not of spins
but for example products of interacting spins. In the critical Ising
model in $d=2$, numerical
simulations~\cite{MayBerGarSol03,MayBerGarSol04} suggest that even
these give the same $X^\infty$, so that an interpretation of
$T/X^\infty$ in terms of an effective temperature appears
plausible. One of the motivations for the current study was to verify
whether this {\em observable-independence} of $X^\infty$ across
different types of observables holds in an exactly solvable model, the
spherical ferromagnet~\cite{BerKac52,Joyce72}. In addition, we will study what
effect different {\em initial conditions} have on $X^\infty$. This is
motivated by our recent study of Ising models in the classical regime
of large $d$, 
where critical fluctuations are irrelevant~\cite{GarSolPagRit05}. It
turned out that magnetized initial do states produce a different value of
$X^\infty$, so that critical coarsening in the presence of a nonzero
magnetization is in a new dynamical universality class even though the
magnetization does decay to zero at long times.

We begin in Sec.~\ref{sec:Gaussian} with a brief review of the
standard setup for the dynamics of the spherical model, as used in
e.g.~\cite{GodLuc00b}. Fluctuations in an effective Lagrange
multiplier enforcing the spherical constraint are neglected, leading
to a theory where all spins are Gaussian random variables. In
Sec.~\ref{sec:finite-range} this is applied to various observables of
finite range, by which we mean correlations and responses probing
lengthscales that can be large but remain small compared to the
overall system size. For spin observables we show that the expected
equality of $X^\infty$ between local and long-range quantities holds
(Sec.~\ref{sec:spin_observables}). We check observable-independence of
$X^\infty$ further by considering bond and spin product observables,
in Sec.~\ref{sec:bond_observables} and~\ref{sec:product_observables}
respectively.

The major part of the paper is then devoted to a study of FDRs for
{\em global} observables, with a focus on the energy, i.e.\ the global
bond observable. Because of the weak infinite-range interaction
generated by the spherical constraint, such observables behave
differently from their long-range analogues in the spherical
model. Calculations of correlation and response functions are
technically substantially more difficult because Lagrange multiplier
fluctuations can no longer be neglected. To account for them we construct
in Sec.~\ref{sec:setup} a systematic expansion of the dynamically
evolving spins in $\rn$. This allows us to calculate the leading
non-Gaussian corrections that we need for global correlations, as
shown for the case of the energy in
Sec.~\ref{sec:energy_general}. After a brief digression to equilibrium
dynamics, we evaluate the resulting expressions in
Sec.~\ref{sec:noneq_large_d} for $d$ above the critical dimension
$d_{\rm c}=4$, and in Sec.~\ref{sec:noneq_small_d} for
$d<4$. Importantly, we will find that in the latter case the
asymptotic FDR is different from those for finite-range
observable. This means that an effective temperature interpretation of
$X^\infty$ is possible at best in a very restricted sense. However, we
will find that our results are in agreement with recent
renormalization group (RG) calculations near $d=4$~\cite{CalGam04} in
the $O(n\to\infty)$-model. This suggests that the non-Gaussian
effects captured in global observables are important for linking the
spherical model to more realistic systems with only short-range
interactions. Finally, we turn in Sec.~\ref{sec:magnetized} to
critical coarsening starting from magnetized initial conditions. Here
already the global {\em spin} observable is affected by non-Gaussian
corrections. Once these are accounted for, we find $X^\infty=4/5$ for
$d>4$ as in the Ising case~\cite{GarSolPagRit05}. For $d<4$ we provide
the first exact values of the asymptotic FDR in the presence of a
nonzero magnetization; these turn out to be highly nontrivial even to
first order in $4-d$ and $d-2$. Our results are summarized and
discussed in Sec.~\ref{sec:conclusion}. Technical details are relegated to two
appendices.

\section{Langevin dynamics and Gaussian theory}
\label{sec:Gaussian}

We consider the standard spherical model Hamiltonian
\be
H = \half \sum_{(ij)} (S_i-S_j)^2
\label{H_spherical}
\ee
The sum runs over all nearest neighbour (n.n.) pairs on a
$d$-dimensional (hyper-)cubic lattice; the lattice constant is taken
as the unit length. At each of the $N$ lattice sites $\rv_i$ there is a
real-valued spin $S_i$. The spherical constraint $\sum_i S_i^2=N$ is
imposed, which can be motivated by analogy with Ising spins $S_i=\pm
1$~\cite{BerKac52}.

The Langevin dynamics for this model can be written as
\be
\partial_t S_i = - \frac{\partial H}{\partial S_i} + \xi_i - \frac{1}{N}
\sum_k S_k\left(-\frac{\partial H}{\partial S_k} + \xi_k\right)S_i
\label{eqn_motion}
\ee
with $\xi_i$ Gaussian white noise with zero mean and covariance $\lav
\xi_i(t)\xi_j(t')\rav = 2T\delta_{ij} \delta(t-t')$. The last term
in\eq{eqn_motion}, i.e.\ the sum over $k$, enforces the
spherical constraint at all times by removing the component of the
velocity vector 
$(\partial_t S_1,\ldots,\partial_t S_N)$ along $(S_1,\ldots,S_N)$. We
use here the Stratonovic
convention for products like $S_k\xi_k$. This allows the ordinary
rules of calculus to be used when evaluating derivatives such as
$\partial_t S_i^2$. Physically it corresponds to the intuitively reasonable
scenario where the noise $\xi_i$ is regarded as a smooth random
process but with a correlation time much shorter than any other
dynamical timescale.

The prefactor of $S_i$ in the last term of\eq{eqn_motion}, being an
average of $N$ contributions, will have fluctuations of $\order(\rn)$.
Conventionally one ignores these and approximates the equation of
motion as
\be
\partial_t S_i = - \frac{\partial H}{\partial S_i} + \xi_i - z(t) S_i
\label{eqn_motion_Gaussian}
\ee
where $z(t)$ can be viewed as an effective time-dependent Lagrange multiplier
implementing the spherical constraint. This approximation works for
local quantities, but as we will see can give incorrect results when
one considers \eg\ fluctuations of the magnetization or the energy,
which involve correlations across the entire system.  One can see
directly that\eq{eqn_motion_Gaussian} is an approximation from the
fact that it corresponds to Langevin dynamics with the effective
Hamiltonian $H+\half z(t)\sum_i S_i^2$. Since the latter is
time-dependent, this dynamics does not satisfy detailed balance. It is
simple to check, on the other hand, that the original equation of
motion\eq{eqn_motion} does satisfy detailed balance and leads to the
correct equilibrium distribution $P\eql(\{S_i\})\propto \exp(-\beta
H)\delta(\sum S_i^2-N)$ where $\beta=1/T$ is the inverse temperature
as usual.

The key advantage of the approximation\eq{eqn_motion_Gaussian} is, of
course, that the spins are Gaussian random variables at all times as
long as the initial condition is of this form. Explicitly, if we
define a matrix $\Om$ with $\Om_{ij}=-1$ for n.n.\ sites $i,j$ and
$\Om_{ii}=2d$, the Gaussian equation of motion is
\be
\partial_t S_i = -\sum_j \Om_{ij}S_j - z(t) S_i + \xi_i
\label{real_space_dotSi}
\ee
We review briefly how this is solved (see e.g.~\cite{GodLuc00b} and
references therein), since these results form the
basis for all later developments. In terms of the Fourier components
$S_\qv = \sum_i S_i \exp(-i\qv\cdot\rv_i)$ of the spins,
equation\eq{real_space_dotSi} reads
\be
\partial_t S_\qv = -(\omega_\qv+z(t))S_\qv + \xi_\qv
\label{dotSq}
\ee
where $\omega_\qv = 2\sum_{a=1}^d (1-\cos q_a)$; we mostly write just
$\omega$. The Fourier mode response function can be read off as
\be
R_\qv(t,\tw) = \exp\left(-\omega(t-\tw)-\int_{\tw}^t dt'\,
z(t')\right) \equiv \sqrt{\frac{g(\tw)}{g(t)}}e^{-\omega(t-\tw)}
\label{Rq}
\ee
where
\be
g(t) = \exp\left(2\int_{0}^t dt'\, z(t')\right)
\label{g_def}
\ee
In terms of this, the time-dependence of the $S_\qv$ becomes
\be
S_\qv(t) = R_\qv(t,0) S_\qv(0) + \int_0^t dt'\,R_\qv(t,t')\xi_\qv(t')
\label{Sqt}
\ee
The equal-time correlator $C_\qv(t,t) = (1/N) \lav
S_\qv(t)S_\qv^*(t)\rav$ follows as
\bea
C_\qv(t,t) &=& C_\qv(0,0) R_\qv^2(t,0) + 2T \int_0^t dt'\,R_\qv^2(t,t') \\
&=& \frac{C_\qv(0,0)}{g(t)}e^{-2\omega t} + 2T \int_0^t dt'\,
\frac{g(t')}{g(t)} e^{-2\omega(t-t')}
\label{Cqtt}
\eea
and we note for later the identity
\be
\partial_t C_\qv(t,t) = 2T -
\left(2\omega+\frac{g'(t)}{g(t)}\right) C_\qv(t,t)
\label{Cqtt_deriv}
\ee
The two-time correlator $C_\qv(t,\tw) = (1/N) \lav
S_\qv(t)S_\qv^*(\tw)\rav$ can be deduced from the analogue of\eq{Sqt}
for initial time $\tw$
\be
S_\qv(t) = R_\qv(t,\tw) S_\qv(\tw) + \int_{\tw}^t dt'\,R_\qv(t,t')\xi_\qv(t')
\ee
as
\be
C_\qv(t,\tw) = R_\qv(t,\tw) C_\qv(\tw,\tw)
\label{C_twotime}
\ee
The position-dependent correlation and response functions
$C_{ij}(t,\tw)$ and $R_{ij}(t,\tw)$ are then just the inverse Fourier
transforms of $C_\qv(t,\tw)$ and $R_\qv(t,\tw)$, respectively, with
$\qv$ conjugate to $\rv_j-\rv_i$.

\subsection{The function $g(t)$}

The calculations outlined above show that the Gaussian dynamics is
fully specified once the function $g(t)$ is known. The latter can be
found from the spherical constraint, which imposes $\dq
C_\qv(t,t)=1$. Here and below we abbreviate $(dq) \equiv
d\qv/(2\pi)^d$, where the integrals runs over the first Brillouin zone
of the hypercubic lattice, i.e.\ $\qv\in[-\pi,\pi]^d$. Using\eq{Cqtt}
this constraint gives an integral equation for $g(t)$:
\be
g(t) =
\dq C_\qv(0,0) e^{-2\omega t} + 2T \int_0^t dt'\,  f(t-t')g(t')
\label{g_eq}
\ee
where 
\be
f(t) = \dq  e^{-2\omega t} = [e^{-4t}I_0(4t)]^d \approx
(8\pi t)^{-d/2}
\label{f_def}
\ee
Here $I_0$ denotes a modified Bessel function and the final expression
gives the asymptotic behaviour for large $t$. In terms of Laplace transforms
$\hat{g}(s)=\int_0^\infty dt\, \exp(-st)g(t)$, eq.\eq{g_eq} then has the
solution
\be
\hat{g}(s) = \frac{1}{1-2T\hat{f}(s)} \dq \frac{C_\qv(0,0)}{s+2\omega}
\label{hat_g_general}
\ee

With the exception of Sec.~\ref{sec:magnetized}, we focus in this
paper on random initial conditions, $C_\qv(0,0)=1$, corresponding to a
quench at time $t=0$ from equilibrium at infinite temperature. In this
case the $\qv$-integral in the last equation is just $\hat{f}(s)$, so
that
\be
\hat{g}(s) = \frac{\hat{f}(s)}{1-2T\hat{f}(s)}
\ee
The asymptotics of the corresponding $g(t)$ are well known; see
\eg~\cite{GodLuc00b,CorLipZan03}. For $T$ above the critical
temperature $\Tc$, which is given by
\be
\Tc^{-1}=2 \hat{f}(0) = \dq \frac{1}{\omega}
\label{Tc}
\ee
there is a pole in $\hat{g}(s)$ at $s=2z\eql$. Here $z\eql$ is found from the
condition $2T\hat{f}(2z\eql)=1$ or
\be
\dq \frac{T}{z\eql+\omega} = 1
\label{zeql_cond}
\ee
The presence of this pole tells us that $g(t)\sim \exp(2z\eql t)$ for
long times, 
implying that the Lagrange multiplier $z(t)$ approaches $z\eql$ for
$t\to\infty$. Correspondingly, the condition\eq{zeql_cond} is just the
spherical constraint at equilibrium, bearing in mind that
$C\eqlup_\qv(t,t)=T/(z\eql+\omega)$ from\eq{dotSq}. Because
$\omega\approx q^2$
for small $q=|\qv|$, the phase space factor in the $\qv$-integrals is
$(dq) \sim d\omega\, \omega^{d/2-1}$ for small $q$ or $\omega$. This
shows that $\Tc$ as given by\eq{Tc} vanishes as $d\to 2$ from above;
consequently we will always restrict ourselves to dimensions $d$ above this
lower critical dimension.

At criticality ($T=\Tc$), $z\eql$ vanishes, and $g(t)$ therefore no
longer grows exponentially; instead one
finds~\cite{GodLuc00b,CorLipZan03}
\be
g(t) \sim t^{-\n}, \quad \n = 
\left\{\begin{array}{lll}
(4-d)/2 & \mbox{for} & 2<d<4\\
0 & \mbox{for} & d>4
\end{array}\right.
\label{g_asympt}
\ee
It is this case, of a quench to the critical temperature, that we will
concentrate on throughout most of this paper. This is because here the
FDR has the most interesting behaviour. 

We note briefly that, in principle, $\dq$ should be written as
$(1/N)\sum_\qv$, with the sum running over all $\qv$ whose components
are integers in the range $-L/2\ldots -1,0,1,\ldots L/2-1$ (assuming
$L$ is even) multiplied by an overall factor $2\pi/L$; there are $N$
such $\qv$. When considering continuous functions of $\qv$ this sum
can be replaced by the integral $\dq$, and this will almost always be
the case in our analysis. Exceptions are situations with a nonzero
magnetization, where the wavevector $\qv=\zv$ is special and has to be
treated separately. This is relevant in equilibrium below $\Tc$, which
we discuss briefly in Sec.~\ref{sec:energy_FDT_eq}, and for
non-equilibrium dynamics starting from magnetized initial states
(Sec.~\ref{sec:magnetized}).

\subsection{Long-time scaling of $C_\qv$}

It will be useful later to have a simplified long-time expression for
$C_\qv(\tw,\tw)$ for the case of a critical quench. At zero wavevector
one has
\be
C_\zv(\tw,\tw) = \frac{1}{g(\tw)}\left(1 + 2T \int_0^{\tw} dt'\,
g(t')\right) \approx \frac{2T\tw}{1-\n}
\label{C0}
\ee
where the last approximation is based on\eq{g_asympt} and is valid for large
$\tw$. For nonzero $\qv$, on the other hand,
\be
C_\qv(\tw,\tw) = \frac{1}{g(\tw)}\left(e^{-2\omega \tw}
+ 2T \int_0^{\tw} dt'\, g(t') e^{-2\omega (\tw-t')}\right)
\approx \frac{T}{\omega}
\label{Cq}
\ee
which is as expected since all nonzero Fourier modes eventually
equilibrate. The crossover between the two limits takes place when
$\omega\tw \sim 1$, or $q\sim \tw^{-1/2}$; physically this represents
the growth of the time-dependent correlation length as $\sim \tw^{1/2}$.
We therefore introduce the scaling variable $\omt=\omega \tw$:
\be
\frac{C_\qv(\tw,\tw)}{C_\zv(\tw,\tw)}
= \frac{e^{-2\omt} + 2T\omega^{-1} \int_0^{\omt} dy\,
g(y/\omega) e^{-2(\omt-y)}}
{1+ 2T\omega^{-1} \int_0^{\omt} dy\,
g(y/\omega)}
\ee
Now keep $\omt$ constant and let $\tw\to\infty$, \ie\ $\omega\to 0$.
Then $g(y/\omega) \sim (y/\omega)^{-\n}$ and the second terms dominate
in denumerator and nominator to give
\be
\frac{C_\qv(\tw,\tw)}{C_\zv(\tw,\tw)} = (1-\n)\int_0^1 dy\,y^{-\n}
e^{-2\omt(1-y)}
\label{Cratio_scaling}
\ee
Combining\eq{Cratio_scaling} with\eq{C0} then gives the desired
long-time scaling form
\be
C_\qv(\tw,\tw) = \frac{T}{\omega}\sc{C}(\omega \tw), \quad
\sc{C}(\omt) = 2\omt\int_0^1 dy\,y^{-\n} e^{-2\omt(1-y)}
\label{C_scaling}
\ee
For $d>4$ ($\n=0$) this simplifies to $\sc{C}(\omt)=1-e^{-2\omt}$. As
the derivation shows, eqs.\eq{Cratio_scaling} and\eq{C_scaling} are
valid whenever $\tw\gg 1$, even for $\omega=\order(1)$. The latter
case corresponds to $\omt\to\infty$ and gives $\sc{C}(\omt)=1$, which
is indeed consistent with\eq{Cq}.

For quantities such as $C_\qv(\tw,\tw)$ that depend only on a single
time variable, what is meant by the long-time limit is
unambiguous. For two-time quantities like $C_\qv(t,\tw)$ we use the
following terminology: the {\em long-time} limit refers to the regime
$t\gg 1$ and $\tw\gg 1$ but without any restriction on $t-\tw$, which
in particular is allowed to be short, i.e.\ of $\order(1)$. The {\em aging}
regime indicates more specifically the limit $\tw\to\infty$ at fixed
ratio $x=t/\tw>1$, which implies that also $t-\tw$ is large, of
$\order(\tw)$. Occasionally we specialize further to the regime of
{\em well-separated} times, which corresponds to $t\gg\tw\gg1$, i.e.\
the asymptotic behaviour of the aging limit for $x\gg 1$.

To illustrate the difference, consider which wavevectors dominate the
integral $\dq C_\qv(t,\tw)$. In the long-time limit at equal times
$t=\tw$, the scaling $C_\qv(\tw,\tw) \sim 1/\omega$ for $\omega\gg
1/\tw$ combined with $(dq) \sim d\omega\, \omega^{d/2-1}$ for small
$\omega$ shows that the integral is divergent at the upper end of the
frequency regime $\omega=\order(\tw^{-1})$ for all $d>2$; in other
words, it is always dominated by values of $\omega$ (and therefore
$q$) of $\order(1)$. This remains true for two-time correlations, as
long as $t-\tw=\order(1)$. In the aging limit, however, we have
$t-\tw=\order(\tw)\gg 1$ and the exponential factor from $R_\qv$
in\eq{C_twotime} then ensures that only values of
$\omega<(t-\tw)^{-1}=\order(\tw^{-1})$ have to be considered in the
integral.

\section{Fluctuation-dissipation relations for finite-range
observables}
\label{sec:finite-range}

In this section we consider FD relations for observables that probe
correlations over a lengthscale that can be much larger than the
lattice spacing, but remains much smaller than the system size. The
latter can then be taken to infinity independently, so that the
$\order(\rn)$-fluctuations of the Lagrange multiplier $z$ become irrelevant.
We begin by considering briefly spin observables, and then discuss
bond observables in some more detail.

\subsection{Spin observables}
\label{sec:spin_observables}

Since all observables that are linear in the spins can be written as
superpositions of the Fourier modes $S_\qv$, the basic ingredient for
understanding the FD behaviour is the FDR for the
latter. Using\eq{Cqtt_deriv}, this follows after a couple of lines as
($C'\equiv \partial_{\tw} C$)
\be
X_\qv(t,\tw) = \frac{TR_\qv(t,\tw)}{C'_\qv(t,\tw)} =
T\left[2T - \left(\omega + \frac{g'(\tw)}{2g(\tw)}\right)
C_\qv(\tw,\tw) \right]^{-1}
\label{Xq}
\ee
This is {\em independent} of the later time $t$, a feature that is
commonly observed in simple non-equilibrium models~\cite{CriRit03}.

The fluctuating magnetization is simply $S_\zv/N$, so setting
$\qv=\zv$ in\eq{Xq} gives directly the FDR for the magnetization
\be
X_\zv(t,\tw) = T\left[2T - \frac{g'(\tw)}{2g^2(\tw)}
\left(1+2T \int_0^{\tw} dt'\, g(t')\right) \right]^{-1}
\ee
As $\tw$ increases this converges on an $\order(1)$ timescale to the
limit-FDR
\be
X^\infty = \frac{T}{2T+(\n/2)[2T/(1-\n)]} = \frac{1-\n}{2-\n} = 
\left\{\begin{array}{lll}
1/2 & (d>4)\\
1-2/d & (d<4)
\end{array}\right.
\label{X_baseline}
\ee
which is identical to the value obtained from the local
magnetization~\cite{GodLuc00b} as one would expect on general
grounds. Without working out the susceptibility explicitly, it is
clear from the $t$-independence of $X_\zv(t,\tw)$ and its fast
convergence to $X^\infty$ that the limiting FD plot is a straight
line. Both of these observations are exactly as in the Ising model in
$d=1$~\cite{MayBerGarSol03}. Simulations have shown that also in the
$d=2$ Ising case the local--global correspondence holds for spin
observables; the limiting FD plot is numerically indistinguishable
from a straight line, though renormalization group arguments suggest
that it should deviate
slightly~\cite{MayBerGarSol04,CalGam02,CalGam05}.

We should clarify that the Gaussian theory above applies directly not
to the FDR for $S_\zv$ but to the one for $S_\qv$ with $q\ll
\tw^{-1/2}$ but $q\gg L^{-1}$, where $L=N^{1/d}$ is the linear system
size. The corresponding physical observable is a ``block''
magnetization, i.e.\ the average of the spins within a block of size
$\ell\sim 1/q$ much larger than the time-dependent correlation length
$\sim\tw^{1/2}$ but still small compared to the overall system
size. For $\qv=\zv$, i.e.\ $\ell=L$, one would in principle need to
account for the non-Gaussian fluctuations. However, it turns out that
these are negligible as long as the system is not magnetized on
average (see Sec.~\ref{sec:magnetized}), so that the above results
remain correct even for the global magnetization itself.

More generally, the FDR for any finite-range spin observable can be
expressed as a superposition of those for the Fourier modes; this can
be seen by arguments parallelling those in the $d=1$ Ising
case~\cite{MayBerGarSol03}. As there, one can then show that the
asymptotic FDR that is approached for well-separated times $t\gg\tw\gg
1$ is dominated by the 
contribution from $\qv=\zv$, and hence identical to $X^\infty$
calculated above~\cite{CalGam05}. At equal times, on the other hand,
equilibrated 
modes with $q=\order(1)$ dominate and give $X=1$. The crossover
between these two regimes takes place when $t-\tw=\order(\tw)$ and
follows (by superposition) from the corresponding crossover at
$q=\order(\tw^{-1})$ in the Fourier mode FDRs. From\eq{C_scaling}
and\eq{Xq} the latter can be expressed as
\be
\fl X_\qv(t,\tw) = \sc{X}(\omega\tw), \quad \sc{X}^{-1}(\omt) = 2-(2\omt-\n)
\int_0^1 dy\,y^{-\n} e^{-2\omt(1-y)}
\label{Xq_scaling}
\ee
in the long-time limit, providing the expected interpolation between
$X=X^\infty=1/[2+\n/(1-\n)]$ for $\omt\to 0$ and $X=1$ for
$\omt\to\infty$.

\subsection{Bond observables}
\label{sec:bond_observables}

Next we consider bond energy observables, $\half(S_i-S_j)^2$, where
$i$ and $j$ are n.n.\ sites. Since all variables are Gaussian, the
connected correlations follow by Wick's theorem. For the correlation
of bond energies one gets
\bea
\fl C_{ij,kl}(t,\tw) &=& 2\frac{1}{4}\lav
[S_i(t)-S_j(t)][S_k(\tw)-S_l(\tw)]\rav^2
= \half\left[C_{ik}-C_{il}-C_{jk}+C_{jl}\right]^2
\label{C_bond}
\eea
where time arguments have been
left implicit. For the local case $(i,j)=(k,l)$, this simplifies to
\be
C_{ij,ij} = 2[1-C_{ij}]^2 
\label{C_bond_local}
\ee
which tends to a nonzero constant for $t=\tw\to\infty$ since $C_{ij}$
then approaches its equilibrium value, which is $<1$.

Next we turn to the response function. In general, if one perturbs the
Hamiltonian $H$ by $-hB\delta(t-\tw)$, then the equation of motion for
$S_i$ acquires an extra term $h(\partial B/\partial S_i)\delta(t-\tw)$. So
the perturbation in $S_i$ is
\be
\delta S_i(t) = h\sum_j R_{ij}(t,\tw)\frac{\partial B}{\partial S_j}(\tw)
\ee
Thus the perturbation of an observable $A$ is
\be \delta A(t) = h\sum_{ij} \frac{\partial A}{\partial S_i}(t)
R_{ij}(t,\tw)\frac{\partial B}{\partial S_j}(\tw) \ee
giving the response function~\cite{CalGam04}
\be
R_{AB}(t,\tw) = \sum_{ij} R_{ij}(t,\tw) \lav \frac{\partial
A}{\partial S_i}(t)\frac{\partial B}{\partial S_j}(\tw)\rav
\ee
For $A=\half(S_i-S_j)^2$, $B=\half(S_k-S_l)^2$ this yields
\be
R_{ij,kl}(t,\tw) =
[R_{ik}-R_{il}-R_{jk}+R_{jl}][C_{ik}-C_{il}-C_{jk}+C_{jl}]
\ee

We now analyse the scaling of correlation, response and the resulting
FDR. In terms of $C_\qv(t,\tw)$, the bond correlation\eq{C_bond} is
\be
\fl C_{ij,kl}(t,\tw) = \half\left[\dq C_\qv
\left(
 e^{i\qv\cdot(\rv_i-\rv_k)}
-e^{i\qv\cdot(\rv_i-\rv_l)}
-e^{i\qv\cdot(\rv_j-\rv_k)}
+e^{i\qv\cdot(\rv_j-\rv_l)}
\right)
\right]^2
\label{C_bond_ijkl}
\ee
We can take out a factor $\exp(i\qv\cdot\dr)$ from all the
exponentials, where $\dr = \half (\rv_i+\rv_j)+\half(\rv_k-\rv_l)$ is
the distance vector between the bond midpoints. In the remaining
exponentials, $\qv$ is multiplied by vectors with lengths of order
unity. 

Now assume $t-\tw\gg 1$. As explained above, integrals of two-time
quantities over $\qv$ are then dominated by the small-$q$ regime,
$q^2\approx \omega<(t-\tw)^{-1}$. We can therefore Taylor expand in
$\qv$ and get, using the equivalence of the $d$ lattice directions,
\be
C_{ij,kl}
= \half\left[d^{-1}(\rv_i-\rv_j)\cdot
(\rv_k-\rv_l) \dq C_\qv q^2 e^{i\qv\cdot \dr}\right]^2
\label{C_bond_Taylor}
\ee
Similarly one finds for the response
\be
\fl R_{ij,kl}
 = \left[d^{-1}(\rv_i-\rv_j)\cdot
(\rv_k-\rv_l)\right]^2 \dq R_\qv q^2 e^{i\qv\cdot \dr}
\dq C_\qv q^2 e^{i\qv\cdot \dr}
\ee
For the {\em local} bond-bond correlation and response one sets
$\dr=\zv$ and has $(\rv_i-\rv_j)\cdot (\rv_i-\rv_j)=1$, which gives
for the FDR
\be
X\bond\loc(t,\tw) = \frac{\dq TR_\qv q^2} {\dq C'_\qv q^2}
= \frac{\dq R_\qv q^2} {\dq X^{-1}_\qv R_\qv q^2 }
\label{X_local_bond}
\ee
So $1/X\bond\loc(t,\tw)$ can be thought of as an average of
$X_\qv^{-1}(t,\tw)$ over $\qv$, with the weight $R_\qv(t,\tw)
q^{-2}$. The factor $R_\qv$ ensures that significant contributions
come only from wavevectors $\qv$ up to length $q\sim
(t-\tw)^{-1/2}$, i.e.\ up to $\omega(t-\tw)\approx 1$. Thus, when
$t-\tw\ll \tw$, the result is dominated by the regime $\omega \tw\gg
1$, where $X_\qv = 1$. For $t-\tw\gg \tw$, meanwhile, one only gets
contributions from $\omega\tw\ll 1$, where $X_\qv =X^\infty$. So the
FDR\eq{X_local_bond} for local bond observables is a scaling function
interpolating between $1$ and $X^\infty$, with the same $X^\infty$ as
for the magnetization. Explicitly one has, using\eq{Rq} and changing
integration variable from $\qv$ to $\omt=\omega\tw$,
\be
X\bond\loc(t,\tw) = \frac{\int_0^{\infty} d\omt\, \omt^{d/2} e^{-(x-1)\omt}}
{\int_0^{\infty} d\omt\, \omt^{d/2} e^{-(x-1)\omt} \sc{X}^{-1}(\omt)}
\label{X_bond_local}
\ee
with $x=t/\tw$ and $\sc{X}$ the scaling form\eq{Xq_scaling} of
$X_\qv$. To find the shape of the FD plot, recall that the equal-time
value of the local-bond correlation\eq{C_bond_local} is a constant in
the long-time limit. For $t-\tw\gg 1$, on the other hand,
eq.\eq{C_bond_Taylor} shows that $C\bond\loc$ scales as
\bea
C\bond\loc(t,\tw) 
&\sim& \frac{g(\tw)}{g(t)}\left[\int d\omega\,
\omega^{d/2-1}\frac{T}{\omega}\sc{C}(\omega\tw) e^{-\omega(t-\tw)}
\omega\right]^2 \\
&\sim& \frac{g(\tw)}{g(t)} \tw^{-d}\left[\int d\omt\,
\omt^{d/2-1}\sc{C}(\omt) e^{-\omt(t-\tw)/\tw}\right]^2
\eea
Since $\sc{C}(\omt)\to 1$ for $\omt\to\infty$, the $\omt$-integral would be
divergent without the exponential cutoff and scales as
$[(t-\tw)/\tw]^{-d/2}$ for $t-\tw\ll \tw$, so that $C\bond\loc(t,\tw) \sim
(t-\tw)^{-d}$ in this regime. The regime $t-\tw>\tw$ where
$X\bond\loc(t,\tw)\neq 1$ is therefore compressed into the region where
$C\bond\loc$ is of order $\tw^{-d}$, so that the long-time limit of the
FD plot is a straight line with equilibrium slope. Qualitatively one
thus has the same behaviour as for local bond observables in the Ising
model~\cite{MayBerGarSol03}.

Next consider {\em long-range} bond observables, where we sum $(ij)$ and $(kl)$
over all bonds. The same proviso as above for the magnetization
applies here, i.e.\ by applying the Gaussian theory we are effectively
considering the bond energies averaged over a block that is large but has to
remain nonetheless small compared to the system size. One can show that the
resulting equal-time correlation again approaches a constant value for
$\tw\to\infty$. (This follows because for large $\dr$, one can use the
small $q$-expansion\eq{C_bond_Taylor} even for equal times. From
$C_\qv(\tw,\tw) \approx T/q^2$ one gets $C(\dr;\tw,\tw)\sim
|\dr|^{2-d}$ for large $\dr$ and so $\dq C_\qv q^2 e^{i\qv\cdot\dr}
\sim \nabla^2 |\dr|^{2-d} \sim |\dr|^{-d}$. The square $|\dr|^{-2d}$
then yields a convergent sum over $\dr$.) So we focus directly on the
regime $t-\tw\gg 1$, where the expansion\eq{C_bond_Taylor} is again
valid. Keeping the bond $(ij)$ fixed, the scalar product
$(\rv_i-\rv_j)\cdot (\rv_k-\rv_l)$ means that only bonds $(kl)$
parallel to $(ij)$ contribute, so that the sum over $(kl)$ becomes a
sum over $\dr$, running over all lattice vectors. (For non-parallel
bonds, $\dr$ could also assume other values not corresponding to
lattice vectors.) The sum over $(ij)$ then just gives an overall
factor of $Nd$. Normalizing by $N$, the block bond correlation function is
\be
C\bond\bl(t,\tw) = \frac{1}{2d}\frac{1}{N}\sum_{\dr} \left[
\dq C_\qv q^2 e^{i\qv\cdot \dr}\right]^2
= \frac{1}{2d} \dq C^2_{\qv} q^4
\label{C_block_bond}
\ee
and similar arguments give for the (normalized) response
\be
R\bond\bl(t,\tw) = \frac{1}{d} \dq R_\qv C_\qv q^4
\ee
so that the FDR becomes
\be
X\bond\bl(t,\tw) = \frac{\dq TR_\qv C_\qv q^4} {\dq
C'_\qv C_\qv q^4} 
= \frac{\dq R_\qv C_\qv q^4} {\dq X^{-1}_\qv R_\qv C_\qv q^4 }
\label{X_bond_long-range}
\ee
Again, this is the inverse of a weighted average of $X^{-1}_\qv$, now
with weight $R_\qv C_\qv q^4$. The same arguments as
for\eq{X_local_bond} then show that $X\bl(t,\tw)$ scales with
$x=t/\tw$ and interpolates between $X=1$ for $x=1$ and
$X=X^\infty$ for $x\to\infty$. The value of the
correlator\eq{C_block_bond} decays from $\order(1)$ at $t=\tw$ to
$\order(\tw^{-d/2})$ at the point $x-1\approx 1$ where aging effects
appear. While this is larger than for the local bond observables, it
still decreases to zero for $\tw\to\infty$, so that the limiting FD
plot is again of pseudo-equilibrium form. This is different to the
case of the Ising model, where the global bond observables give
nontrivial limiting FD plots~\cite{MayBerGarSol03}.

In more detail, the scaling of the block bond correlator\eq{C_block_bond}
in the aging regime $t-\tw\gg 1$ is
\bea
C\bond\bl(t,\tw)
&\sim& \frac{g(\tw)}{g(t)}
\int\!d\omega\, \omega^{d/2-1} \frac{\Tc^2}{\omega^2}
\sc{C}^2(\omega\tw) e^{-2\omega(t-\tw)}\omega^2
\\
&\sim& \left(\frac{t}{\tw}\right)^{\n} \tw^{-d/2}
\int\!d\omt \,\omt^{d/2-1} \sc{C}^2(\omt) e^{-2(x-1)\omt}
\eea
The integral scales as $(x-1)^{-d/2}$ for $x\approx 1$, so
$C\bond\bl(t,\tw) \sim (t-\tw)^{-d/2}$ there. For $x\gg 1$, on the other
hand, the integral becomes $\sim x^{-(d+4)/2}$, so $C\bond\bl(t,\tw)\sim 
\tw^{-\n+2} t^{\n-(d+4)/2}$. Explicitly, in this $t\gg\tw$ regime,
$C\bond\bl \sim \tw^2 t^{-(d+4)/2}$ for $d>4$, and $C\bond\bl \sim
\tw^{d/2} t^{-d}$ for $d<4$. The response function scales in the same
way as $\partial_{\tw}C\bond\bl$, because $X$ is everywhere of order unity.

\subsection{Product observables}
\label{sec:product_observables}

Instead of the bond observables $\half(S_i-S_j)^2$ we could consider
the spin products $A=S_i S_j$, $B=S_k S_l$. The correlations are then
\bea
\fl C_{ij,kl}(t,\tw) 
&=& \lav S_i(t)S_j(t)S_k(\tw)S_l(\tw)\rav - \lav
S_i(t)S_j(t)\rav \lav S_k(\tw)S_l(\tw)\rav\\
\fl &=& C_{ik}(t,\tw)C_{jl}(t,\tw) + 
C_{il}(t,\tw)C_{jk}(t,\tw)
\label{C_prod}
\eea
The local equal-time correlation function $C_{ij,ij}(\tw,\tw)$ thus
approaches $2$ for $\tw\to\infty$. The corresponding response function
is
\be
R_{ij,kl}(t,\tw) = R_{ik}C_{jl} + R_{il}C_{jk} + R_{jk}C_{il} + R_{jl}C_{ik}
\ee
In the {\em local} case, one can replace all
functions by local ones in the aging regime -- there are no
cancellations leading to extra 
factors of $q^2$ as was the case for bond observables, compare
e.g.\eq{C_bond_ijkl} and\eq{C_bond_Taylor} -- so that the FDR
\be
X\product\loc(t,\tw) = \frac{4\Tc R_{ii}C_{ii}}{4C'_{ii}C_{ii}}
= \frac{\Tc R_{ii}}{C'_{ii}}
\label{X_prod_local}
\ee
becomes identical to the one for the spin autocorrelation and
response. In particular, one again gets the same $X^\infty$.

For the global (block) case, we can write
\bea
\fl C\product\bl(t,\tw) &=& \sum_{(ij),(kl)}C_{ij,kl}(t,\tw) 
\\
\fl &=& \frac{1}{4}\sum_{ijkl} \nnn_{ij} \nnn_{kl} \left[
C_{ik}(t,\tw)C_{jl}(t,\tw) + 
C_{il}(t,\tw)C_{jk}(t,\tw)\right]
\\
\fl &=& \frac{1}{2}\sum_{ijkl} \nnn_{ij} \nnn_{kl}
C_{ik}(t,\tw)C_{jl}(t,\tw)
= \half \dq \nnn^2_\qv C_\qv^2(t,\tw)
\label{C_global_product}
\eea
where $\nnn_{ij}=1$ if $i$ and $j$ are nearest neighbours and 0
otherwise, and $\nnn_\qv$ is its Fourier transform. For the response
one has similarly
\be
R\product\bl(t,\tw) = \dq \nnn^2_\qv R_\qv(t,\tw) C_\qv(t,\tw)
\ee
In the aging regime, where $t-\tw\gg 1$, the integrals are dominated
by small $q$, where $\nnn_\qv$ can be approximated by the constant
$\nnn_\zv=2d$. This cancels from the FDR, which becomes
\be
X\product\bl(t,\tw) = \frac{\dq \Tc R_\qv C_\qv}{\dq C'_\qv C_\qv}
\label{Xbl_large_d}
\ee
This is the inverse of the average of $X_\qv^{-1}$ with weight $R_\qv
C_\qv$. Again, this is a scaling function of $t/\tw$ interpolating
between $1$ and the same $X^\infty$ as for spin observables.

The scaling of the block product correlation function\eq{C_global_product}
itself is a little more complicated than for the bond observables and
depends on dimensionality. Focussing again on $t-\tw\gg 1$ one has
$C\product\bl(t,\tw) \approx 2d^2 \dq C_\qv^2(t,\tw)$. The integral defines
the function $\CC(t,\tw)$ discussed in Sec.~\ref{sec:noneq_large_d}
for $d>4$ and Sec.~\ref{sec:noneq_small_d} for $d<4$. In the former
case, one has from (\ref{CC_decomp}--\ref{CC_scaling}) that
$\CC(t,\tw) = \CC\eql(t-\tw)\sc{\CC}(t/\tw)$ where $\CC\eql(t-\tw)\sim
(t-\tw)^{(4-d)/2}$ asymptotically; this equilibrium contribution
governs the behaviour of $C\product\bl(t,\tw)$ for $t-\tw\ll \tw$.  Where
aging effects appear ($t-\tw \sim \tw$), $C\product\bl\sim \tw^{(4-d)/2}$
and so one gets a limiting pseudo-equilibrium FD plot. In the regime
of well-separated times $x\gg 1$, the scaling function $\sc{\CC}(x)$ decays as
$x^{-2}$ so that $C\product\bl(t,\tw) \sim \tw^{2} t^{-d/2}$. These
scalings, though not the overall magnitude of $C\product\bl$, are the same as
for the energy correlation function $C_E$ in\eq{C4_final} below: both
functions are proportional to $\CC(t,\tw)$ in the aging regime.

In the opposite case $d<4$, the equal-time value $\CC(t,t)$ (and
therefore $C\product\bl(t,t)$) diverges as $t^{(4-d)/2}$,
see\eq{CC_equal_time}. The normalized correlator $\CC(t,\tw)/\CC(t,t)$
is a scaling function $\G(x)$ of $x=t/\tw$, implying that the
normalized FD plot will approach a nontrivial limit form, with
asymptotic slope $X^\infty$ as shown above. Quantitatively, because
$\G(x)\sim x^{-d/2}$ for $x\gg 1$, one has $C\product\bl(t,\tw) \sim
t^{(4-d)/2}(\tw/t)^{d/2} \sim \tw^{d/2} t^{2-d}$ for $t\gg\tw$.

\section{Correlation and response for global observables}
\label{sec:setup}

We now ask what happens if we go from block observables to
truly global ones, which reflect properties averaged over the entire
system; the total energy is an important example. We anticipate that
here non-Gaussian fluctuations are important. Indeed, the results
above show that this must be case. Otherwise we could directly extend
the Gaussian theory results from block to global observables, with no
change to correlation and response functions. The global {\em bond}
observable is just the energy. Using the spherical constraint, this
can be written as
\be
E=\sum_{(ij)}\half(S_i-S_j)^2=N-\sum_{(ij)}S_i S_j
\ee
and so is identical to the global {\em spin product} observable, up to
a trivial additive constant and sign. So the global bond and product
observables must have identical correlation and response functions;
but we saw above that this requirement is not satisfied by the
Gaussian theory. Thus, non-Gaussian fluctuations are essential to get
correct results for global observables.

Physically, the origin of the distinction between block observables
and global ones is the effective infinite-range interaction induced by the
spherical constraint. In a model with short-range interactions, block
observables will show identical behaviour to global ones whenever the
block size $\ell$ is larger than any correlation length in the system,
whether or not $\ell\ll L$: the behaviour of any large subsystem is
equivalent to that of the system as a whole. In the spherical model,
the infinite-range interaction breaks this connection, and global
correlation and response functions cannot be deduced from those for
block observables.

\subsection{Non-Gaussian fluctuations}

To make progress, we need to return to the original equation of
motion\eq{eqn_motion}. This can be written as
\be
\partial_t S_i = -\sum_{j} \Om_{ij}S_j + \xi_i - (z(t)+\rn\dz)S_i
\label{eqn_motion2}
\ee
where the notation emphasizes that the fluctuating contribution to the
Lagrange multiplier is of $\order(N^{-1/2})$. The latter induces
non-Gaussian fluctuations in the $S_i$ of the same order. This shows
quantitatively why the Gaussian theory works for block observables: as
long as one considers correlations of a number of spins that is $\ll
N$, fluctuations of $\order(\rn)$ can be neglected. For global
observables, on the other hand, we require the correlations of all $N$
spins and the Gaussian approximation then becomes invalid.

To account systematically for non-Gaussian effects we represent the
spins as $S_i = \X_i + N^{-1/2} \Y_i$, where $\X_i$ gives the limiting
result for $N\to\infty$, which has purely Gaussian statistics, and
$N^{-1/2}\Y_i$ is a leading-order fluctuation correction which will be
non-Gaussian. Inserting this decomposition into\eq{eqn_motion2} and
collecting terms of $\order(1)$ and $\order(\rn)$ gives $\partial_t \X_i =
-\Om_{ij} \X_j -z(t) \X_i + \xi_i$ as expected; to lighten the
notation we use the summation convention for repeated indices from now
on. For the non-Gaussian corrections one gets the equation of motion
\be
\partial_t \Y_i = -\Om_{ij} \Y_j -z(t)\Y_i - \dz\, \X_i
\ee
with solution
\be
\Y_i(t) = R_{ij}(t,0)\Y_j(0) - \int_0^t dt'\, R_{ij}(t,t')\X_j(t')\dz(t')
\label{yt}
\ee
The properties of $\dz(t')$ can now be determined from the requirement
that, due to the spherical constraint, $N^{-1}\sum_i S^2_i(t)=1$ at
all times. Inserting $S_i = \X_i + N^{-1/2} \Y_i$ and expanding to the
leading order in $\rn$ gives the condition
\be
\frac{1}{N} \sum_i \X_i(t)\Y_i(t) = -\half \rn \sum_i (\X_i^2(t)-1)
\equiv -\half \del(t)
\label{norm_condition}
\ee
where the last equality defines $\del(t)$, a fluctuating quantity of
$\order(1)$ that describes the (normalized) fluctuations of the
squared length of the Gaussian spin vector $\X_i$. At $t=0$, the
condition\eq{norm_condition} is solved 
to leading order by setting $\Y_i(0) = -\half \del(0)\X_i(0)$, since
$(1/N) \sum_i \X_i^2(0)=1+\order(\rn)$. With this assignment, and
setting $a(t)=2\dz(t)+\del(0)\delta(t)$, eq.\eq{yt} reads
\be
\Y_i(t) = - \half \int dt'\, R_{ij}(t,t')\X_j(t')a(t')
\label{yt2}
\ee
We have left the integral limits unspecified here: the factor $R_{ij}$
automatically enforces $t'<t$, and we use the convention $a(t')=0$ for
$t'<0$. The spherical constraint condition\eq{norm_condition} then becomes
\be
 \int dt'\, \frac{1}{N} \X_i(t)R_{ij}(t,t')\X_j(t')a(t')
= \del(t)
\label{norm_condition2}
\ee
Now, up to fluctuations of $\order(\rn)$ which are negligible to
leading order (even if they are correlated with $a(t')$), we can
replace $(1/N) \X_i(t)R_{ij}(t,t')\X_j(t')$ by its average
\be
\fl \p(t,t') \equiv \frac{1}{N} \lav\X_i(t)R_{ij}(t,t')\X_j(t')\rav = 
\frac{1}{N} R_{ij}(t,t')C_{ij}(t,t')
= \dq R_\qv(t,t')C_\qv(t,t')
\label{p_def}
\ee
If we then define the inverse operator, $L$, of $\p$ via
\be
\int dt'\, \p(t,t')L(t',\tw) = \delta(t-\tw)
\label{pinv_def}
\ee
for $\tw\geq 0$, then the solution to\eq{norm_condition2} is
\be
a(t) = \int dt'\, L(t,t')\del(t')
\ee
where for consistency we adopt the convention $\del(t')=0$ for
$t'<0$. With\eq{yt2} we then get an explicit expression for the
non-Gaussian $\order(\rn)$-corrections to the spins,
\be
\Y_i(t) = - \half \int dt' dt''\, R_{ij}(t,t')\X_j(t')L(t',t'')\del(t'')
\label{yt3}
\ee
in terms of the properties of the uncorrected Gaussian spins $\X_i$.

\subsection{The functions $\p$ and $L$}
\label{sec:KandL}

Before proceeding, we analyse the properties of $\p$ and
$L$. From\eq{p_def}, $\p(t,t')$ vanishes for $t<t'$ while its
limit value for $t\to t'^{+}$ is
$(1/N)\delta_{ij}C_{ij}(t,t)=(1/N)C_{ii}(t,t)=1$. Inserting\eq{Rq},
\eq{Cqtt_deriv} and\eq{C_twotime} into\eq{p_def}, one also finds that
the equal-time slope has the simple value $\partial_{t'} \p
\left.(t,t')\right|_{t=t'^{+}}=2T$. From these properties and the
definition\eq{pinv_def} it follows that
\be
\fl L(t,t') = \delta'(t-t') + \Lone(t,t'), \qquad
\Lone(t,t') = 2T\delta(t-t') - \Ltwo(t,t')
\label{pinv_structure}
\ee
where $\Ltwo(t,t')$ vanishes for $t<t'$ and jumps to a finite value at
$t=t'^{+}$; otherwise it is smooth and, as we will later see,
positive. The structure of\eq{pinv_structure} can be easily verified
e.g.\ for the limit of equilibrium at high temperature $T$, where
$z\eql=T$ and all $\omega$ can be neglected compared to $z\eql$. One then 
has $\p(t,t')=\exp(-2T(t-t'))$ and the inverse\eq{pinv_def} can be
calculated by Laplace transform. Since
the Laplace transform of $\p(t,t')$ is
$\hat\p(s)=1/(s+2T)$ this gives $\hat{L}(s)=s+2T$, which
corresponds to\eq{pinv_structure} with $\Ltwo\equiv 0$.

We next determine the long-time forms of $\p$ and $\Ltwo$ for
quenches to criticality.  In both cases it is useful to factor out the
equilibrium contribution. For $\p$ this is, from\eq{p_def} and
using\eq{Rq} and\eq{C_twotime},
\be
\p\eql(t-\tw) = \dq e^{-2\omega(t-\tw)} \frac{\Tc}{\omega}
\label{p_eql}
\ee
Apart from the factor of 2 in the time argument, this is just the
(critical) equilibrium spin-spin autocorrelation function. One can
also write $\p(t)=2\Tc\int_t^\infty dt'\, f(t')$ from\eq{f_def} and this
shows that $\p\eql(t)\sim t^{(2-d)/2}$ for large time differences. The
ratio $\p(t,\tw)/\p\eql(t-\tw)$ will show deviations from 1 when aging
effects appear, i.e.\ when $t-\tw\sim \tw$. The form of these
deviations can be worked out by using the scaling form\eq{C_scaling} of
$C_\qv(\tw,\tw)$ and recalling that only the small $q$-regime
contributes, where $(dq)\sim d\omega\, \omega^{d/2-1}$. Changing
integration variable to $\omt=\omega \tw$ gives
\bea
\p(t,\tw)&=&\p\eql(t-\tw)\sc{\p}(t/\tw)
\label{p_scaling_general}
\\
\sc{\p}(x) &=& x^\n \frac{\int d\omt \, \omt ^{(d-4)/2} e^{-2(x-1)\omt
} \sc{C}(\omt)}
{\int d\omt \, \omt ^{(d-4)/2} e^{-2(x-1)\omt }}
\label{p_scaling}
\eea
where the first factor in\eq{p_scaling} arises from the two factors of
$[g(\tw)/g(t)]^{1/2}$ contributed by $R_\qv(t,\tw)$ and
$C_\qv(t,\tw)$, respectively. By construction, $\sc{\p}(x)$ should
approach $1$ for $x\to 1$; indeed, in this
limit the $\omt$-integrals are dominated by large values of $\omt\sim 1/(x-1)$,
for which $\sc{C}=1$. The decay for large $x$ follows from
$\sc{C}(\omt)\sim \omt$ for small $\omt$ as $\sc{\p}(x)\sim
x^{\n-1}$. Explicitly, one finds by using\eq{C_scaling} and carrying
out the $\omt$-integrals that 
\be
\sc{\p}(x) = \frac{d-2}{2}(x-1)^{(d-2)/2}x^{\n} \int_0^1 dy\,
y^{-\n}(x-y)^{-d/2}
\label{p_scaling_fn_y_integral}
\ee
For $d>4$, where $\n=0$, this gives
\be
\sc{\p}(x)=1-\left(\frac{x-1}{x}\right)^{(d-2)/2} \qquad (d>4)
\label{FKd_gt_4}
\ee
while for $d<4$ the required indefinite integral is $[(d-2)/2]\int dy\,
y^{(d-4)/2}(x-y)^{-d/2} = -x^{-1}(x/y-1)^{(2-d)/2}$ and one gets simply
\be
\sc{\p}(x)
=x^{(2-d)/2} \qquad (d<4)
\label{FKd_lt_4}
\ee

Next we determine $\Ltwo$. Combining\eq{pinv_def}
and\eq{pinv_structure}, the defining equation is
\be
\int dt'\, \p(t,t')\Ltwo(t',\tw) = 2T \p(t,\tw) -
\partial_{\tw}\p(t,\tw)
\label{ptwo_def}
\ee
Again it makes sense to extract the equilibrium part of $\Ltwo$. This
is defined by
\be
\int dt'\, \p\eql(t-t')\Ltwo\eql(t'-\tw) = 2T \p\eql(\dt) + \p\eql'(\dt)
\label{ptwo_eql_condition}
\ee
where $\dt=t-\tw$. Solving by Laplace transform gives
\be
\hat\Ltwo\eql(s) =
\frac{2T\hat\p\eql(s)+(s\hat\p\eql(s)-1)}{\hat\p\eql(s)} = 
s + 2T - \frac{1}{\hat\p\eql(s)}
\label{ptwo_eql}
\ee
where from\eq{p_eql}, at criticality,
\be
\hat\p\eql(s) = \Tc \dq \frac{1}{\omega(s+2\omega)}
\label{p_eql_LT}
\ee
The leading small-$s$ behaviour of this is
$\hat\p\eql(0)-\hat\p\eql(s) \sim s^{(d-4)/2}$ for $d>4$ (plus, for
$d>6$, additional analytic terms of integer order in $s$ which are
irrelevant for us). For $d<4$, on the other hand, $\hat\p\eql(s) \sim
s^{(d-4)/2}$ is divergent for $s\to 0$. Inserting these scalings
into\eq{ptwo_eql} and inverting the Laplace transform gives for the
asymptotic behaviour of $\Ltwo\eql$
\be
\Ltwo\eql(t) \sim \left\{ \begin{array}{ll}
t^{(2-d)/2} & \mbox{($d>4$)} \\
t^{(d-6)/2} & \mbox{($d<4$)}
\end{array}\right.
\label{ptwo_eql_asymptotics}
\ee
It will be important below that, for $d>4$, $\p\eql(t)$ and
$\Ltwo\eql(t)$ both decay asymptotically as $t^{(2-d)/2}$. The ratio
between them can be worked out from\eq{ptwo_eql}, by expanding
for small $s$ as $1/\hat\p\eql(s) \approx 1/[\hat\p\eql(0)-cs^{(d-4)/2}] =
1/\hat\p\eql(0) + cs^{(d-4)/2}/\hat\p^2\eql(0)$ where $c$ is some
constant; comparing with $\hat\p\eql(s) \approx
\hat\p\eql(0)-cs^{(d-4)/2}$ gives
\be
\Ltwo\eql(t)=\p\eql(t)/\hat\p\eql^2(0)
\label{ptwo_p_link}
\ee
for large time differences $t$.

The integral of $\Ltwo\eql(t)$ over all times follows
from\eq{ptwo_eql} as 
\be
\fl \hat\Ltwo\eql(0) = \int_0^\infty dt\,\Ltwo\eql(t) = \left\{ 
\begin{array}{ll}
2\Tc - \hat\p\eql^{-1}(0) = 2\Tc[1-1/\dq (\Tc/\omega)^{2}]
& \mbox{($d>4$)}\\
2\Tc & \mbox{($d<4$)}
\end{array}
\right.
\label{L2hat0}
\ee
Using the fact that $\dq (\Tc/\omega)=1$, one has $\dq
(\Tc/\omega)^2>1$ so that $\hat\Ltwo\eql(0)$ is positive independently
of $d$. This is consistent with the intuition that, with the sign as
chosen in\eq{pinv_structure}, the function $\Ltwo$ is positive.

With the equilibrium part of $\Ltwo$ determined we make a long-time
scaling ansatz for $\Ltwo$,
\be
\Ltwo(t,\tw) = \Ltwo\eql(t-\tw)\sc{L}(t/\tw)
\label{L_scaling}
\ee
so that\eq{ptwo_def} becomes
\bea
\fl\lefteqn{\int dt'\,
\p\eql(t-t')\Ltwo\eql(t'-\tw) \sc{\p}(t/t')
\sc{L}(t'/\tw) = }
\nonumber\\
&=&2\Tc \p\eql(\dt)\sc{\p}(x)+
\p\eql'(\dt)\sc{\p}(x)+
\frac{t}{\tw^2}\p\eql(\dt)\sc{\p}'(x)
\label{ptwo_scaling_cond}
\eea
where $\dt=t-\tw$ and $x=t/\tw$ as before. We now take the aging limit
of large 
$\tw$ with $\dt=\order(\tw)$ to determine $\sc{L}$. The second and
third terms on the r.h.s.\ are then smaller by factors of order
$1/\tw$ than the first, and can be neglected to leading order. The
second term on the r.h.s.\ of\eq{ptwo_eql_condition} is likewise
subdominant, and this can be used to rewrite the dominant first term
on the r.h.s.\ of\eq{ptwo_scaling_cond}, giving
\bea
\fl \sc{\p}\left(x\right)
&=& \frac{\int dt'\,\p\eql(t-t')\Ltwo\eql(t'-\tw)
          \sc{\p}(t/t')\sc{L}(t'/\tw)}
         {\int dt'\, \p\eql(t-t')\Ltwo\eql(t'-\tw)}
\label{unscaled_ptwo}
\eea
We consider first $d>4$. Then both the functions $\p\eql(\dt)$ and
$\Ltwo\eql(\dt)$ have finite integrals $\hat\p\eql(0)$ and
$\hat\Ltwo\eql(0)$, respectively, over $\dt=0\ldots\infty$. In the
aging limit, the factors 
$\p\eql(t-t')$ and $\Ltwo\eql(t'-\tw)$ therefore act to concentrate
the mass of the integrals appearing in\eq{unscaled_ptwo} around
$t'=\tw$ and $t'=t$. This can be seen more formally by changing to
$y=t'/\tw$ as the integration variable and taking $\tw$
large. Then the factors $\p\eql(\tw(x-y))$ and $\Ltwo\eql(\tw(y-1))$
produce singularities $\sim(x-y)^{(2-d)/2}$ for $y\to x$ and $\sim
(y-1)^{(2-d)/2}$ for $y\to 1$, respectively, and because these are
non-integrable they dominate the integral for $\tw\to\infty$. All
other factors in the integrals are slowly varying near the relevant
endpoints and can be replaced by their value there. In the aging limit we
can therefore write\eq{unscaled_ptwo} as
\be
\fl \sc{\p}(x)
= \frac{\hat\p\eql(0)\Ltwo\eql(\dt) \sc{\p}(1)\sc{L}(x)+
\p\eql(\dt)\hat\Ltwo\eql(0) \sc{\p}(x)\sc{L}(1)}
{\hat\p\eql(0)\Ltwo\eql(\dt)+\p\eql(\dt)\hat\Ltwo\eql(0)}
\ee
Eq.\eq{ptwo_p_link} tells us that the $\dt$-dependent factors cancel,
giving together with\eq{L2hat0} and $\sc{\p}(1)=\sc{L}(1)=1$
\be
\fl \sc{\p}(x)
= \frac{\hat\p\eql^{-1}(0) \sc{L}(x)+\hat\Ltwo\eql(0) \sc{\p}(x)}
{\hat\p\eql^{-1}(0)+\hat\Ltwo\eql(0)}
= \frac{\sc{L}(x)+[2\Tc\hat\p\eql(0)-1]\sc{\p}(x)}
{2\Tc\hat\p\eql(0)}
\label{ptwo_simple_scaling}
\ee
In $d>4$, where $\hat\p\eql(0)$ is finite, we therefore have the
simple result that the scaling functions for $\p$ and $\Ltwo$ are
identical,
\be
\sc{L}(x) = \sc{\p}(x)
\label{same_sc_fn}
\ee
But in the limit $d\to 4$ from above, $\hat\p\eql(0)$ diverges
and\eq{ptwo_simple_scaling} gives no information about
$\sc{L}$. For $d<4$ a different approach is therefore needed
to determine $\sc{L}$. One subtracts from\eq{ptwo_scaling_cond} 
the first and second terms on its r.h.s., using\eq{ptwo_eql_condition} to
rewrite them as an integral and changing integration variable from
$t'$ to $y=t'/\tw$. This gives
\bea
\fl\lefteqn{
\tw\int_1^x dy\,
\p\eql(\tw(x-y))\Ltwo\eql(\tw(y-1))\left[\sc{\p}\left({x}/{y}\right)
\sc{L}\left(y\right)-\sc{\p}(x)\right]
= }\nonumber\\
&=&\frac{t}{\tw^2}\p\eql(\dt)\sc{\p}'(x)
\label{L_condition_almost}
\eea
For $y\to x$, $\p\eql(\tw(x-y))$ contributes a singularity $\sim
(x-y)^{(2-d)/2}$ which is integrable in $d<4$. For $y\to 1$, the terms
in square brackets vanish as $\sim y-1$ since $\sc{L}(y)$ is
smooth at $y=1$ as we will see below, in the sense that
$\sc{L}'(1)$ is finite. These terms combine with the $\sim
(y-1)^{(d-6)/2}$ from $\Ltwo\eql$ to give an integrable $\sim
(y-1)^{(d-4)/2}$. The contributions from the short time behaviour of
$\p\eql$ and $\Ltwo\eql$ are therefore unimportant in the aging
limit and we can replace these functions by their power-law
asymptotes. Up to overall $d$-dependent numerical factors the
condition\eq{L_condition_almost} then becomes
\bea
\fl\lefteqn{
\tw^{-1}\int_1^x dy\,
(x-y)^{(2-d)/2}(y-1)^{(d-6)/2}\left[\sc{\p}\left({x}/{y}\right)
\sc{L}\left(y\right)-\sc{\p}(x)\right]
= }\nonumber\\
&=&\frac{t}{\tw^2}\dt^{(2-d)/2}\sc{\p}'(x)
\eea
In the aging limit $\dt$ scales as $\tw$, and so the l.h.s.\ of this
equation ($\sim\tw^{-1}$) becomes large compared to the r.h.s.\ ($\sim
\tw^{-d/2})$ unless the $y$-integral vanishes. The required
condition for $\sc{L}$ is therefore
\be
\fl
\int_1^x dy\,
(x-y)^{(2-d)/2}(y-1)^{(d-6)/2}\left[\sc{\p}\left({x}/{y}\right)
\sc{L}\left(y\right)-\sc{\p}(x)\right] = 0
\label{pinv_cond}
\ee
This is in principle an integral equation for
$\sc{L}$. Fortunately, however, the solution is the naive
extension of\eq{same_sc_fn} to $d<4$: with $\sc{\p}(x)=x^{(2-d)/2}$
from\eq{FKd_lt_4} one sees that for
$\sc{L}(x)=\sc{\p}(x)=x^{(2-d)/2}$ the square bracket
in\eq{pinv_cond} vanishes identically. The identity\eq{same_sc_fn}
therefore holds both for $d<4$ and for $d>4$.

In summary, we have determined long-time scaling forms for $\p$ and $L$ for
quenches to criticality. For $\p$, the result is\eq{p_scaling_general}
with\eq{p_eql} and (\ref{FKd_gt_4},\ref{FKd_lt_4}); for $L$, we
have\eq{L_scaling} with\eqq{ptwo_eql}{ptwo_eql_asymptotics}
and\eq{same_sc_fn}. Combined with\eq{yt3}, this fully determines the
leading non-Gaussian corrections to the spherical model dynamics (at
long times, and after a quench to criticality from a random initial state).

\section{General expressions for energy correlation and response}
\label{sec:energy_general}

In this section we derive general expressions for the two-time correlation and
response functions of the energy, taking into account non-Gaussian
fluctuations. The results will be valid for arbitrary quenches since
we will leave $\p$ and $L$ unspecified.

\subsection{Energy correlation function}
\label{sec:energy_corr}

We can write the energy\eq{H_spherical} as $H=\half S_i\Om_{ij}
S_j$. Inserting $S_i = \X_i + N^{-1/2} \Y_i$, the energy correlation
function (normalized by $N$) is to leading order
\be
\fl C_E(t,\tw)=\frac{1}{4N}\lav \left.\left(
\X_i \Om_{ij} \X_j + 2\rn \X_i \Om_{ij} \Y_j
\right)\right|_{t}
\left.\left(
\X_k \Om_{kl} \X_l + 2\rn \X_k \Om_{kl} \Y_l
\right)\right|_{\tw} \rav'
\label{CE_basic}
\ee
Using\eq{yt3}, all quantities involved can be expressed in terms of
the Gaussian variables $\X_i$ so that the average can be performed using
Wick's theorem, i.e.\ by taking products of all possible pairings. We use
the prime on the average to indicate the connected correlation
function. This just means that in the Wick expansion all terms not
containing any pairings of a variable at $t$ with one at $\tw$ have to
be discarded, since these terms give the disconnected contribution $\lav
\left. (\ldots)\right|_{t}\rav \lav\left. (\ldots)\right|_{\tw}
\rav$. Multiplying out\eq{CE_basic} one obtains four
contributions. The first one is
\be
\fl 4C_E^{(1)} = \frac{1}{N}\Om_{ij}\Om_{kl}\lav
\X_i(t)\X_j(t)\X_k(\tw)\X_l(\tw)\rav' =
\frac{2}{N}\Om_{ij}\Om_{kl}C_{jk}(t,\tw)C_{il}(t,\tw)
\label{C1_first}
\ee
To eliminate one of the factors of $\Om$, note from\eq{Rq} that
$(\partial_t+z(t)) R_\qv(t,\tw) = -\omega R_\qv(t,\tw)$ for
$t>\tw$. In real space, this reads
\be
\left(\partial_t+z(t)\right)R_{ik}(t,\tw) =
-\Om_{ij}R_{jk}(t,\tw) = 
-R_{ij}(t,\tw)\Om_{jk}
\label{Om_elimination}
\ee
and from\eq{C_twotime} an exactly analogous relation holds for
$C_{ij}(t,\tw)$. Thus
\bea
\fl 4C_E^{(1)}
&=&
- \frac{2}{N}[(\partial_t+z(t))C_{ik}(t,\tw)]\Om_{kl}C_{il}(t,\tw)
\\
\fl &=&
- \frac{1}{N}(\partial_t+2z(t))C_{ik}(t,\tw)\Om_{kl}C_{il}(t,\tw) \equiv
- (\partial_t+2z(t))\COC(t,\tw)
\label{C1}
\eea
The last equality defines $\COC$, which is just the normalized trace
of the product of the matrices $C_{ik}(t,\tw)$, $\Om_{kl}$ and
$C_{il}=C_{li}$; in Fourier space, $\COC(t,\tw) = \dq \omega
C_\qv^2(t,\tw)$.

\begin{figure}
\centerline{\includegraphics[width=7.0cm,clip=true]{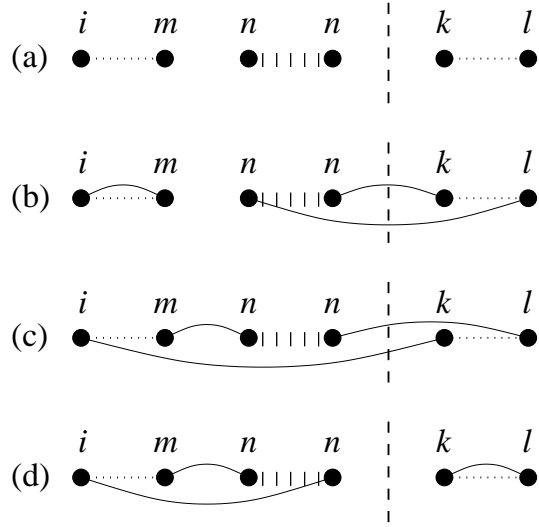}}
\caption{Illustration of Wick pairings for $C_E^{(2)}$. (a) The six
Gaussian spins that need to be paired in\eq{C2_raw} are indicated by
circles with site labels. Spins arising from the expansion of $H(t)$
and $H(\tw)$ are to the left and right of the vertical dashed line,
respectively; any pairing contributing to the connected correlator
must have links across this line. Dotted lines connect spins that
are already coupled in space: $s_i$ and $s_m$ via the factor
$\Omega_{ij}R_{jm}$, and $s_k$ and $s_l$ via $\Omega_{kl}$. The vertical
lines between $s_n$ and $s_n$ indicate that pairings which couple these
spins are not allowed. (b) The solid lines show the only Wick pairing
that contributes to
leading order in $1/N$: it gives two independent groups of spins. (c)
only gives one group and so is subleading. (d) has two groups but is
excluded from the connected correlator because there are no pairs
across the dashed line.
\label{fig:C2}
}
\end{figure}
The second contribution to $C_E$ is
\bea
4C_E^{(2)} &=& \frac{2}{N^{3/2}}\Om_{ij}\Om_{kl}\lav
\X_i(t)\Y_j(t)\X_k(\tw)\X_l(\tw)\rav' \\
&=&
-\frac{1}{N} \int dt' dt''\, \Om_{ij}\Om_{kl} R_{jm}(t,t')L(t',t'')
\nonumber\\
& &\times \lav \X_i(t)\X_m(t')\rn\del(t'') \X_k(\tw)\X_l(\tw) \rav'
\eea
where we have inserted\eq{yt3}. Writing $\rn\del(t'')$ explicitly this
becomes
\bea
4C_E^{(2)} &=&
-\frac{1}{N^2} \int dt' dt''\, 
\Om_{ij}\Om_{kl}R_{jm}(t,t')L(t',t'') \nonumber\\
& &\times
\lav \X_i(t)\X_m(t')\left[\X_n^2(t'')-\lav
\X_n^2(t'')\rav\right] \X_k(\tw)\X_l(\tw) \rav'
\label{C2_raw}
\eea
We now need to perform the Wick expansion of the average. The
subtraction $\X_n^2-\lav \X_n^2\rav$ means that all terms which would
pair $\X_n$ with $\X_n$ are excluded; $\X_k$ and $\X_l$ also cannot be
paired because we are considering the connected correlation. We can
reduce the number of pairings further by considering that we need to
get an overall result of $\order(1)$. The index $j$ does not need to
be considered further: after summing over $j$, $\Om_{ij}R_{jm}$ is
some translationally invariant function of the distance vector
between spins $s_i$ and $s_m$. If the remaining indices $i,k,l,m,n$
where unrestricted, then together with the $1/N^2$ prefactor we would
maximally get an $\order(N^3)$ result. Each of the factors $\Om_{ij}R_{jm}$
and $\Om_{kl}$ couples two indices and so reduces the order of
the result by $1/N$. Having already got two such couplings outside
the average, we can only ``afford'' one extra coupling from the Wick
pairings to get a contribution of $\order(1)$. After some reflection
one sees that this only leaves the pairing $[im][kn][ln]$: $[im]$
introduces no further coupling beyond $\Om_{ij}R_{jm}$, and
$[kn][ln]$ gives only one further coupling beyond
$\Om_{kl}$. Alternatively, we can think of this pairing as having the
indices $i,m$ and $k,l,n,n$ in two independent groups; each group
gives a factor of $N$ and this just cancels the $1/N^2$ prefactor. All
other allowed Wick pairings give smaller terms, as illustrated
graphically in Fig.~\ref{fig:C2}a,b. For example,
$[ik][mn][ln]$ together with $\Om_{ij}R_{jm}$ and $\Om_{kl}$
couples {\em all} indices into a single group and thus gives a term of
only $\order(1/N)$ (Fig.~\ref{fig:C2}c). The pairing
$[in][mn][kl]$ would give two 
independent groups and thus an $\order(1)$ term, but is excluded
because $k$ and $l$ cannot be paired in the connected correlator
(Fig.~\ref{fig:C2}d). Bearing in mind that our 
dominant pairing has a symmetry factor of 2 because the $\X_n$'s in
$[kn][ln]$ can be swapped we have thus finally
\bea
\fl 4C_E^{(2)} &=&
-2 \int dt' dt''\, 
\frac{1}{N} \Om_{ij} R_{jm}(t,t') C_{im}(t,t') L(t',t'') 
\nonumber\\
\fl & &\times
\frac{1}{N}\Om_{kl}C_{nk}(t'',\tw)C_{nl}(t'',\tw) \\
\fl &=&
\int dt' dt''\, 
[(\partial_t+2z(t))\p(t,t')-\delta(t-t')]
 L(t',t'') \COC(t'',\tw)
\eea
In going to the last line we have exploited\eq{Om_elimination} to
eliminate $\Om_{ij}$. Since $t'$ is an integration variable, we have
also been careful here to subtract off with the $\delta(t-t')$ the
spurious contribution which the $\partial_t$ applied to the step
discontinuity in $\p(t,t')$ would otherwise give. The $t'$-integration
can now be carried out using\eq{pinv_def} and we get
\bea
4C_E^{(2)} 
&=&
(\partial_t+2z(t)) \COC(t,\tw)-\int dt' L(t,t') \COC(t',\tw)
\\
&=&
2z(t) \COC(t,\tw)-\int dt' \Lone(t,t') \COC(t',\tw)
\label{C2}
\eea
Note that we could use the same trick here as
in\eqq{Om_elimination}{C1} to write
$\COC(t',\tw)=-(\half\partial_{t'}+z(t'))\CC(t',\tw)$, with
$\CC(t,\tw)$ defined in the obvious manner as $\CC = N^{-1}
C_{ik}C_{ki} = \dq \omega C_\qv^2$.  However, this reduction from
$\COC$ to $\CC$ applies only for $t'>\tw$, while for $t'<\tw$ one
would need to take a time derivative w.r.t.\ $\tw$ instead of $t'$.
This case distinction would make evaluation of\eq{C2} awkward, so we 
retain $\COC$ here and below.

The third contribution to $C_E$ is obtained by simply swapping the
roles of $t$ and $\tw$ in\eq{C2} and remembering that $\COC$ is
symmetric in its time arguments,
\bea
4C_E^{(3)} &=&
\frac{2}{N^{3/2}}\Om_{ij}\Om_{kl}\lav \X_i(t)\X_j(t)\X_k(\tw)\Y_l(\tw)\rav' \\
&=&
2z(\tw) \COC(t,\tw)-\int dt' \Lone(\tw,t') \COC(t,t')
\label{C3}
\eea

\begin{figure}
\centerline{\includegraphics[width=9.0cm,clip=true]{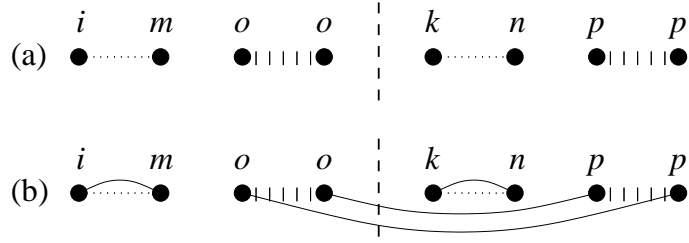}}
\caption{Wick pairings for $C_E^{(4)}$. (a) represents the constraints
for the possible pairings in\eq{C4_raw}. (b) is the only pairing that
contributes to leading order, forming three independent groups of
spins.
\label{fig:C4}
}
\end{figure}
The fourth and last contribution to $C_E$ is, using again\eq{yt3} and
writing out $\del(t'')$ and $\del(\tw'')$
\bea
\fl 4C_E^{(4)} &=& \frac{4}{N^2}\Om_{ij}\Om_{kl}\lav
\X_i(t)\Y_j(t)\X_k(\tw)\Y_l(\tw)\rav' \\
\fl &=&
\frac{1}{N^3} \int dt' dt'' d\tw' d\tw''\, 
\Om_{ij}\Om_{kl}
R_{jm}(t,t')L(t',t'')R_{ln}(\tw,\tw')L(\tw',\tw'')
\nonumber\\
\fl & & \times \lav 
\X_i(t)  \X_m(t')  \X_o^2(t'') 
\X_k(\tw)\X_n(\tw')\X_p^2(\tw'')
\rav'
\label{C4_raw}
\eea
where it is understood that, because of subtractions which we have not
written explicitly, pairings of $\X_o$ with itself and of $\X_p$ with
itself are to be excluded. The only pairing that gives an overall
$\order(1)$ contribution turns out to be $[im][kn][op][op]$ and gives
the required 3 independent index groups ($i,m$; $k,n$; $o,o,p,p$) to
cancel the $1/N^3$ prefactor; see Fig.~\ref{fig:C4}. With the symmetry
factor 2 for the possible swap of the $[op][op]$ pairings one gets
\bea
\fl 4C_E^{(4)} 
&=&
2 \int dt' dt'' d\tw' d\tw''\, 
\frac{1}{N}\Om_{ij}R_{jm}(t,t')C_{im}(t,t')
\nonumber\\
\fl & & \times
\frac{1}{N}\Om_{kl}R_{ln}(\tw,\tw')C_{kn}(\tw,\tw')
L(t',t'')L(\tw',\tw'')\frac{1}{N} C_{op}^2(t'',\tw'')\\
\fl &=&
\frac{1}{2} \int dt' dt'' d\tw' d\tw''\, 
[(\partial_t+2z(t))\p(t,t')-\delta(t-t')]L(t',t'')
\nonumber\\
\fl & & \times
[(\partial_{\tw}+2z(\tw))\p(\tw,\tw')-\delta(\tw-\tw')]
L(\tw',\tw'') \CC(t'',\tw'')
\eea
Using\eq{pinv_def} one can carry out
two of the time integrations to get
\bea
4C_E^{(4)} 
&=&
\frac{1}{2} \int dt' d\tw' \, 
[2z(t)\delta(t-t')-\Lone(t,t')]
\nonumber\\
& & \times
[2z(\tw)\delta(\tw-\tw')-\Lone(\tw,\tw')]
\CC(t',\tw')
\label{C4}
\eea

\subsection{Energy response function}

To find the energy response, consider perturbing the Hamiltonian by a
term $-h\delta(t-\tw)H=-(h/2)\delta(t-\tw)S_i\Om_{ij}S_j$, where $h$
is the field conjugate to the energy. The equation of motion in the
presence of the perturbation is therefore
\be
\partial_t S_i = - \Om_{ij}S_j - (z(t)+h\dz(t))S_i +h\delta(t-\tw)\Om_{ij}S_j
\ee
where $h \dz$ is now the change in the Lagrange multiplier induced by
the perturbation. The fluctuating component of $z$ of $\order(\rn)$
is in principle still present, but negligible compared to $h\dz$ for
field strengths that are $\order(N^0)$. Inserting a corresponding
expansion of the spins, $S_i=\X_i+h\Y_i$, gives for $\X_i$ the unperturbed
equation of motion and for the perturbed component $\Y_i$
\be
\partial_t\Y_i = -\Om_{ij}\Y_j - z(t)\Y_i -\dz(t)\X_i +\delta(t-\tw)\Om_{ij}\X_j
\label{dot_ri}
\ee
The solution of this is
\be
\Y_i(t) = R_{ik}(t,\tw)\Om_{kl}\X_l(\tw) - \int_{\tw}^t dt'\,
R_{ik}(t,t') \dz(t') \X_k(t')
\label{yt_resp}
\ee
One now needs to determine $\dz$. This can be done by noting that the
normalized length of $S_i$ is
\be
\frac{1}{N}\sum_i S_i^2 = \frac{1}{N}\sum_i \X_i^2 + 2h \frac{1}{N}
\sum_i \X_i \Y_i + \order(h^2)
\ee
The change to first order in $h$ must vanish to preserve the spherical
constraint, giving the condition $(1/N)\sum_i \lav \X_i \Y_i\rav=0$
or, using\eq{yt_resp},
\be
\frac{1}{N} R_{ik}(t,\tw)\Om_{kl}C_{il}(t,\tw) =
\int_{\tw}^t dt'\, \frac{1}{N} R_{ik}(t,t')C_{ik}(t,t')\dz(t') 
\ee
In the integrand one recognizes the definition\eq{p_def} of $\p$, so
that one can write the solution of this as
\be
\dz(t) = \int dt'\, L(t,t') \ROC(t',\tw) 
\ee
with obvious notation for $\ROC$. 

Now we can find the change in the energy, $1/(2N) \lav S_i\Om_{ij}
S_j\rav$, which is given by $(h/N) \lav \Y_i \Om_{ij}\X_j\rav$ to linear order
in $h$. Dividing by $h$ and using\eq{yt_resp} then gives the energy
response function
\bea
R_E(t,\tw) &=& \frac{1}{N}\lav  \Y_i \Om_{ij}\X_j\rav\\
&=& 
\frac{1}{N} R_{ik}(t,\tw)\Om_{kl}\Om_{ij}C_{jl}(t,\tw)
\nonumber\\
& &
{} - {}\int dt' dt''\,
\ROC(t,t')L(t',t'') \ROC(t'',\tw)
\label{RE_general}
\eea
One can eliminate one of the time integrals by
using\eq{Om_elimination}, being careful to remove the unwanted
contribution from differentiating the step discontinuity in
$\ROC(t,\tw)$. Using 
also\eq{pinv_def} and\eq{pinv_structure} then gives
\bea
\fl 2R_E
&=& 
(-\partial_t-2z(t))\ROC(t,\tw) + \delta(t-\tw)\ROC(\tw^{+},\tw)
\nonumber\\
\fl & &{} - {}\int dt' dt''\,
[(-\partial_t-2z(t))\p(t,t')+\delta(t-t')]L(t',t'') \ROC(t'',\tw)
\\
\fl &=& 
- \int dt' L(t,t') \ROC(t',\tw)
+ \delta(t-\tw)\ROC(\tw^{+},\tw)
\label{RE}
\eea

\subsection{Equilibrium}
\label{sec:energy_FDT_eq}

Above, we derived general expressions for the energy two-time
correlation and response, in terms of the known correlation and
response functions for the Gaussian spins which in turn determine $\p$ and
$L$. Before looking at non-equilibrium, we consider briefly the
equilibrium situation; even here the results for the dynamics are new
as far as we are aware.

For the response function one uses that at equilibrium $R\Om^aC(t) =
\dq e^{-2(z\eql+\omega)t}\omega^a [T/(z\eql+\omega)]$ for
$a=1,2$. Here we have retained a possible nonzero equilibrium value $z\eql$ of
the Lagrange multiplier, to include the
case of equilibrium at $T\neq \Tc$. Inserting into\eq{RE_general} and
taking LTs gives
\bea
\hat R\Eql(s) &=& \dq
\frac{T\omega^2}{(z\eql+\omega)[s+2(z\eql+\omega)]} \nonumber\\
& & {} - {} \frac{1}{\hat\p\eql(s)} \left(\dq
\frac{T\omega}{(z\eql+\omega)[s+2(z\eql+\omega)]}\right)^2
\eea
where $\hat\p\eql$ is generalized from\eq{p_eql_LT} to
\be
\hat\p\eql(s) = T \dq \frac{1}{(z\eql+\omega)[s+2(z\eql+\omega)]}
\label{phat_eql}
\ee
Using the spherical constraint condition
$T\dq (z\eql+\omega)^{-1} = 1$ one then shows, after a few lines of
algebra, that
\be
\hat R\Eql(s)=\frac{1}{4}\left(s + 2T -
\frac{1}{\hat\p\eql(s)}\right) = \frac{1}{4}\hat\Ltwo\eql(s)
\label{R_Eql}
\ee
Remarkably, therefore, the equilibrium energy response function
$R\Eql(t)$ is directly proportional to the inverse kernel
$\hat\Ltwo\eql(t)$. Its asymptotics are then given
by\eq{ptwo_eql_asymptotics}. The long-time equilibrium susceptibility
which encodes the response to a
step change in the field is, using $\hat\p\eql(0)=\dq T/[2(z\eql+\omega)^2]$
and the generalization of\eq{L2hat0} to $T\neq \Tc$,
\be
\fl \chi\Eql=\int_0^\infty dt\, R\Eql(t) = \frac{1}{4}\hat
\Ltwo\eql(0) = \frac{T}{2}\left(1-\frac{1}{\dq [T/(z\eql+\omega)]^2}\right)
\label{chi_Eql}
\ee
It is easily shown that this is consistent with the known result for
the temperature dependence of the equilibrium energy, $E=\langle
H\rangle = \dq \half\omega[T/(z\eql+\omega)]$: one confirms $\chi\Eql
= T\,dE/dT$ as it should be. The factor $T$ arises because our field
$h$ is introduced via $H\to H-hH=(1-h)H$ and so corresponds to a
temperature change of $T/(1-h)-T=hT$ to linear order in $h$. The
inclusion of subleading non-Gaussian fluctuations is crucial for achieving this
consistency; as discussed at the beginning of Sec.~\ref{sec:setup},
the Gaussian theory does not even give the same answers for the
fluctuations of $H$ in its two representations as a bond or spin
product observable. The same phenomenon occurs in a purely static
calculation of the energy fluctuations and response~\cite{Joyce72}.

The temperature-dependence of the susceptibility\eq{chi_Eql} deserves
some comment. As $T$ approaches $\Tc$ from above, one has $z\eql\to
0$. For $d<4$, the denominator of the second term in\eq{chi_Eql} then
diverges, and $\chi\Eql/T$ smoothly approaches the value $1/2$ and
remains constant for $T<\Tc$. For $d>4$, the denominator has a finite
limit for $z\eql\to 0$. This produces the well-known discontinuity in
$\chi\Eql/T$ at $T=\Tc$, since for $T<\Tc$ the second term
in\eq{chi_Eql} again does not contribute. To see this explicitly, one
notes that for $T<\Tc$ the $\omega=0$ term in the spherical constraint
condition, with its weight $1/N$, has to be treated separately:
\be
T\left(\dq \frac{1}{z\eql+\omega} + \frac{1}{Nz\eql} \right)=1
\ee
For $z\eql\to 0$ (on the scale $\order(N^0)$) the first integral is
$\dq 1/\omega = 1/\Tc$ and so $z\eql = (1/N)T\Tc/(\Tc-T)$. This then
gives for the denominator integral in\eq{chi_Eql}
\be
\dq \frac{T^2}{(z\eql+\omega)^2} + \frac{T^2}{Nz\eql^2} \approx
N\frac{(\Tc-T)^2}{\Tc^2}
\ee
which diverges for $N\to\infty$ at $T<\Tc$ as claimed.

An interesting question is how the discontinuity at $T=\Tc$ in $d>4$
of the susceptibility $\chi\Eql$, i.e.\ the {\em integrated} response, 
is related to the temperature variation of the {\em time-dependent}
$R\Eql(t)$. In\eq{phat_eql} one notes that, for $T<\Tc$,
$\hat\p\eql(s)$ acquires a distinct contribution from the $\omega=0$
mode
\bea
\hat\p\eql(s) &=& 
\dq \frac{T}{\omega(s+2\omega)} +
\frac{1}{N}\frac{T}{z\eql(s+2z\eql)}\\
& = & \dq \frac{T}{\omega(s+2\omega)} + \frac{\Tc-T}{\Tc s}
\eea
where in the second line we have neglected $z\eql=\order(1/N)$
against $s$. From\eq{R_Eql} one sees that the additional contribution
to $\hat\p\eql(s)$ produces an extra pole in $\hat R\Eql(s)$, which
for small $(\Tc-T)/\Tc$ is located at $s=1/t_0$ with
$t_0=[T\Tc/4(\Tc-T)]\dq \omega^{-2}$. Transforming to the time domain,
one finds that this gives an extra contribution of $\sim 1/t_0
\exp(-t/t_0)$ to $R\Eql(t)$. Crucially, this has a finite integral
even in the limit $T\to \Tc$, where $t_0$ diverges, and the appearance
of this term causes the discontinuity in $\chi\Eql$ at
$T=\Tc$. Translating these results to the time-dependent susceptibility
\be
\chi\Eql(t) = \int_0^t dt' R\Eql(t')
\label{chi_Eql_t}
\ee
one has that, for fixed finite $t$, $\chi\Eql(t)$ depends smoothly on
temperature around $\Tc$. For large $t$, $\chi\Eql(t)$ approaches a
plateau value equal to the equilibrium susceptibility at $T=\Tc^{+}$;
the approach to this plateau is as $\sim t^{(4-d)/2}$. For $T<\Tc$,
however, $\chi\Eql(t)$ eventually increases further on a diverging
timescale $t\sim t_0 \sim 1/(\Tc-T)$ to approach a higher plateau
value given by the susceptibility at $T=\Tc^{-}$. By FDT (see below)
the energy correlation function correspondingly shows a power-law
decay to a plateau for $t\ll t_0$, from which it falls to zero only
for $t>t_0$.

A final check on our results is that in equilibrium the energy
response and correlation function should be related by the usual
FDT. This is indeed the case. One combines the results\eq{C1},
\eq{C2}, \eq{C3}, \eq{C4} for the constituent parts of the correlation
function and decouples all the convolution integrals using temporal
Fourier transforms. It is easiest to do this starting from expressions
where all factors of $\Om$ have been preserved, e.g.\ for $C^{(1)}_E$
one uses\eq{C1_first} rather than\eq{C1}. After some straightforward
but lengthy algebra one then indeed finds the Fourier domain version of
equilibrium FDT, $\int_{-\infty}^{\infty} dt\,C\Eql(t)e^{-i\nu t}=T[\hat
R\Eql(-i\nu)-\hat R\Eql(i\nu)]/(i\nu)$.

\section{Energy correlation and response: Non-equilibrium, $d>4$}
\label{sec:noneq_large_d}

We now evaluate the behaviour of the energy correlation and response
for the out-of-equilibrium dynamics after a quench to criticality. For
the correlation function, we need the long-time behaviour of
\be
\COC(t,\tw) = \dq \frac{\Tc^2}{\omega^2}\sc{C}^2(\omega \tw)
\frac{g(\tw)}{g(t)} e^{-2\omega(t-\tw)}\omega
\ee
At equilibrium, where $\sc{C}=1$ and $g(t)=$ const., this gives
$\Tc\p\eql(t-\tw)$. In the non-equilibrium case one gets a correction
factor which becomes relevant in the aging regime, where  $t-\tw\sim
\tw\gg 1$:
\bea
\COC(t,\tw) &=& \Tc\p\eql(t-\tw)\sc{\COC}\left(t/\tw\right)
\label{COC}
\\
\sc{COC}(x) &=& \left(\frac{t}{\tw}\right)^{\n}
\frac{\int d\omt \, \omt ^{(d-4)/2} e^{-2(x-1)\omt } \sc{C}^2(\omt)}
{\int d\omt \, \omt ^{(d-4)/2} e^{-2(x-1)\omt }}
\eea
This is valid for all dimensions $d>2$, but in the rest of this
section we focus on the case $d>4$, where $\n=0$. The regime $2<d<4$
is more complicated and treated separately in the next section.

Before assembling the four parts of our expression for $C_E(t,\tw)$,
it is useful to have a guide on what to expect overall. As discussed
above, the equilibrium energy fluctuations remain finite at
criticality. One therefore expects that, in the short-time regime
$t-\tw\sim\order(1)$, $C_E(t,\tw)$ will decay as in
equilibrium. From\eq{ptwo_eql_asymptotics}, \eq{R_Eql} and FDT it
follows that this decay becomes a power law, $\sim(t-\tw)^{(4-d)/2}$
as $t-\tw$ becomes large. Aging effects should then appear when
$t-\tw\sim \tw$ and manifest themselves through a scaling function of
$t/\tw$. Overall, one expects in the aging regime
\be
C_E(t,\tw)=(t-\tw)^{(4-d)/2}\times[\mbox{scaling function of
}t/\tw]
\label{CE_expect}
\ee
and this is indeed what we will find. Consider now $C_E^{(1)}$, for
which we obtained in\eq{C1} that $4C_E^{(1)} = -(\partial_t+2z(t))\COC(t,\tw)$.
Now from\eq{g_def} and the fact that $g(t)\to$ const.\ for
$t\to\infty$ and $d>4$ it follows that $z(t)=g'(t)/[2g(t)]$
decreases more quickly than $1/t$. Using also\eq{COC}, we see that the
$z(t)$ term is negligible for large times and that $4C_E^{(1)}$
decays as $(t-\tw)^{-d/2}$ times an aging correction. This is
negligible compared to\eq{CE_expect}, so we do not need to analyse
this contribution further.

For $C_E^{(2)}$, we have from\eq{C2}
\be
\fl 4C_E^{(2)} = 2z(t) \COC(t,\tw)-2\Tc \COC(t,\tw) + \int dt' \Ltwo(t,t')
\COC(t',\tw)
\label{C2_large_d}
\ee
Because of the factor $\COC(t',\tw)$ it is useful to split the
$t'$-integral into the regimes $t'>\tw$ and $t'<\tw$. The first regime
gives 
\be
\Tc\int_{\tw} dt' \Ltwo\eql(t-t')
\p\eql(t'-\tw)
\sc{L}(t/t') 
\sc{\COC}(t'/\tw)
\label{C2_part}
\ee
As in the analysis of\eq{ptwo_scaling_cond} one can now argue that,
for large $t-\tw$, the integral will be dominated by the regions
$t'\approx t$ and $t'\approx \tw$. The weights contributed by these
regions are $\p\eql(t-\tw)\int_0^{\infty} dt'\Ltwo\eql(t')$ and
$\Ltwo\eql(t-\tw)\int_0^{\infty} dt'\p\eql(t')$, respectively; the
integrals are both finite, so that both terms scale as
$(t-\tw)^{(2-d)/2}$. These weights are then multiplied by the relevant
values of scaling functions in the integrand of\eq{C2_part}. The
overall result is smaller by $\sim 1/(t-\tw)$ than the leading
term\eq{CE_expect} in $C_E$, and can be neglected. The $t'<\tw$ part
of the integral from\eq{C2_large_d} is
\be
\Tc\int^{\tw} dt' \Ltwo\eql(t-t')
\p\eql(\tw-t') \sc{L}(t/t') \sc{\COC}(\tw/t')
\ee
The factor $\p\eql(\tw-t')$ concentrates the weight of the integrand
near $t'=\tw$, so that for $t-\tw$ and $\tw\gg 1$ all other factors
can be replaced by their values at $t'=\tw$, giving the result
$\Tc\hat\p\eql(0) \Ltwo\eql(t-\tw)\sc{L}(x)$. This scales as
$(t-\tw)^{(2-d)/2}$ times a scaling function of $x$ and so is also
negligible compared to the leading contribution\eq{CE_expect}.
From\eq{COC} the second term on the r.h.s.\ of\eq{C2_large_d} has the
same subleading scaling, and the first term is even smaller. Overall,
$C_E^{(2)}$ therefore gives a subleading contribution to $C_E$ in the
aging regime. By very similar arguments one shows that $C_E^{(3)}$
from\eq{C3} is also subleading.

In the aging regime we therefore only need to consider $C_E^{(4)}$
from\eq{C4}. The long-time behaviour of the function $\CC(t,\tw)$ is
easily worked out as
\bea
\fl\CC(t,\tw) &=& \CC\eql(t-\tw)\sc{\CC}(t/\tw)
\label{CC_decomp}
\\
\fl\CC\eql(t-\tw) &=& \dq \frac{\Tc^2}{\omega^2}e^{-2\omega(t-\tw)} = 2\Tc
\int_{t-\tw}^\infty dt'\, \p\eql(t') \sim (t-\tw)^{(4-d)/2}
\label{CC_eql}
\\
\fl\sc{\CC}(x) &=& \frac{\int d\omt \, \omt ^{(d-6)/2} e^{-2(x-1)\omt }\sc{C}^2(\omt )}
{\int d\omt \, \omt^{(d-6)/2} e^{-2(x-1)\omt}}
\label{CC_scaling}
\eea
where the scaling function $\sc{\CC}(x)\sim x^{-2}$ for $x\gg 1$. Now
consider the $t'$-integral from\eq{C4},
\be
I(t,\tw') = \int dt' \, 
[(2z(t)-2\Tc)\delta(t-t')+\Ltwo(t,t')]\CC(t',\tw')
\label{Idef}
\ee
The equilibrium part $\Ltwo\eql(t-t')$ of $\Ltwo(t,t')$ ensures that
this has its mass concentrated near $t'=t$. Because we are looking at
the aging regime, we have $t-\tw'(>t-\tw)\gg 1$, and\eq{CC_eql} then
shows that the function $\CC(t',\tw')$ is slowly varying near
$t'=t$. It can thus be replaced by its value there, $\CC(t,\tw')$, to
give to leading order
\be
I(t,\tw') = (2z(t)-2\Tc+\hat\Ltwo\eql(0))\CC(t,\tw') =
\frac{1}{\hat\p\eql(0)}\CC(t,\tw')
\label{I_approx}
\ee
where in the last step we have used $z(t)\ll 1$. One might suspect
that in the $t'$-integral of\eq{Idef}, the region around $t'\approx
\tw'$ also contributes, since this is where $\CC(t',\tw')$ is
largest. However, it can be shown that this region only gives a
subleading contribution compared to\eq{I_approx}.

In terms of $I(t,\tw')$ we can write $C_E^{(4)}$ using\eq{C4} as
\be
4C_E^{(4)} =
\frac{1}{2} \int d\tw' [(2z(\tw)-2\Tc)\delta(\tw-\tw')+\Ltwo(\tw,\tw')]I(t,\tw')
\label{C4_I}
\ee
Again, the integral is dominated by the region $\tw'\approx \tw$
because of the factor $\Ltwo\eql(\tw-\tw')$ and we thus get our final
expression
\be
4C_E^{(4)} =
\frac{1}{2\hat\p\eql(0)}I(t,\tw) = 
\frac{1}{2\hat\p\eql^2(0)}\CC(t,\tw)
\label{C4_final}
\ee
As argued above, in the aging regime the full energy correlator will
be identical to this.

We can now turn to the evaluation of the out-of-equilibrium response
function. We will not write explicitly the second term in\eq{RE},
which just removes the unwanted $\delta(t-\tw)$ contribution arising
from the derivative applied to the first term's step
discontinuity. Thus,
\be
2R_E = (-\partial_t-2\Tc)\ROC(t,\tw) + \int dt' \Ltwo(t,t') \ROC(t',\tw)
\label{RE_again}
\ee
The long-time scaling of the function $\ROC$ is
\bea
\fl \ROC(t,\tw) &=& \ROC\eql(t-\tw)\sc{\ROC}(t/\tw)
\\
\fl \ROC\eql(t-\tw) &=& \Tc \dq e^{-2\omega(t-\tw)} = \Tc f(t-\tw) \sim 
(t-\tw)^{-d/2}
\label{ROC_eql}
\\
\fl \sc{\ROC}(x) &=& \frac{\int d\omt\, \omt^{(d-2)/2} e^{-2(x-1)\omt}\sc{C}(\omt)}
{\int d\omt\, \omt^{(d-2)/2} e^{-2(x-1)\omt}}
\label{ROC_scaling}
\eea
The integrand in\eq{RE_again} has its mass concentrated near $t'=t$,
from the factor $\Ltwo\eql(t-t')$, and near $t'=\tw$, from the factor
$\ROC\eql(t'-\tw)$. This gives
\be
\fl 2R_E = (-\partial_t-2\Tc)\ROC(t,\tw) + \hat\Ltwo\eql(0) \ROC(t,\tw) +
\half\Ltwo\eql(t-\tw)\sc{L}(t/\tw)
\label{RE_aux}
\ee
using that $\int_0^\infty dt'\, \ROC\eql(t') = \half 2\Tc \int_0^\infty
dt'\, f(t') = \half \p\eql(0) = \half$ (see after\eq{p_eql}). The last
term in\eq{RE_aux} 
scales as $(t-\tw)^{(2-d)/2}$ times a function of $x=t/\tw$, while the
other terms scale at most as $(t-\tw)^{-d/2}$ and so are
subdominant. Using also that $\p\eql$ and $\Ltwo\eql$ are
asymptotically proportional for $d>4$ from\eq{ptwo_p_link} and that
$\sc{L}=\sc{\p}$ then gives the final aging regime expression
\be
R_E = 
\frac{1}{4}\Ltwo\eql(t-\tw)\sc{L}(t/\tw) = 
\frac{1}{4}\Ltwo(t,\tw)
\label{RE_final}
\ee
Interestingly, the simple relation\eq{R_Eql} between the energy
response and $\Ltwo$ therefore holds not only at equilibrium but also in
the aging regime of the non-equilibrium dynamics, and thus overall
across the entire long-time regime.

We can now, finally, find the FDR for the energy correlation and
response. In the aging regime of interest here, it is useful to
rewrite the response function\eq{RE_final} as
\be
R_E(t,\tw) = \frac{1}{4\hat\p\eql^2(0)}\p\eql(t-\tw)\sc{\p}(x) =
\frac{1}{4\hat\p\eql^2(0)}\p(t,\tw)
\label{RE_final_K}
\ee
using\eq{ptwo_p_link} and $\sc{L}=\sc{\p}$. Recalling the
definition\eq{p_def}, and the fact that $C_E$ is given by\eq{C4_final}
to leading order in the aging limit, we then obtain
\be
X_E(t,\tw) = \frac{\Tc R_E(t,\tw)}{C'_E(t,\tw)} = \frac{\dq \Tc R_\qv
C_\qv}{\dq C'_\qv C_\qv} 
\label{XE}
\ee
Remarkably, this is in fact {\em identical} to the FDR\eq{Xbl_large_d}
for large blocks of spin product observables. In particular, $X_E$
tends to the limit value $X^\infty=1/2$ for $t\gg\tw$, and this is
identical to the one we found for spin and bond observables: for
$d>4$, all observables we have considered lead to a unique value of
$X^\infty=1/2$.

The FD plot for the energy has a limiting pseudo-equilibrium
form; this follows from\eq{CC_eql} and\eq{C4_final}, which show that
$C_E(t,\tw)$ has decayed to a small value $\sim\tw^{(4-d)/2}$ at the
point where aging effects begin to appear. More generally, the
correlation function scales as $C_E \sim (t-\tw)^{(4-d)/2}$ for
$x\approx 1$, and for $x\gg1$, where $\sc{\CC}(x) \sim x^{-2}$
from\eq{CC_scaling}, as $C_E \sim t^{(4-d)/2}x^{-2} =
\tw^{2}t^{-d/2}$.

It should be pointed out that while the FDR for the energy matches
that for blocks of product observables at long times, the correlation
and response functions themselves do not. This follows from the
nontrivial proportionality factors $1/{\hat\p\eql^2(0)}$
in\eq{C4_final} and\eq{RE_final_K}. The latter are required to match
the limiting behaviour for $t-\tw\ll\tw$ of the aging regime results
to the asymptotics of the equilibrium results for $t-\tw\gg
1$. Indeed, by combining the effective equilibrium behaviour for
$t-\tw=\order(1)$ with the above aging expressions we can write
\bea
C_E(t,\tw) &=& C\Eql(t-\tw)\sc{\CC}(x)
\label{CE_d_gt_4_longtime}
\\
R_E(t,\tw) &=& R\Eql(t-\tw)\sc{\p}(x)
\label{RE_d_gt_4_longtime}
\eea
and these are now valid throughout the long-time regime, i.e.\ for
large $\tw$ but independently of whether $t-\tw$ is large or not. As
promised they match with the aging expressions for $1\ll \dt
\ll\tw$. For the response this is obvious from\eq{R_Eql}. For $C_E$,
\eq{ptwo_p_link} and\eq{R_Eql} show that $-\partial_{\dt} C\Eql(\dt) =
(\Tc/4)\Ltwo\eql(\dt)\approx [\Tc/4\hat\p\eql^2(0)]\p\eql(\dt)$ for
large $\dt$; from\eq{CC_eql} this agrees with the corresponding
derivative of the result in\eq{C4_final},
$-\partial_{\dt}\CC\eql(\dt)/[8\hat\p\eql^2(0)] =
[2\Tc/8\hat\p\eql^2(0)]\p\eql(\dt)$. Note from the discussion
after\eq{chi_Eql_t} that the function $C\Eql(\dt)$ discontinuously
acquires an additive constant (non-decaying) part as $T$ crosses $\Tc$ from
above in $d>4$. What we mean in\eq{CE_d_gt_4_longtime} is the limiting
form for $T\searrow \Tc$ {\em from above}, which does not contain this
plateau. That this is the correct choice can be seen as follows: the
non-decaying part of $C\Eql(\dt)$ at $T=\Tc^{-}$ arises from the
fluctuations of the $\qv=\zv$ Fourier mode, i.e.\ of the magnetization,
which are larger by $\order(N)$ than those of the other Fourier
modes. In the context of our non-equilibrium calculation, all Fourier
modes have fluctuations of comparable order (in system size), so that
this contribution is absent; it would appear only for times $\tw$ that
scale with system size $N$.

Finally, the long-time
expressions\eqq{CE_d_gt_4_longtime}{RE_d_gt_4_longtime} also show that
an energy FD plot would have a pseudo-equilibrium form at long times,
because e.g.\ in $C_E(t,\tw)$ the equilibrium factor has already
decayed to $\sim \tw^{(4-d)/2}$ of its initial value when aging
effects appear around $\dt\sim \tw$.

\section{Energy correlation and response: Non-equilibrium, $d<4$}
\label{sec:noneq_small_d}

In dimension $d<4$, the evaluation of the energy correlation function in the
aging regime is somewhat more complicated. One can nevertheless show
that, as before, the dominant contribution to $C_E$ is $C_E^{(4)}$,
with the other three terms being subleading. We thus again need to
consider the function $I(t',\tw)$ defined in\eq{Idef}, and then work
out $C_E$ using\eq{C4_I}. One can see clearly that the approach
leading above to\eq{I_approx} no longer works: in $d<4$,
$\hat\Ltwo\eql(0)=2\Tc$, so the leading terms in\eq{I_approx} actually
cancel. To treat this cancellation accurately, we
use\eq{ptwo_eql_condition} to write the coefficient $2\Tc$ in the
following way
\be
2\Tc = - \frac{\p\eql'(t)}{\p\eql(t)} + \int dt'\,
\frac{\p\eql(t')}{\p\eql(t)} \Ltwo\eql(t-t')
\ee
This allows us to express\eq{Idef} as
\bea
\fl I(t,\tw') &=& \left[2z(t)+\frac{\p\eql'(t)}{\p\eql(t)}\right]
\CC(t,\tw')
\label{I_subtract}
\\
\fl & &  {}+{} \int_0^t
dt'\,\Ltwo\eql(t-t') \left[\sc{L}(t/t')\CC(t',\tw')
-\frac{\p\eql(t')}{\p\eql(t)}\CC(t,\tw')\right]
\nonumber
\eea
To make progress, we need the behaviour of $\CC$ for $d<4$. For long
times we have
\be
\fl \CC(t,\tw) = \dq C^2_\qv(t,\tw) = \Tc^2 \frac{g(\tw)}{g(t)}\dq
\omega^{-2} \sc{C}^2(\omega\tw) e^{-2\omega(t-\tw)}
\label{CC_gen_dlt4}
\ee
The equal-time value thus scales as 
\be
\CC(t,t) \sim \int d\omega\, \omega^{d/2-3} \sc{C}^2(\omega t) =
t^{(4-d)/2}\int d\omt \, \omt ^{(d-6)/2} \sc{C}^2(\omt )
\label{CC_equal_time}
\ee
hence $\CC(t,t)=\CCd t^{(4-d)/2}$ with some constant $\CCd$. A scaling
function is then obtained if we normalize $\CC(t,\tw)$ by this
equal-time value:
\be
\frac{\CC(t,\tw)}{\CC(t,t)} = 
\frac{g(\tw)}{g(t)}
\frac{\tw^{(4-d)/2}}{t^{(4-d)/2}}
\frac{\int d\omt \, \omt ^{(d-6)/2} \sc{C}^2(\omt ) e^{-2\omt (t-\tw)/\tw}}
     {\int d\omt \, \omt ^{(d-6)/2} \sc{C}^2(\omt )}
\label{sc_CC_ord}
\ee
This is a function $\G(x)$ of $x=t/\tw$; note that the first two
fractions cancel since $g(t)\sim t^{-\kappa}=t^{(d-4)/2}$. So far this
scaling expression holds for $t>\tw$. To be able to use it also for
non-ordered times, note that for $t<\tw$,
\be
\frac{\CC(t,\tw)}{\CC(t,t)}=\left(\frac{\tw}{t}\right)^{(4-d)/2}
\frac{\CC(\tw,t)}{\CC(\tw,\tw)}=x^{(d-4)/2}
\G(\tw/t)
\label{sc_CC_dis}
\ee
So we have overall
\be
\fl\frac{\CC(t,\tw)}{\CC(t,t)} =\G(t/\tw),
\quad
\G(x)=\left\{
\begin{array}{ll}
{\displaystyle\frac{\int d\omt\, \omt^{(d-6)/2} \sc{C}^2(\omt) e^{-2(x-1)\omt}}
{\int d\omt\, \omt^{(d-6)/2} \sc{C}^2(\omt)}} & \mbox{for $x\geq 1$} \\
x^{(d-4)/2}\G(1/x) & \mbox{for $x\leq 1$}
\end{array}
\right.
\label{CC_d_lt_4}
\ee
One easily works out the asymptotics of $\G$: $\G(x)\sim x^{-d/2}$ for
$x\to\infty$, $\G(x)\sim x^{d-2}$ for $x\to 0$. For $|x-1|\ll1$, on
the other hand, $1-\G(x)\sim |x-1|^{(4-d)/2}$. This corresponds to
$\CC(t,t)-\CC(t,\tw)\sim |t-\tw|^{(4-d)/2}$ for $1\ll |t-\tw|\ll
\tw$. (The behaviour of this difference for smaller time intervals
$t-\tw=\order(1)$ is not captured accurately by the scaling
form\eq{CC_d_lt_4}, but integrals over this regime contribute
only subleading corrections to the results below.)

We can now insert the scaling form\eq{CC_d_lt_4} of $\CC$
into\eq{I_subtract}. The non-integral terms turn out to be subleading
(see below), so
\bea
\fl I(t,\tw') &=& \int dt' \, \Ltwo\eql(t-t')
\nonumber\\ 
\fl & &\times\left[\sc{L}(t/t')\CC(t',t')
\G(t'/\tw')
-\frac{\p\eql(t')}{\p\eql(t)}\CC(t,t)\G(t/\tw')\right]
\\
\fl &=& \CC(t,t) \int dt' \, 
\Ltwo\eql(t-t')\left[\frac{t'}{t}\G(t'/\tw') - \frac{\p\eql(t')}{\p\eql(t)}
\G(t/\tw')\right]
\label{I_sub2}
\eea
where we have used that, from\eq{FKd_lt_4} and\eq{same_sc_fn},
$\sc{L}(t/t')\CC(t',t')/\CC(t,t) = (t'/t)^{(d-2)/2}
(t'/t)^{(4-d)/2} = t'/t$. The remaining ``subtracted'' integral now no
longer has its mass concentrated near $t'=t$ because the terms in
square brackets give a factor $t-t'$ there. The whole integration
range contributes, so that we can replace $\Ltwo\eql(t-t')$ by its
asymptotic form\eq{ptwo_eql_asymptotics}, $\Ltwo\eql(t-t')\approx\Ld
(t-t')^{(d-6)/2}$ with a $d$-dependent constant $\Ld$. Similarly, the
ratio ${\p\eql(t')}/{\p\eql(t)}$ can be replaced by $(t'/t)^{(2-d)/2}$
to leading order. Scaling the integration variable as $y=t'/\tw'$ then
gives
\be
\fl I(t,\tw') = I(x')
= \CCd\Ld x'^{-2}\int_0^{x'} dy\,(1-y/x')^{(d-6)/2}y\left[\G(y) - 
y^{-d/2}\G(x')x'^{d/2}\right]
\label{I_sub3}
\ee
This shows that in the aging regime $I$ depends on $x'=t/\tw'$
only. One can now also see that the neglected terms
from\eq{I_subtract} are indeed subleading: they scale as
$t^{-1}\CC(t,\tw')\sim t^{(2-d)/2}\G(t/\tw')$.

We can now proceed to simplifying\eq{C4_I} in the aging regime. Using
the same subtraction method as above, and remembering that
$C_E=C_E^{(4)}$ to leading order, one finds by analogy with\eq{I_subtract}
\be
\fl 8C_E(t,\tw) = \int d\tw' \, 
\Ltwo\eql(\tw-\tw')\left[\sc{L}(\tw/\tw')
I(t,\tw') - \frac{\p\eql(\tw')}{\p\eql(\tw)}I(t,\tw)\right]
\label{C4_sub}
\ee
Here subleading terms similar to those in the first line
of\eq{I_subtract} have already been neglected. With the scaled
integration variable $y=\tw'/\tw$, and using the asymptotic forms of
$\Ltwo\eql(\tw-\tw')$ and $\p\eql(\tw')$ as in\eq{I_sub3} one gets
\be
\fl 8C_E(t,\tw') = \Ld \tw^{(d-4)/2}
\int_0^1 dy \, (1-y)^{(d-6)/2}\left[y^{(d-2)/2} I(x/y) - y^{(2-d)/2}I(x)\right]
\label{CE_sub}
\ee
This shows that $C_E$ scales as $\tw^{(d-4)/2}$ times a function of
$x=t/\tw$. It is difficult to work out the whole functional dependence
on $x$. We therefore focus below on the asymptotic behaviour for
$x\to\infty$, which gives the asymptotic FDR. First, though, we check that in
the limit $x\to 1$, where aging effects are unimportant, our
expression matches with the equilibrium result
$C\Eql(t-\tw)$. From\eq{R_Eql} the latter behaves as $C\Eql(\dt) =
(\Tc/4)\int_{\dt}^\infty dt'\,\Ltwo\eql(t') \approx
[\Ld\Tc/2(4-d)]\dt^{(d-4)/2}$ for large $\dt$. To compare with the
prediction from\eq{CE_sub}, one uses that $I(x') \sim \ln(x'-1)$ for
$x'\approx1$; this follows from the behaviour of $\G(x)$ for $x\approx
1$. (Note that the proportionality constant in $I\sim \ln(x'-1)$ is
{\em positive}, so that $I$ itself is -- in contrast to the case $d>4$
-- {\em negative}.) Inserting into\eq{CE_sub} one then finds that for
$x\approx 1$ the integral scales as $(x-1)^{(d-4)/2}$. Overall, one
gets $C_E(t,\tw) \sim \tw^{(d-4)/2}(x-1)^{(d-4)/2} =
(t-\tw)^{(d-4)/2}$, consistent with the expectation from the
equilibrium result. One can work out the prefactor of the power law
and finds that this, too, agrees as it should.

Turning now to the behaviour of\eq{CE_sub} for large $x$, we need the
asymptotics of $I(x')$.
The leading tail $\sim y^{-d/2}$ of $\G(y)$ is subtracted off in the
expression in square brackets in\eq{I_sub3}, leaving as the next term
$y^{-(d+2)/2}$. Even together with the additional factor $y$ this
makes the integral convergent at the upper end for $x'\to\infty$. In
the limit we thus get $I(x') \approx -\Ad\CCd\Ld x'^{-2}$, with
\be
\Ad = \int_0^\infty dy\,y\left[\Gd y^{-d/2}-\G(y)\right]
\label{Ad}
\ee
and $\Gd=\lim_{x\to\infty} \G(x)x^{d/2}$. 

Inserting this inverse-square asymptotic behaviour of $I(x')$ gives
for the integral in\eq{CE_sub} the scaling $\Ad\Bd\CCd\Ld x^{-2}$ with
\be
\fl \Bd = \int_0^1 dy \, 
(1-y)^{(d-6)/2}\left(y^{(2-d)/2}-y^{(d+2)/2}\right)
= -
\frac{\Gamma\left(\frac{d-4}{2}\right)\Gamma\left(\frac{d+4}{2}\right)}
{\Gamma(d)}
\label{CE_factor}
\ee
It then follows, finally, that $C_E(t,\tw)=(\Ad\Bd\CCd\Ld^2/8)
\tw^{d/2} t^{-2}$ for $t\gg\tw$, and
\be
C'_E(t,\tw) = (\Ad\Bd\CCd\Ld^2 d/16) \tw^{(d-2)/2} t^{-2}
\label{CE_prime}
\ee

To get the FDR, we now need the response function $R_E$. This can be
evaluated very similarly to the case $d>4$ and one finds that the last
term in\eq{RE_aux} is again the dominant one, giving
\be
R_E(t,\tw) = \frac{1}{4}\Ltwo\eql(t-\tw)\sc{L}(t/\tw) =
\frac{1}{4}\Ltwo(t,\tw)
\label{RE_d_lt_4}
\ee
(The $\ROC(t,\tw)$ terms in\eq{RE_aux} look dangerous: they scale as
$(t-\tw)^{-d/2}$ and are thus {\em larger} than $\Ltwo\eql(t-\tw)\sim
(t-\tw)^{(d-6)/2}$ in $d<3$. However, their prefactors
$-2\Tc+\hat\Ltwo\eql(0)$ cancel. Treating this cancellation more
carefully then shows that these terms do remain subleading compared
to\eq{RE_d_lt_4}.) As before, $4R_E$ equals $\Ltwo$ in the aging
regime, and this then holds across the whole long-time
regime since for $t-\tw=\order(1)$ it matches the equilibrium
behaviour $4R_E(t,\tw)=\Ltwo\eql(t-\tw)$. For $t\gg\tw$, on the other
hand, the response\eq{RE_d_lt_4} becomes
\be
R_E(t,\tw) = \frac{1}{4}\Ld t^{(d-6)/2}
\left(\frac{t}{\tw}\right)^{(2-d)/2} = \frac{\Ld}{4} \tw^{(2-d)/2} t^{-2}
\ee
Comparing with\eq{CE_prime} then finally gives for the asymptotic FDR
for the energy in $d<4$, 
\be
X_E^\infty = \frac{4\Tc}{\Ad\Bd\CCd\Ld d}
\label{Xinf_dlt4}
\ee
After evaluating the various numerical factors (see
\ref{sec:X_E_infty}) this can be written as
\be
X_E^\infty = \frac{2}{d\Atd}
\frac{\Gamma(d)\Gamma{\ts\left(\frac{4-d}{2}\right)}}
{\Gamma\left(\frac{d+4}{2}\right)}
\label{Xinf_dlt4_explicit}
\ee
where
\be
\fl\Atd =  \frac{\pi}{\sin[\pi(4-d)/2]}
+\int_1^\infty dx 
\int_0^1 dy\, y^{(d-4)/2} \frac{(x-y)^{(2-d)/2}}{1+x-y}
\left(1-y-x^{-(d+2)/2}\right)
\label{Atd}
\ee
Near $d=4$, one can expand in $\epsilon=4-d$.  It can be shown by
explicit calculation that the integral in\eq{Atd} is exactly zero for
$d=4$, giving $\Atd = 2/\epsilon+\order(\epsilon)$ and so 
\be
X_E^\infty = \frac{\epsilon
\Gamma(4-\epsilon)\Gamma(\epsilon/2)}{(4-\epsilon)\Gamma(4-\epsilon/2)}
+ \order(\epsilon^2) = \frac{1}{2}-\frac{\epsilon}{3} +
\order(\epsilon^2)
\ee
Notably, this is {\em different} from the FDR
$X^\infty=1-2/d=1/2-\epsilon/8+\order(\epsilon^2)$ for all the other,
finite-range, observables that we considered previously in $d<4$. It is,
however, consistent with RG calculations to
$\order(\epsilon)$ for the $O(n)$ model in the limit $n\to\infty$, for
an analogous choice of observable~\cite{CalGam04}. Remarkably,
therefore, the 
non-Gaussian fluctuations induced by the weak infinite-range interaction
in the spherical model seem to mimick precisely the effects that are
seen in more realistic models such as the $O(n)$, even though in the
latter all interactions are short-ranged and there is no difference
between the behaviour of block observables and global ones.

\begin{figure}
\centerline{\includegraphics[width=11.0cm,clip=true]{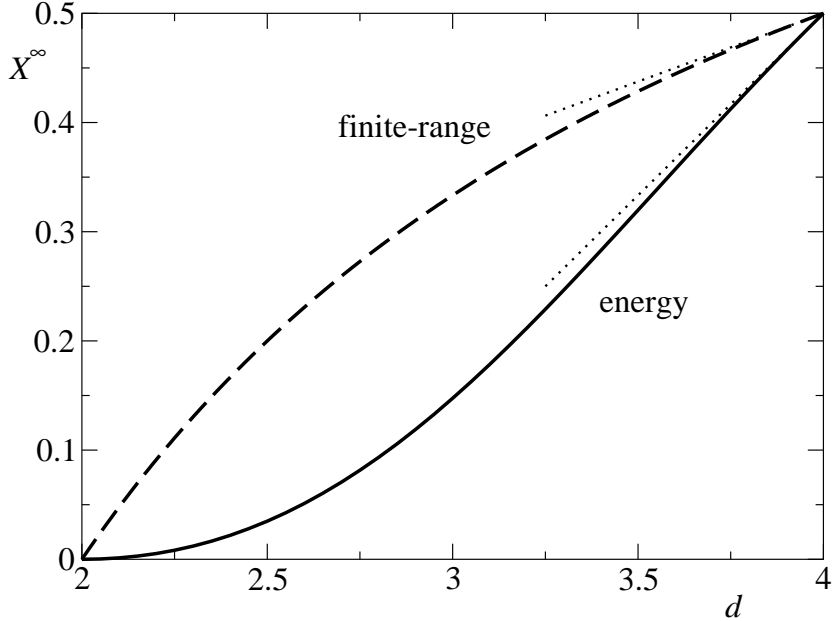}}
\caption{Asymptotic FDR $X^\infty$ vs dimension $d$ for finite-range
observables (equation~(\protect\ref{X_baseline}), dashed) and energy
(equation~(\protect\ref{Atd}), solid). Dotted lines indicate the
corresponding linear expansions $1/2-\epsilon/8$ and $1/2-\epsilon/3$ in
$\epsilon=4-d$. 
\label{fig:X_normal_and_E}
}
\end{figure}
We show in Fig.~\ref{fig:X_normal_and_E} the dependence of the
asymptotic energy-FDR $X_E^\infty$ on dimension $d$ and compare with
the result for finite-range observables. They agree in $d\geq 4$, but the
difference between them grows as $d$ decreases towards $d=2$, with the
energy FDR always having the lower value. In the limit $d\to 2$, both
FDRs converge to zero, but while the finite-range FDR
$X^\infty=\epsilon'/2+\order(\epsilon'^2)$ does so linearly in
$\epsilon'=d-2$, the energy FDR vanishes quadratically as 
$X_E^\infty=\epsilon'^2/8+\order(\epsilon'^3)$, due to the divergence
of $\Atd=4/\epsilon'^2+\order(1/\epsilon')$.

As in the case $d>4$, an energy FD plot would have a
pseudo-equilibrium form which hides all non-equilibrium effects at
long times. Indeed, one could write\eq{CE_sub} in the form
$C_E(t,\tw)=C\Eql(t-\tw)\sc{{C_E}}(x)$, and the decay of the
equilibrium factor $C\Eql$ squeezes all details about the aging factor
$\sc{{C_E}}(x)$ into a vanishingly small region of the FD plot for
long times. While we have not calculated $\sc{{C_E}}$ explicitly, the
discussion after\eq{CE_sub} shows that $\sc{{C_E}}(1)=1$ as it should
be. For large $x$, on the other hand, we saw in \eq{CE_prime} that
$C_E(t,\tw)\sim \tw^{d/2} t^{-2}$, implying $\sc{{C_E}}(x)\sim
\tw^{d/2}t^{-2}(t-\tw)^{(4-d)/2} \sim x^{-d/2}$. This matches
continuously at $d=4$ with the prediction\eq{CE_d_gt_4_longtime} for
$d>4$, where the aging correction decays as $\sc{\CC}(x) \sim x^{-2}$.

\section{Magnetized initial states}
\label{sec:magnetized}

In this final section we consider the dynamics for initial
configurations with nonzero 
magnetization, focussing as before on the non-equilibrium
dynamics that results when the system is subsequently brought to the
critical temperature $\Tc$. Physically, such a situation could arise
in an ``up-quench'', where the system is equilibrated at $T<\Tc$ and
temperature is then increased to $T=\Tc$. As explained in the
introduction, our interest in this
scenario arises from recent results~\cite{GarSolPagRit05} which show
that such initial conditions produce FDRs that differ nontrivially
from those for zero magnetization. The analysis
of~\cite{GarSolPagRit05} was limited to high dimensions $d$ or
infinite-range interactions, however; the calculation below will allow us
to see explicitly how the results change in finite dimension. In
particular, we will obtain exact FDRs for magnetized coarsening
below the upper critical dimension, i.e.\ $d<4$ in the spherical
model.

We will continue to use the notation $C_\qv(t,\tw)=(1/N) \lav
S_\qv(t)S_\qv^*(t)\rav$ for Fourier mode correlations. For $\qv=\zv$
this is now a full, unsubtracted correlator, with
$C_\zv(t,t)=Nm^2(t)+\order(1)$ and $m(t)=(1/N)\lav S_\zv(t)\rav$ the
time-dependent magnetization. The difference $\Cz(t,\tw) =
C_\zv(t,\tw)-Nm(t)m(\tw)$ is the connected correlator, which has
values of $\order(1)$ and is the relevant quantity for analysing the
FD behaviour. For $\qv\neq\zv$, connected and full
correlators coincide:
\be
\Cc_\qv(t,\tw) = C_\qv(t,\tw)-(1/N)\lav S_\qv(t)\rav \langle
S^*_\qv(\tw)\rangle = C_\qv(t,\tw)
\ee

We now need to check how the analysis in the previous sections is
modified by the 
presence of a nonzero magnetization. The Fourier space equation of
motion\eq{dotSq} remains valid, and so do the resulting expressions
for the response function $R_\qv(t,\tw)$\eq{Rq} and the full
correlator $C_\qv(t,\tw)$\eqq{Cqtt}{C_twotime}.
The expression\eq{hat_g_general} for the Laplace transform of $g(t)$ that
results from the spherical constraint also still holds, but the
solution is now different. In the $\qv$-integral,
the $\qv=\zv$ contribution $(1/N)C_\zv(0,0)/s =
m^2(0)/s$ has to be treated separately. In fact, one sees that for
$s\to 0$ this term always dominates the rest of the integral, which
diverges less strongly. At criticality, where $T=\Tc=[\dq
1/\omega]^{-1}=[2\hat{f}(0)]^{-1}$, one thus has for $s\to 0$
\be
2\Tc \hat{g}(s) = \frac{m^2(0)}{s} \left[\dq
\left(\frac{1}{2\omega}-\frac{1}{s+2\omega}\right)\right]^{-1}
\label{gs}
\ee
which using\eq{p_eql_LT} can be rearranged into
\be
\frac{\hat{g}(s)}{m^2(0)}=s^{-2}\left[\int (dq)
\frac{\Tc}{\omega(2\omega+s)}\right]^{-1} = [s^{2}\hat{K}\eql(s)]^{-1}
\label{gs2}
\ee 
For $d>4$, $\hat{K}\eql(0)$ is finite so this scales as $s^{-2}$; for
$d<4$, on the other hand, $\hat{K}\eql(s)$ diverges as $s^{(d-4)/2}$
so that $\hat{g}(s) \sim s^{-d/2}$. Translating back to the time
domain, $g(t)$ behaves for large $t$ as
\be
g(t)\sim m^2(0)\,t^\alpha, \qquad \alpha = \left\{\begin{array}{lll}
1 & \mbox{ } & (d>4)\\ (d-2)/2 & \mbox{ } & (d<4) \end{array}\right.
\label{g_scaling_magn}
\ee
Note that this asymptotic behaviour is independent of any details of
the initial condition except for the presence of a nonzero $m(0)$; it
depends on the actual value of $m(0)$ only through the prefactor
$m^2(0)$. For the time-dependence of $m(t)$, one gets by taking an
average of\eq{Sqt}
\be
m(t) = R_\zv(t,0) m(0) = \frac{m(0)}{\sqrt{g(t)}}
\label{m_in_terms_of_g}
\ee
Because of the proportionality of $g(t)$ to $m^2(0)$ for large $t$,
the asymptotic decay of $m(t)$ is independent of the initial
conditions, in terms of {\em both} the decay exponent {\em and} the prefactor.

\subsection{Finite-range spin observables}

We first analyse the correlation and response functions for
observables that relate to a number of spins that is much smaller
than $N$. As for the unmagnetized case, the fluctuations of the
Lagrange multiplier $z$ can then be neglected. To understand the
magnetized case, it is useful to shift the spin variables by $m(t)$.
We will see that the equations of motion then acquire the same form as
before, so that we can directly transfer the main results from the
unmagnetized 
case. Explicitly, we consider the following decomposition of the spin
variables
\be
S_i=m+U_i
\ee
where $U_i$ is a zero-mean variable. The equation of
motion\eq{real_space_dotSi} for $S_i$ then gives
\be
\partial_t m+\partial_t{U_i}=-\Om_{ij}(m+U_j)+\xi_i-z(m+U_i)
\ee
From\eq{m_in_terms_of_g} and the definition\eq{g_def}, $\partial_t\ln m(t) = 
-(1/2)\partial_t\ln g(t) = -z(t)$, so $\partial_t m = -zm$. Also $\sum_j
\Om_{ij}=0$, giving
\be
\partial_t{U_i} = -\Omega_{ij}U_j + \xi_i - zU_i
\label{dot_Ui}
\ee
This is the same as the equation for $S_i$ in the unmagnetized case,
and so one can deduce directly the solutions for the correlation
functions of the $U_i$; these are the connected correlations
$\Cc$. The initial values again become unimportant for long times,
allowing us to work out the scaling of the $\Cc$, and then together
with the response $R$ also the FDR $X$. It is clear from the
description in terms of the subtracted spins $U_i$ that there is
nothing special about the case $\qv=\zv$, and all results will have a
smooth limit as $\qv\to \zv$. Because we are neglecting the
fluctuations of the Lagrange multiplier $z$, this limit again has to be
understood as that of a block magnetization calculated over a
region much larger than the correlation length (in the time regime
being considered) but much smaller than the linear system size, so
that $q\gg 1/L$.

Applying\eq{Cqtt}, we can now write down directly the connected
correlation function at equal times as
\be
\Cc_\qv(\tw,\tw)=\Cc_\qv(0,0)\frac{e^{-2\omega\tw}}{g(\tw)}
+2\Tc\frac{e^{-2\omega\tw}}{g(\tw)}\int_0^{\tw}dt'e^{2\omega t'}g(t')
\label{Cqtt_magn}
\ee
At long times, the first term is subleading due to the
scaling\eq{g_scaling_magn}, and one has the behaviour
\be
\Cc_\qv(\tw,\tw) = \frac{\Tc}{\omega}\sc{C}(\omega \tw), \quad
\sc{C}(\omt) = 2\omt\int_0^1 dy\,y^{\alpha} e^{-2\omt(1-y)}
\label{C_scaling_magn}
\ee
This result is of course the same as\eq{Xq_scaling}, except for the
replacement of $-\n$ by $\alpha$ which reflects the difference in the
asymptotic behaviour of $g(t)$.

The two-time connected correlations are
$\Cc_\qv(t,\tw)=R_\qv(t,\tw)\Cc_\qv(\tw,\tw)$ with $R_\qv(t,\tw)$
given by\eq{Rq} as before. As a consequence, the expression\eq{Xq} for
the FDR $X_\qv$ also remains valid, and one finds the scaling form
$X_\qv(t,\tw) = \sc{X}(\omega\tw)$ with
\be
\sc{X}^{-1}(\omt) = 2-(2\omt+\alpha)\int_0^1 dy\
y^{\alpha}e^{-2\omt(1-y)}
\ee
which is directly analogous to\eq{Xq_scaling}. In the limit
$\omt=\omega\tw\rightarrow\infty$, $X_\qv=1$, which means that all
modes with $\qv\neq 0$ equilibrate once $\tw\gg 1/\omega$. In the
opposite limit $\omt\rightarrow 0$, corresponding to $\tw\ll
1/\omega$,
\be
X_\qv=\frac{\alpha+1}{\alpha+2} = \left\{\begin{array}{lll}
2/3 & \mbox{ } & (d>4) \\
d/(d+2) & \mbox{ } & (d<4)
\end{array}
\right.
\label{Xqneq}
\ee
For $\qv\to\zv$ this result applies independently of the value of
$\tw$ as long as $\tw\gg 1$, so that the FDR for the block
magnetization will be a straight line with slope\eq{Xqneq}. This is as
for the unmagnetized case, but the actual value of the FDR is now {\em
different}. It is also different from the value $X^\infty=4/5$
predicted for Ising models in the limit of large
$d$~\cite{GarSolPagRit05}; we will see below that the latter value is
obtained for the {\em global} magnetization, which is affected by
local spin fluctuations of $\order(\rn)$.

For later reference we write down the long-time forms of the
correlation and response functions for $\qv\to\zv$. By setting
$\omega=0$ in\eq{Cqtt_magn} and taking the long-time limit where the
first term becomes negligible, we find for the connected equal-time
correlator
\be
\Cc_\zv(\tw,\tw) = \frac{2\Tc\tw}{1+\alpha}
\ee
The response function is, from\eq{Rq} and\eq{m_in_terms_of_g},
\be
R_\zv(t,\tw) = \sqrt{\frac{g(\tw)}{g(t)}} = \frac{m(t)}{m(\tw)}
= \left(\frac{\tw}{t}\right)^{\alpha/2}
\label{R_simplification_magn}
\ee
where the last equality holds for long times. The two-time correlator
is therefore
\be
\Cc_\zv(t,\tw)=\frac{2\Tc\tw}{1+\alpha}\left(\frac{\tw}{t}\right)^{\alpha/2}
\label{Cc_0}
\ee
From these results one of course retrieves the long-time FDR
$X_\zv(t,\tw)={\Tc R_\zv(t,\tw)}/{\Cc_\zv'(t,\tw)}=({\alpha+1})/({\alpha+2})$,
obtained in\eq{Xqneq} via the limit $\omega\tw\to 0$. As explained
above, these results apply in the regime $1/L\ll q\ll
1$. For $\qv=\zv$ itself, they capture only the Gaussian part of the
spin-fluctuations, and non-Gaussian corrections become relevant as
discussed in the next section.

\subsection{General expressions for magnetization correlation and response}

We now turn to the FD behaviour of the global magnetization,
corresponding to $\qv=\zv$ rather than $q\gg {1}/{L}$. All $N$ spins
are now involved and one needs to account for the fluctuating
contribution of the Lagrange multiplier, which we write as
$z+N^{-1/2}\Delta z$ as before. To understand why this is necessary in
the magnetized case, but was not in the unmagnetized scenario,
consider the equation of motion\eq{dotSq} for the zero-wavevector
Fourier component of the spins,
\be
\partial_t S_\zv = -(z+N^{-1/2}\Delta z)S_\zv + \xi_\zv
\ee
In the unmagnetized case, $S_\zv$ is a zero-mean quantity of
$\order(N^{1/2})$. The $\Delta z$-term then contributes only subleading
$\order(1)$ fluctuations. For nonzero magnetization, on the other hand, the
mean of $S_\zv$ is $Nm$, with fluctuations around this of
$\order(N^{1/2})$. The coupling of $\Delta z$ with $m$ then gives an
$\order(N^{1/2})$ contribution to $\partial_t S_\zv$, which is no
longer negligible.

To find the resulting non-Gaussian fluctuations in $S_\zv$ explicitly,
we make the decomposition $S_i=\X_i+N^{-1/2}\Y_i$ as before.  The
discussion in Sec.~\ref{sec:setup} then goes through, and we
retrieve\eq{yt3} for the $\order(N^{-1/2})$-corrections $\Y_i$ to the
spins. For the zero Fourier mode, in particular, we have
\be
\Y_\zv(t)=-\half \int dt' dt''R_\zv(t,t') \X_\zv(t')
L(t',t'')\Delta(t'')
\label{r0}
\ee 
To simplify the calculation of connected correlations, we now
decompose the Gaussian part of the spins into $\X_i = m + u_i$, so
that the $u_i$ are zero-mean Gaussian variables. This corresponds to a
decomposition $U_i=u_i+\sqrt{N}\Y_i$ of the fluctuating parts of the
spins into leading Gaussian terms and small non-Gaussian corrections,
in analogy to the representation $S_i=\X_i+\sqrt{N}\Y_i$ in the
unmagnetized case. The $u_i$ obey the equation of motion\eq{dot_Ui},
and their correlation and response functions are the $\Cc_\qv$ and
$R_\qv$ calculated previously.

We will write the connected correlation function for the global
magnetization which {\em includes} non-Gaussian corrections as
$\Dc(t,\tw)=(1/N)\lav U_\zv(t)U_\zv(\tw)\rav$. Making the
decomposition into Gaussian and non-Gaussian parts, this reads
\be
\Dc(t,\tw)=\frac{1}{N}
\langle[u_\zv(t)+N^{-1/2}r_\zv(t)][u_\zv(\tw)+N^{-1/2}r_\zv(\tw)]\rangle
\label{concor}
\ee
with, from\eq{r0},
\bea
r_\zv(t) &=& -\frac{1}{2}\int dt'\,dt''\, 
R_\zv(t,t')[Nm(t')+u_\zv(t')]L(t',t'')\Delta(t'') \\
         &=& -\frac{N}{2}\int dt'\,M(t,t')\Delta(t')
\label{r}
\eea
Here we have defined 
\be
M(t,\tw)=\int dt'\,R_\zv(t,t') m(t') L(t',\tw) = 
m(t) \int^t dt'\,L(t',\tw)
\label{Mdef}
\ee
where the second form follows from\eq{R_simplification_magn}. $M$ is,
like $R$ and $L$, causal and so vanishes for $t<\tw$. In\eq{r}
we have also discarded the Gaussian fluctuation term $u_\zv$, which is
of $\order(N^{1/2})$ and so negligible against the term $Nm$. This is
in line with the intuition discussed earlier that non-Gaussian
fluctuations arise only from the coupling of $\Delta z$ to $m$. Note
also that $r_\zv$ is $\order(N)$, so that in\eq{concor} the
non-Gaussian correction $N^{-1/2}r_\zv$ is of the same order as the
Gaussian fluctuation $u_\zv$, again as expected.

Substituting\eq{r} into\eq{concor}, we see that we need the two-time
correlations of $u_\zv$ and $\Delta$. In the presence of a nonzero $m$
the latter becomes
\be
\Delta=\frac{1}{\sqrt{N}}\sum_i(s_i^2-1) = \frac{1}{\sqrt{N}}\sum_i
(m^2-1+2u_i m+u_i^2) 
\label{Delta_m_nonzero}
\ee
The required correlations are therefore $\lav u_\zv(t)u_\zv(\tw)\rav =
N\Cc_\zv(t,\tw)$ and 
\bea
\fl \langle u_\zv(t)\Delta(t')\rangle &=& \frac{1}{\sqrt{N}}\sum_{ij}
\langle
u_i(t)[m^2(t')-1+2u_j(t')m(t')+u_j^2(t')]\rangle
\\
\fl
&=& \frac{2}{\sqrt{N}} m(t')
\sum_{ij}\langle u_i(t)u_j(t')\rangle= 2m(t')\sqrt{N}\Cc_\zv(t,t')
\eea
For the autocorrelation of $\Delta$, we can exploit the fact
that $\lav \Delta\rav = 0$ to write\eq{Delta_m_nonzero} as $\Delta = N^{-1/2}
\sum_i (2u_i m+u_i^2 - \lav u_i^2\rav)$. This gives
\bea
\fl \langle \Delta(t')\Delta (\tw')\rangle
&=&
\frac{1}{N}\sum_{ik}\langle[2u_i(t')m(t')+u_i^2(t')-\langle
u_i^2(t')\rangle][\cdots
t'\rightarrow \tw' \cdots]\rangle 
\\
\fl &=& \frac{4}{N}m(t')m(\tw')\sum_{ij}\langle u_i(t')u_j(\tw')\rangle
+\frac{2}{N}\sum_{ij}\langle u_i(t')u_j(\tw')\rangle^2
\\
\fl &=& 4m(t')m(\tw')\Cc_\zv(t',\tw')+2\dq\Cc_\qv^2(t',\tw')
\eea
where we have used Wick's theorem to simplify the fourth-order average
$\langle (u_i^2-\langle u_i^2\rangle)(u_j^2-\langle
u_j^2\rangle)\rangle = \langle u_i^2 u_j^2\rangle - \langle u_i^2
\rangle \langle u_j^2\rangle = 2\langle u_i
u_j\rangle^2$. Abbreviating the $q$-integral as $\CCt(t',\tw')$, the
full connected correlation function\eq{concor} can thus be written as
\bea
\fl \Dc(t,\tw)&=&
\Cc_\zv(t,\tw)
-\frac{1}{2\sqrt{N}} \int dt'\,M(t,t')\langle u_\zv(\tw)\Delta(t')\rangle
\nonumber
\\
\fl &&{}-{}
\frac{1}{2\sqrt{N}} \int dt'\, M(\tw,t')
\langle u_\zv(t)\Delta(t')\rangle
\nonumber
\\
\fl &&{}+{}\frac{1}{4}\int\,dt'\,d\tw'\,
M(t,t') M(\tw,\tw') \langle \Delta(\tw')\Delta(t')\rangle
\\
\fl &=&
\Dc^{(1)}(t,\tw)+\Dc^{(2)}(t,\tw)
\label{DM}
\eea
where
\bea
\fl \Dc^{(1)}(t,\tw)&=&\Cc_\zv(t,\tw)-\int dt'\,
    [M(t,t')\Cc_\zv(\tw,t')+M(\tw,t')\Cc_\zv(t,t')]m(t')
\nn
\fl & &{}+{}\int\, dt'\,d\tw'\, M(t,t')M(\tw,\tw')m(t')m(\tw')\Cc_\zv(t',\tw')
\label{c_m1_unscaled}
\\
\fl &=& \int\, dt'\,d\tw'\, [\delta(t-t')-M(t,t')m(t')]
\nn
\fl & & \times [\delta(\tw-\tw')-M(\tw,\tw')m(\tw')]\Cc_\zv(t',\tw')
\label{DM_part}
\eea
and
\be
\fl \Dc^{(2)}(t,\tw)=\half\int\, dt'\,d\tw'\, M(t,t')M(\tw,\tw')\CCt(t',\tw')
\label{c_m2}
\ee

Next we derive an expression for the corresponding magnetization response
function. To this purpose we expand the spins for small fields as
\be
S_i=s_i+hr_i 
\ee
where $s_i$ are the unperturbed spins and we neglect the
$\order(N^{-1/2})$ non-Gaussian corrections as irrelevant, as in the
unmagnetized case. The Lagrange multiplier is similarly written as
$z+h\Delta z$. By collecting the $\order(h)$ terms from the equation
of motion for the $S_i$, we then find by analogy with\eq{dot_ri} that
the $r_i$ obey
\be
\partial_t{r_i}=-\Om_{ij}r_j-zr_i-\Delta z\, s_i+\delta(t-\tw)
\label{dot_ri_magn}
\ee
Here the last term represents a field impulse at time $\tw$, uniform
over all sites as is appropriate for the field conjugate to the
global magnetization. Since $r_i(t)=0$ before the field is applied,
i.e.\ for $t<\tw$, this impulse perturbation gives
$r_i(\tw^+)=1$. Starting from this value we can then
integrate\eq{dot_ri_magn} forward in time to get
\be
r_i(t)=\sum_j R_{ij}(t,\tw)-\int_{\tw}^t dt'\, R_{ij}(t,t')\Delta z(t')s_j(t')
\label{ri_sol_magn}
\ee
The condition we need to impose in order to get $\Delta z$ is
that the spherical constraint $(1/N)\sum_i(s_i+hr_i)^2=1$ needs to be
satisfied to linear order in $h$, giving the condition
$({1}/{N})\sum_i\langle r_is_i\rangle=0$. Inserting\eq{ri_sol_magn} into
this yields
\be
R_\zv(t,\tw)m(t)=\int_{\tw}^{t}dt'\,K(t,t')\Delta z(t')
\ee
where we have used the definition\eq{p_def} of $K(t,t')$. Applying the
inverse kernel $L$ gives
\be
\Delta z(t)=\int_{\tw}^t dt'\,L(t,t')m(t')R_\zv(t',\tw)
\ee
Note that this result vanishes when $m=0$, consistent with the fact
that we did not need to consider perturbations of $z$ in our
calculation of the magnetization response in the unmagnetized case. We
can now write down the magnetization response function, which we
denote by $\Q(t,\tw)$. It is given by $\Q=({1}/{N})\sum_i\langle r_i
\rangle$; inserting the result for $\Delta z$ into\eq{ri_sol_magn}, we
get explicitly
\bea
\fl \Q(t,\tw) &=& R_\zv(t,\tw)-\int
dt''\,dt'\,R_\zv(t,t'')m(t'')L(t'',t')m(t')R_\zv(t',\tw)
\\
\fl &=& \int dt'\, [\delta(t-t')-M(t,t')m(t')]R_\zv(t',\tw)
\label{response_magn}
\eea

This completes the derivation of the general expressions for the
magnetization correlation and response. To make progress, we need to
find the kernel $M(t,t')$. This requires $L(t,t')$, which is the
inverse of $K(t,t')=\dq R_\qv(t,t')C_\qv(t,t')$. As is clear from the
discussion in Sec.~\ref{sec:setup}, the correlator occurring here is
the {\em unsubtracted} one. Because $C_\zv(t,t')=Nm(t)m(t')$ is
$\order(N)$, the $\qv=\zv$ term needs to be treated separately in
spite of its vanishing weight $1/N$. It makes a contribution
$(1/N)R_\zv(t,t')Nm(t)m(t')= m^2(t)\theta(t-t')$, where we have
simplified using\eq{R_simplification_magn}. We can thus write
\be
K(t,t') = \Kc(t,t') + m^2(t)\theta(t-t')
\label{K_decomp}
\ee
with the $\qv\neq \zv$ contribution
\be
\Kc(t,t') = \dq R_\qv(t,t')\Cc_\qv(t,t')
\ee
We have switched to the connected correlator here; this makes no
difference for $\qv\neq 0$, but allows us to include $\qv=\zv$ in the
integral because $\Cc_\zv=\order(1)$. To say more, we will need to
distinguish between dimensions $d>4$ and $d<4$.

\subsection{Magnetization correlation and response: Non-equilibrium, $d>4$}

The scaling of the connected part $\Kc$ of $K$ can be analysed exactly
as in the case of zero magnetization: it consists of the same equilibrium
time dependence modulated by an aging function,
$\Kc(t,t')=K\eql(t-t')\sc{\Kc}(t/t')$. The aging part can be worked out
exactly as in\eq{p_scaling} with the only difference arising from the
changed asymptotic behaviour of $g(t)\sim t^{\alpha}$ rather than
$t^{-\n}$. For $\sc{\Kc}$ we can therefore use
directly\eq{p_scaling_fn_y_integral}, with $\n$ replaced by $-\alpha$:
\bea
\sc{\Kc}(x)&=&
\frac{d-2}{2}(x-1)^{(d-2)/2}x^{-\alpha} \int_0^1 dy\,
y^{\alpha}(x-y)^{-d/2}
\label{sc_Kc}
\eea  
In $d>4$, where $\alpha=1$, the integral can be computed explicitly
to give
\be
\sc{\Kc}(x)=1-\frac{d-2}{d-4}\left(\frac{x-1}{x}\right)+
\frac{2}{d-4}\left(\frac{x-1}{x}\right)^{\frac{d-2}{2}}
\ee
We will see below that the precise behaviour of this function does not
affect the results. Briefly though, for $x-1\ll 1$ the second term on the
r.h.s.\ is leading so that $\sc{\Kc}$ decreases linearly with $x-1$,
while for large $x$ one finds by expanding in $1/x$ that $\sc{\Kc} \approx
[(d-2)/4]x^{-2}$.

To find the inverse kernel $L$, consider how $K(t,t')$ varies with
$t$. The first part in\eq{K_decomp} starts off close to unity and
decays on $\order(1)$ timescales $t-t'$ as $K\eql(t-t')$, with a
modulation by the aging factor $\sc{\Kc}(t/t')$ once $t-t'$ becomes
comparable to $t'$. The second part, on the other hand, is small
initially but only decays on aging timescales. Comparing
$\p\eql(t-t')\sim (t-t')^{(2-d)/2}$ to $m^2(t)\sim 1/t$, this second
term therefore eventually becomes dominant, for $t-t'\sim
t^{2/(d-2)}$.
This discussion
suggests that also the inverse kernel $L$ should be composed of two
parts with distinct long-time behaviour. We therefore write
\be
L = \Lc+\dL
\ee
where $\Lc$ is the inverse of $\Kc$ and $\dL$ arises from the
zero-wavevector part of $K$. The continuous part of $L$ is then
similarly decomposed as $\Ltwo = \Lctwo + \dLtwo$. 

We proceed by writing the defining equations for $\Ltwo$ and
$\Lctwo$. The full inverse $L$ is defined by\eq{pinv_def} and as before has
singular parts which are related to the behaviour of $K(t,t')$ for
$t\to t'$. One can show directly from the definition of $K$, and
exactly as in the unmagnetized case, that 
\be
K(t'^+,t') = 1, \qquad
\left.\partial_{t'} K(t,t')\right|_{t\to t'^+} = 2\Tc
\label{K_initial}
\ee
The decomposition\eq{pinv_structure} of the inverse kernel $L$
therefore also remains valid, and from\eq{pinv_def} and\eq{K_decomp}
we get the following equation for its continuous part $\Ltwo$
\be
\fl \int dt'\,[\Kc(t,t')+m^2(t)]\Ltwo(t',\tw)
=2\Tc\Kc(t,\tw)+2\Tc m^2(t) -\partial_{\tw} \Kc(t,\tw)
\label{N.6}
\ee
This is the analogue of the relation\eq{ptwo_def} for the case $m=0$. We can
argue similarly for $\Lc$, which is defined by
\be
\int dt'\,\Kc(t,t')\Lc(t',\tw)=\delta(t-\tw)
\label{defi}
\ee
From\eq{K_decomp} and\eq{K_initial},
\be
\Kc(t'^+,t') = 1-m^2(t), \qquad
\left.\partial_{t'}\Kc(t,t')\right|_{t\to t'^+} = 2\Tc
\ee
and this initial behaviour implies that $\Lc$ can be decomposed as
\be
\Lc(t',\tw)=\frac{\delta'(t'-\tw)}{1-m^2(\tw)}
+\frac{2\Tc}{[1-m^2(\tw)]^2}\delta(t'-\tw)-\Lctwo(t',\tw)
\label{lc}
\ee
Inserting into the definition\eq{defi} gives for the continuous part
$\Lctwo$ 
\be
\fl \int dt'\,\Kc(t,t')\Lctwo(t',\tw)=
\frac{2\Tc}{[1-m^2(\tw)]^2} \Kc(t,\tw)
-\frac{1}{1-m^2(\tw)}\partial_{\tw}\Kc(t,\tw)
\label{lc2}
\ee
Now for long times, we can approximate $1-m^2(\tw)\approx
1$. Then\eq{lc2} becomes identical to the relation\eq{ptwo_def} which
determined $\Ltwo$ in the unmagnetized case. Since $\Kc$ has the same
scaling form as $K$ in\eq{ptwo_def}, except for the replacement of
$\sc{K}$ by $\sc{\Kc}$, the solution for $\Lctwo$ can be found in
exactly the same way. In particular, the scaling functions describing
the aging corrections in $\Kc$ and $\Lctwo$ are again identical, and
we can write directly
\be
\Lctwo(t,t')=\Ltwo\eql(t-t')\sc{\Kc}(t/t')
\ee
as the long-time form of $\Lctwo$. Here $\Ltwo\eql$ is the same function
as in the unmagnetized case, with Laplace transform\eq{ptwo_eql}.

It now remains to find $\dLtwo$. Subtracting\eq{lc2} from\eq{N.6} gives
\bea
\fl\lefteqn{
\int_{\tw}^t dt'\,\Kc(t,t')\dLtwo(t',\tw)+m^2(t)\int_{\tw}^t
dt'\,\dLtwo(t',\tw) = 
2\Tc \frac{-2m^2(\tw)+m^4(\tw)}{[1-m^2(\tw)]^2} \Kc(t,\tw) }
\nonumber
\\
& & 
{}+{}m^2(t)\left[2\Tc-\int_{\tw}^t dt'\,\Lctwo(t',\tw)\right]
+\frac{m^2(\tw)}{1-m^2(\tw)}\partial_{\tw}\Kc(t,\tw)
\label{deltal2}
\eea
To make progress we assume that, by analogy with the
structure\eq{K_decomp} of $K$, $\dLtwo$ varies only on aging
timescales; we will find this confirmed {\em a posteriori}. We can
then concentrate on the aging regime $t-\tw\sim\tw\gg 1$. In this
regime, the integral $\int_{\tw}^t dt'\,\Lctwo(t',\tw) = \int_{\tw}^t dt'\,
\Ltwo\eql(t'-\tw) \sc{\Kc}(t'/\tw)$ on the r.h.s.\ of\eq{deltal2}
becomes to leading order $\int_{\tw}^\infty
dt'\,\Ltwo\eql(t'-\tw)=\hat\Ltwo\eql(0)$; the aging correction
$\sc{\Kc}$ is unimportant because the integral converges for
$t'-\tw=\order(1)$, and the upper integration limit can be set to
infinity for the same reason. From\eq{ptwo_eql},
$\hat\Ltwo\eql(0)=2\Tc-1/\hat\p\eql(0)$, and so the square bracket on
the r.h.s.\ of\eq{deltal2} becomes simply the constant
$1/\hat\p\eql(0)$. In the first and third term on the r.h.s., on the
other hand, $\Kc$ and $\partial_{\tw}\Kc$ scale as
$(t-\tw)^{-(d-2)/2}$ and $(t-\tw)^{-d/2}$, respectively, and are
negligible compared to the second term in the aging regime.
Disregarding these subleading terms, equation\eq{deltal2}
is transformed to
\be
\int_{\tw}^t 
dt'\,\Kc(t,t')\dLtwo(t',\tw)+m^2(t)\int_{\tw}^t dt'\,\dLtwo(t',\tw)
=\frac{m^2(t)}{\hat{K}\eql(0)}
\ee
In the first integral, if as assumed $\dLtwo$ varies only on aging
timescales, the integral is dominated by the region $t'\approx t$
because of the factor $K\eql(t-t')$ in $\Kc(t,t')$. This factor again
makes the integral convergent within a region $t-t'=\order(1)$, and so
we can approximate it by $\hat{K}\eql(0)\dLtwo(t,\tw)$. This gives
\be
\int_{\tw}^{t} dt'\,\dLtwo(t',\tw) = 
\frac{1}{\hat{K}\eql(0)}-
\frac{\hat{K}\eql(0)}{m^2(t)}\dLtwo(t,\tw)
\label{dl2}
\ee
and specifically in the limit $t/\tw\to 1$
\be
\dLtwo(\tw,\tw)=\hat{K}\eql^{-2}(0)m^2(\tw)
\label{dLtwo_initial}
\ee
To find $\dLtwo$ for $t/\tw>1$, we use that $m^2(t)=\mu_d/t$ for large
$t$ and in $d>4$, see\eqq{g_scaling_magn}{m_in_terms_of_g},
with some dimension-dependent coefficient $\mu_d$.
Taking a derivative of\eq{dl2} with respect to $t$ then gives
\be
\dLtwo(t,\tw)\left[1+\mu_d^{-1}\hat{K}\eql(0)\right]=
-t {\mu_d}^{-1}\hat{K}\eql(0)\partial_t\dLtwo(t,\tw)
\ee
This implies $\partial(\ln\dLtwo)/\partial(\ln t) =
-[1+\mu_d/\hat{K}\eql(0)]$ and so together with\eq{dLtwo_initial}
\be
\dLtwo(t,\tw) = \hat{K}\eql^{-2}(0)
\frac{\mu_d}{\tw}\left(\frac{t}{\tw}\right)^{-[1+\mu_d/\hat{K}\eql(0)]}
\label{incremento}
\ee
This can be simplified because in fact $\mu_d=\hat{K}\eql(0)$. To see
this, note from\eq{gs2} that $\hat{g}(s)/m^2(0)=1/[\hat{K}\eql(0)s^2]$
for small $s$; here we have used that $\hat{K}\eql(0)$ is finite for
$d>4$. Transforming back to the time domain gives
$g(t)/m^2(0)=m^{-2}(t)=t/\hat{K}\eql(0)$ for large $t$, i.e.\
$m^2(t)=\hat{K}\eql(0)/t$.
Thus\eq{incremento} simplifies to
\be
\dLtwo(t,\tw) = \frac{\tw}{\mu_d t^2}
\label{dLtwo}
\ee
This result is consistent with our assumption that $\dLtwo$
only varies on aging timescales. Overall, we have thus found for
$\Ltwo(t,\tw)$ the following long-time form
\be
\Ltwo(t,\tw)=\Ltwo\eql(t-\tw)\sc{\Kc}\left({t}/{\tw}\right)
+\frac{\tw}{\mu_d t^2}
\label{Ltwo_high_d}
\ee
Of course, for time differences $t-\tw=\order(1)$, $\dLtwo(t,\tw)$
will deviate from the form\eq{dLtwo} derived for the aging
regime. However, one can verify by expanding both sides of\eq{deltal2} to
$\order(t-\tw)$ that, even for $t=\tw$, $\dLtwo$ remains of order $1/\tw$,
so that these small deviations are always negligible compared to the
first term in\eq{Ltwo_high_d}.

We can now proceed to find $M$ as defined in\eq{Mdef}, and from there
finally the explicit forms of the magnetization correlation and
response functions. Using the general structure\eq{pinv_structure} of
$L$ we have
\be
m^{-1}(t)M(t,\tw)=\delta(t-\tw)+2\Tc-\int_{\tw}^{t}dt'\, \Ltwo(t',\tw)
\label{M_general}
\ee
The integral can be separated into the contributions from the two
parts of $\Ltwo$ as given in\eq{Ltwo_high_d}. The first part yields an
integral that converges for $t'-\tw=\order(1)$; for $t-\tw\gg 1$, it
therefore gives $\hat{L}\eql(0)=2\Tc-1/\hat{K}\eql(0)=2\Tc-1/\mu_d$ to leading
order. The second part, on the other hand, yields explicitly
$\int_{\tw}^{t}dt' (\tw/\mu_d t'^2) = \mu_d^{-1}(1-\tw/t)$, so that
\be
\fl m^{-1}(t)M(t,\tw)=\delta(t-\tw)+\frac{1}{\mu_d} -
\frac{1}{\mu_d}\left(1-\frac{\tw}{t}\right) = \delta(t-\tw) +
\frac{1}{\mu_d}\frac{\tw}{t}
\label{M}
\ee
This result applies for $t-\tw\gg 1$. For $t-\tw=\order(1)$ it is not
accurate; e.g.\ at $t=\tw$ the continuous part of $m^{-1}(t)M(t,\tw)$
is, from\eq{M_general}, $2\Tc$ rather than $1/\mu_d$. However, this
deviation over an $\order(1)$ time-range only gives subleading corrections in
the integrals over $M$ that we need below, as indeed does the
$\delta(t-\tw)$-term.
This can be seen in\eq{DM_part} and\eq{response_magn}, where only the
combination $\delta(t-\tw)-M(t,\tw)m(\tw)$ occurs; the latter can be written as
\be
\fl\delta(t-\tw)-M(t,\tw)m(\tw) = \delta(t-\tw)[1-m(t)m(\tw)] - 
\frac{\tw^{1/2}}{t^{3/2}} = \delta(t-\tw) - \frac{\tw^{1/2}}{t^{3/2}}
\label{1_minus_Mm}
\ee
The second form holds for long times, where the $m(t)m(\tw)$ term that
originated from the $\delta$-term in\eq{M} is negligible.

We can now work out the expression\eq{DM} for the full connected
correlation function. One can show that the contribution $\Dc^{(2)}$
involving $\CCt$ is negligible in the long-time limit.
In the expression\eq{c_m1_unscaled} for the remainder $\Dc^{(1)}$,
let us call the second and third term $I_2$ and $I_3$.
We need $\Cc_\zv$, which
from\eq{Cc_0} reads $\Cc_\zv(t',\tw')=\Tc \tw'^{3/2}t'^{-1/2}$ for $t'>\tw'$;
because $\Cc_\zv$ is symmetric in time this then implies
$\Cc_\zv(t',\tw')=\Tc t'^{3/2}\tw'^{-1/2}$ for $t'<\tw'$. Paying
due attention to this temporal ordering of the arguments of $\Cc_\zv$ and
using\eq{1_minus_Mm}, one finds
\bea
\fl I_2(t,\tw) &=& \Tc\left[\int_0^{\tw}dt'\, \frac{t'^{1/2}}{t^{3/2}}
\frac{t'^{3/2}}{\tw^{1/2}}+\int_{\tw}^{t}dt'\,
\frac{t'^{1/2}}{t^{3/2}}\frac{\tw^{3/2}}{t'^{1/2}}
\right] +  \Tc \int_0^{\tw}dt'\, \frac{t'^{1/2}}{\tw^{3/2}}
\frac{t'^{3/2}}{t^{1/2}}
\\
\fl &=& \Tc\left(\frac{\tw}{t}\right)^{3/2}\left(\frac{4}{3}t-\frac{2}{3}\tw\right)
\eea
Similarly, the double integral in\eq{c_m1_unscaled} can be evaluated as
\bea
\fl I_3(t,\tw)&=&\int_0^t dt'\int_0^{\tw}\!\!d\tw'\,
\frac{t'^{1/2}}{t^{3/2}}\frac{\tw'^{1/2}}{\tw^{3/2}}\,\Cc_\zv(\tw',t')
\\
\fl &=&\Tc\int_{\tw}^t dt'\int_0^{\tw}\!\!d\tw'\, \frac{t'^{1/2}}{t^{3/2}}
\frac{\tw'^{1/2}}{\tw^{3/2}}\frac{\tw'^{3/2}}{t'^{1/2}}
+2\Tc\int_0^{\tw}\!\! dt'\int_0^{t'}\!\!d\tw'\, \frac{t'^{1/2}}{t^{3/2}}
\frac{\tw'^{1/2}}{\tw^{3/2}}\frac{\tw'^{3/2}}{t'^{1/2}}
\\
\fl &=&\frac{\Tc}{3}\left(\frac{\tw}{t}\right)^{3/2}(t-\tw)
+ \frac{2\Tc}{3}\int_0^{\tw} dt'\frac{t'^3}{t^{3/2}\tw^{3/2}}
\\
\fl &=&\frac{\Tc}{6}\left(\frac{\tw}{t}\right)^{3/2}(2t-\tw)
\eea
Our final long-time result for the connected magnetization correlator
including non-Gaussian corrections is then
\bea
\fl \Dc(t,\tw)&=&
\Tc\frac{\tw^{3/2}}{t^{1/2}}+I_2(t,\tw)+I_3(t,\tw)\\
\fl &=&
\Tc\left(\frac{\tw}{t}\right)^{3/2}\left(t-\frac{4}{3}t+\frac{2}{3}\tw
+\frac{1}{3}t-\frac{1}{6}\tw\right)
\ = \ \frac{\Tc\tw}{2}
\left(\frac{\tw}{t}\right)^{3/2}
\label{Cm_dgt4}
\eea
For the conjugate magnetization response function we get
from\eq{response_magn} and\eq{M}, together with
$R_\zv(t,\tw)=(\tw/t)^{1/2}$,
\bea
\fl \Q(t,\tw) &=& 
R_\zv(t,\tw)-\int_{\tw}^t dt'\,
\frac{t'^{1/2}}{t^{3/2}}R_\zv(t',\tw)
\\
\fl &=&\left(\frac{\tw}{t}\right)^{1/2}-\frac{\tw^{1/2}}{t^{3/2}}(t-\tw)
=\left(\frac{\tw}{t}\right)^{3/2}
\label{Rm_dgt4}
\eea
The FDR follows finally as
\be
X_{\rm m}(t,\tw)=\frac{\Tc\Q(t,\tw)}{\Dc'(t,\tw)}=\frac{4}{5}
\label{X_m_d_gt_4}
\ee
Interestingly, this agrees exactly with the result for the Ising
ferromagnet in the limit of large dimensionality
$d$~\cite{GarSolPagRit05}. As in the unmagnetized case, we see
therefore that it is the {\em global} observables in the spherical
model, which are strongly affected by non-Gaussian fluctuations, that
behave like their analogues in short-range models. In fact, even the
expressions for the correlation and response functions we find here
are identical to those in the large-$d$ Ising case, implying
``universality'' at a more detailed level than one might have expected.

It is worth noting
that the effect of the non-Gaussian corrections is very large:
compared to the Gaussian result
$\Cc_\zv(t,\tw)\sim \tw(\tw/t)^{1/2}$, the corrections increase the
decay exponent 
to $\Dc(t,\tw)\sim\tw(\tw/t)^{3/2}$, so that $\Dc/\Cc_\zv \sim \tw/t
\ll 1 $ for $t\gg \tw$: there is an almost perfect cancellation of
Gaussian terms and non-Gaussian corrections for well-separated
times. Similar comments apply to the response. The
overall effect of the non-Gaussian corrections on the FD relation is
to leave this as a straight line (since $X$ is constant), but to
increase the slope from 2/3 to 4/5.

\subsection{Magnetization correlation and response: Non-equilibrium, $d<4$}

We now consider systems below the upper critical dimension, $d<4$;
here there are no predictions yet from other models for the
non-equilibrium FD behaviour following a quench of a magnetized
initial state to $\Tc$ (but see Sec.~\ref{sec:conclusion}). As in the
case of the energy correlations for 
unmagnetized initial states, leading order cancellation effects have
to be taken care of for these low values of $d$.

We again need to know the scaling of $K$ to determine $\Ltwo$; from
this we then get $M$ and finally the correlation and response
functions. The long-time scaling of the connected part of $K$ is
$\Kc(t,t')=K\eql(t-t')\sc{\Kc}({t}/{t'})$ as before, with $\sc{\Kc}$
given by \eq{sc_Kc} but now $\alpha=(d-2)/2$.  The contribution to $K$
from $\qv=\zv$, given by second term in\eq{K_decomp}, is negligible
relative to $\Kc(t,t')$ for $t-t'=\order(1)$. However, for $t-t'\gg
1$, it becomes comparable and has the same overall time scaling as
$\Kc(t,t')$ in the aging regime. To see this explicitly, recall
from\eqq{g_scaling_magn}{m_in_terms_of_g} that
the square of the magnetization decays asymptotically as $m^2(t)=\mu_d
t^{-\alpha}=\mu_d t^{(2-d)/2}$ with some constant $\mu_d$. Similarly,
the equilibrium part of $\Kc$ behaves as $K\eql(t-t') =
k_d(t-t')^{(2-d)/2}$ for $t-t'\gg 1$ (see after\eq{p_eql}). This gives
\be 
\frac{m^2(t)}{K\eql(t-t')}
=
\frac{\mu_d(t-t')^{(d-2)/2}}{k_dt^{(d-2)/2}}=
\frac{\mu_d}{k_d}\left(\frac{t/t'-1}{t/t'}\right)^{(d-2)/2}
\label{scaling}
\ee
in the aging regime where both $t'$ and $t-t'$ are large.
(For $t-t'=\order(1)$ this
expression is not accurate but this is irrelevant because there the
term $m^2(t)$ is subleading compared to $\Kc(t,t')$ anyway.)
We thus have the overall scaling of $K$
\be
\fl K(t,t')=K\eql(t-t')\sc{K}(t/t'), 
\quad
\sc{K}(x) = \sc{\Kc}(x)+\frac{\mu_d}{k_d}
\left(\frac{x-1}{x}\right)^{(d-2)/2}
\label{Ktot}
\ee
To simplify the scaling function, we integrate the
expression\eq{sc_Kc} by parts and rescale $y\to xy$, bearing in mind
that $\alpha=(d-2)/2$:
\bea
\sc{\Kc}(x)&=&\frac{d-2}{2} \left(\frac{x-1}{x}\right)^{(d-2)/2}
\left[\frac{2}{d-2}(x-1)^{(2-d)/2}\right.   
\nn
& & {}-{}\left.\int_0^{1/x}dy\,y^{(d-4)/2}(1-y)^{(2-d)/2}\right]
\eea
For $x\to 1$ the integral becomes a Beta-function which evaluates to
$\Gamma((d-2)/2) \Gamma((4-d)/2)$, and extracting this term gives
\bea
\fl \sc{\Kc}(x)&=& x^{(2-d)/2}-\Gamma\left({\ts\frac{d}{2}}\right)
\Gamma\left({\ts\frac{4-d}{2}}\right) \left(\frac{x-1}{x}\right)^{(d-2)/2}
\nn
\fl & & {}+{} \frac{d-2}{2}
\left(\frac{x-1}{x}\right)^{(d-2)/2}\int_{1/x}^1 dy\,
y^{(d-4)/2}(1-y)^{(2-d)/2}
\eea
The second term has the same dependence on $x$ as the additional
contribution from zero wavevector in\eq{Ktot}. In fact, it turns out
that these two terms cancel exactly: from our
definitions of $k_d$ and $\mu_d$ we have $K\eql(t)=k_dt^{(2-d)/2}$ for
large $t$, while $m^{-2}(t)=g(t)/m^2(0)=\mu_d^{-1}t^{(d-2)/2}$. Laplace
transforming gives $\hat{K}\eql(s)=k_d \Gamma((4-d)/2) s^{(d-4)/2}$
and $\hat{g}(s)/m^2(0) = \mu_d^{-1}\Gamma(d/2)s^{-d/2}$ to leading
order for small $s$. But then\eq{gs2} shows that
$\mu_d^{-1}\Gamma(d/2)=[k_d\Gamma((4-d)/2)]^{-1}$, or $\mu_d/k_d =
\Gamma(d/2)\Gamma((4-d)/2)$, proving the cancellation anticipated
above. Overall, we thus have for the scaling function of $K$
\be
\fl\sc{K}(x)=x^{(2-d)/2}+\frac{d-2}{2}\left(\frac{x-1}{x}\right)^{(d-2)/2}
\int_{1/x}^1 dy\,y^{(d-4)/2}(1-y)^{(2-d)/2}
\label{Ktot_final}
\ee
Expanding for $x\approx 1$, one sees that the leading order variation
is linear in $x-1$:
\bea
\fl \sc{K}(x)&\approx&1+\frac{2-d}{2}(x-1)+
\frac{d-2}{2}\left(\frac{x-1}{x}\right)^{(d-2)/2}
\int_{1/x}^1dy\, (1-y)^{(2-d)/2}
\nn
\fl &\approx&1+\half\frac{(d-2)^2}{4-d}(x-1)
\eea
Note that the prefactor is positive, so that $\sc{K}(x)$ {\em
increases} with $x$ in the current scenario. This trend persists for
all $x$, not just $x\approx 1$, and the scaling function monotonically
approaches a limit value for $x\to\infty$. The latter
follows from\eq{Ktot} as $\mu_d/k_d=\Gamma(d/2)\Gamma((4-d)/2)$,
since the connected contribution $\sc{\Kc}(x)$ decays to zero for
$x\to\infty$.

With the single overall scaling\eq{Ktot} of $K$ we no longer need to
decompose $\Ltwo$ into $\Lctwo$ and $\dLtwo$ as we did in $d>4$;
instead the long-time behaviour of $\Ltwo$ will have the same 
structure as in the unmagnetized case,
\be
\Ltwo(t,\tw)=\Ltwo\eql(t-\tw)\sc{L}(t/\tw)
\label{Ltot}
\ee
One can then follow exactly the discussion in Sec.~\ref{sec:KandL} to
arrive at the integral equation\eq{pinv_cond}
for $\sc{L}(x)$. Solving the latter
looks a rather formidable task, given that $\sc{K}$ itself has the
relatively complicated form\eq{Ktot_final}. Remarkably, however,
the solution can be found in closed form and is given simply by
\be
\sc{L}(x)=\frac{2}{4-d}\,x^{(2-d)/2}+\frac{2-d}{4-d}\,x^{-d/2}
\label{solution_L}
\ee
We were led to this result initially by a systematic series expansion
of both $\sc{K}(x)$ and $\sc{L}(x)$ in terms of $(x-1)/x$. We do not
detail this here, but verify in~\ref{sec:L_solution} by
direct calculation that\eq{solution_L} does indeed
solve\eq{pinv_cond}.

We next calculate the kernel $M(t,\tw)$. One inserts the
scaling\eq{Ltot} into\eq{M_general}, subtracting off and adding back
on the contribution from the equilibrium part of $\Ltwo$:
\bea
m^{-1}(t)M(t,\tw)&=&\delta(t-\tw)+2\Tc-\int_{\tw}^t
dt'\,\Ltwo\eql(t'-\tw)
\nn
& &{}+{}\int_{\tw}^{t} dt'\,\Ltwo\eql(t'-\tw)
\left[1-\sc{L}\left(\frac{t'}{\tw}\right)\right]
\label{Mt}
\eea
This is done to account for the leading order cancellation of the
second and third terms:
\be
\fl 2\Tc-\int_{\tw}^t\!\!
dt'\,\Ltwo\eql(t'-\tw) = 
2\Tc-\hat\Ltwo\eql(0)+\int_t^\infty \!\!dt'\,\Ltwo\eql(t'-\tw) = 
\frac{2\lambda_d}{4-d}
(t-\tw)^{({d-4})/{2}}
\ee
where in the last step we have restricted ourselves to the aging
regime $t-\tw\gg 1$ and used the asymptotic
behaviour\eq{ptwo_eql_asymptotics}, $\Ltwo\eql(t'-\tw)=\lambda_d
(t'-\tw)^{(d-6)/2}$. In the remaining integral in\eq{Mt}, the factor
$1-\sc{L}(t'/\tw)\sim (t'-\tw)/\tw$ ensures that the integration no
longer has its weight concentrated near $t'=\tw$; put differently,
after scaling the integration variable by $\tw$ to $y=t'/\tw$ the
integral is convergent at the lower end. We thus obtain in the aging
regime
\bea
\fl m^{-1}(t)M(t,\tw)&=&\delta(t-\tw) + \lambda_d\tw^{(d-4)/2}
\left\{\frac{2}{4-d}(x-1)^{(d-4)/2} \right. 
\nn
& & {}+{} \left. \int_1^x dy\,(y-1)^{(d-6)/2}[1-\sc{L}(y)]\right\}
\label{initial_M}
\eea
Inserting\eq{solution_L} for $\sc{L}$, the integral in\eq{initial_M}
can be done explicitly to give
$[2/(4-d)]x^{(2-d)/2}(x-1)^{(d-4)/2}$ for the sum of the terms in
curly brackets. A little care is needed
here because the separate integrals over $(y-1)^{(d-6)/2}$ and
$(y-1)^{(d-6)/2}\sc{L}(y)$ are divergent at the lower end.
One can avoid this by analytical continuation from $d>4$, where these
divergences are absent, or by integrating from $1+\epsilon$ to $x$ and
taking $\epsilon\to 0$ at the
end.
The $\delta$-term is again subleading for long times in the relevant
combination\eq{1_minus_Mm}, which we can write as
\bea
\fl \delta(t-\tw)-M(t,\tw)m(\tw) &=& \delta(t-\tw) -
 \frac{2}{4-d}\frac{\lambda_d\mu_d\tw^{(d-4)/2}}{\tw^{(d-2)/4}t^{(d-2)/4}}x^{(2-d)/2}(x-1)^{(d-4)/2}
\nn
\fl &=& \delta(t-\tw)-\frac{1}{t}\sc{M}(x)
\label{combination}
\eea
with
\be
\sc{M}(x) = \frac{d-2}{2}\, x^{(2-d)/4}\left(\frac{x-1}{x}\right)^{(d-4)/2}
\label{M_scaling_final}
\ee
Here we have eliminated the constants $\lambda_d$ and
$\mu_d$ using the following argument. From\eq{gs2},
$\hat{g}(s)/m^2(0)=\hat{L}\eql(s)/s^2$ for small
$s$, i.e.\ $\hat{L}\eql(s)=s^2\hat{g}(s)/m^2(0)$. In the time domain
this gives at long times
$\Ltwo\eql(t)=-L\eql(t)=-(\partial_t)^2m^{-2}(t)=-\mu_d^{-1}(\partial_t)^2
t^{(d-2)/2} = -\mu_d^{-1}[(d-2)/2][(d-4)/2]t^{(d-6)/2}$, so that
$\lambda_d\mu_d=(d-2)(4-d)/4$.

Reassuringly, for $d\to 4$ the result\eq{M_scaling_final} tends to
$\sc{M}(x)=x^{-1/2}$,
matching smoothly onto the result\eq{1_minus_Mm} we found earlier in $d>4$.
For $d<4$, the scaling function diverges as $x\to 1$. Since
from\eq{M_general} the continuous part of $M(t,\tw)$ is exactly given
by $2\Tc m(t)$ for equal times, this indicates that the above aging
regime expression must break down eventually when $t-\tw$ becomes
small, as expected. In the integrals where $M$ appears below such
effects can be neglected, however, because they only give subleading
corrections.

With the scaling of $\sc{M}$ in hand we can now compute the connected
correlation function $\Dc(t,\tw)$ including non-Gaussian corrections,
given by\eq{DM}. After rescaling the integration variables
the first part\eq{DM_part} can be written as
\bea
\fl \Dc^{(1)}(t,\tw)&=& \int_0^x dy\,
[\delta(y-x)-x^{-1}\sc{M}(x/y)]
\nn
\fl & & \times \int_0^1 d\yw\,
[\delta(\yw-1)-\sc{M}(1/\yw)] \Cc_\zv(\tw y,\tw y')
\label{Ione}
\eea
where the Gaussian magnetization correlator is
\be
\fl \Cc_\zv(\tw y,\tw y')
= ({4\Tc\tw}/{d})\min\left\{y(y/\yw)^{(d-2)/4},\yw(\yw/y)^{(d-2)/4}\right\}
\ee
from\eq{Cc_0}.
At first sight\eq{Ione} suggests, e.g.\ from the $\delta(y-x)$ term,
an asymptotic decay of $\Dc^{(1)}\sim \tw x^{(2-d)/4}$
for large $x$. But this would not match continuously with the
result\eq{Cm_dgt4} we found for $d>4$. A cancellation of such leading
order terms must therefore occur for $x\to\infty$. To show this
explicitly, we verify from\eq{M_scaling_final} the identity
\bea
\fl\lefteqn{\int_0^x dy\,
  \left[\delta(y-x)-x^{-1}\sc{M}\left({x}/{y}\right)\right]y^{-(d-2)/4} = }
\nn 
&=& x^{(2-d)/4}-\frac{d-2}{2x}\int_0^x
 dy\,(x/y)^{(2-d)/4}(1-y/x)^{(d-4)/2}y^{(2-d)/4}\
\\
&=& x^{(2-d)/4}-\frac{d-2}{2}x^{(2-d)/4}\int_0^1 dz\,(1-z)^{(d-4)/2} =  0
\label{identity}
\eea
Multiplying this by $({4\Tc\tw}/d) \int_0^1 d\yw\, [\delta(\yw-1)-
\sc{M}(1/\yw)]\yw^{(d+2)/4}$ and subtracting from\eq{Ione} exactly
cancels all contributions in the range $y>\yw$, giving
\bea
\fl \Dc^{(1)}(t,\tw)&=&\frac{4\Tc \tw}{dx} \int_0^1 d\yw\,
\int_0^{\yw} dy\, \sc{M}\left({x}/{y}\right)
\left[\delta(\yw-1)-\sc{M}\left({1}/{\yw}\right)\right] 
\nn
\fl & & \times \left[\yw\left(\frac{\yw}{y}\right)^{(d-2)/4}
- y\left(\frac{y}{\yw}\right)^{(d-2)/4}\right] 
\label{C_m1}
\eea
This shows that $\Dc^{(1)}/\tw$ is a scaling function of
$x=t/\tw$. Its full $x$-dependence has to be found numerically
from\eq{C_m1} or via series
expansion~\cite{Annibale_thesis},
but we can obtain the large-$x$ behaviour that is required for the
asymptotic FDR $X^\infty$ in closed form. For $x\rightarrow\infty$
one can replace the function 
$\sc{M}(x/y)$ with its asymptotic form
$[(d-2)/2](x/y)^{(2-d)/4}$ from\eq{M_scaling_final} to get 
\bea
\fl \frac{\Dc^{(1)}(t,\tw)}{\Tc\tw}
&=&\frac{2(d-2)}{d} x^{-(d+2)/4}
\int_0^1 d\yw\, \int_0^{\yw} dy\,y^{(d-2)/4}
\left[\delta(\yw-1)-\sc{M}\left(\frac{1}{\yw}\right)\right]
\nn
\fl& & \times \left[\yw\left(\frac{\yw}{y}\right)^{(d-2)/4}
-y\left(\frac{y}{\yw}\right)^{(d-2)/4}\right]
\\
\fl&=&\frac{2(d-2)}{d+2} x^{-(d+2)/4}
\left[1-\frac{d-2}{2} \frac{\Gamma\left(\frac{d+4}{2}\right)
\Gamma\left(\frac{d-2}{2}\right)}{\Gamma(d+1)}\right]
\label{C_m1inf}
\eea
This exhibits the expected leading order cancellation for large
$x$, which gives an additional factor of $1/x$ compared to the naive
result $x^{(2-d)/4}$.

To complete the calculation of the correlation function we need to
evaluate $\Dc^{(2)}$ from\eq{c_m2}, which cannot be neglected for
$d<4$.  This requires the long-time behaviour of $\CCt(t,\tw)$, which is
given by\eq{sc_CC_ord} and\eq{sc_CC_dis} for $t>\tw$ and
$t<\tw$, respectively, as for the unmagnetized case. The only
modification arises from the different behaviour of $g(t)$. One thus finds
\be
\fl \frac{\CCt(t,\tw)}{\CCt(t,t)}={\mathcal{G}}(t/\tw),
\quad
{\mathcal{G}}(x)=\left\{
\begin{array}{ll}
{\displaystyle \frac{\int d\omt\, \omt^{(d-6)/2} \sc{C}^2(\omt) e^{-2(x-1)\omt}}
{x\int d\omt\, \omt^{(d-6)/2} \sc{C}^2(\omt)}} & \mbox{for $x\geq 1$} \\
x^{(d-6)/2}\G(1/x) & \mbox{for $x\leq 1$}
\end{array}
\right.
\label{CCt_scaling}
\ee
The scaling of the equal-time value of $\CCt$ 
is, from\eq{sc_CC_ord}, $\CCt(t,t)=\CCtd t^{(4-d)/2}$ with
$\CCtd = \sigma_d \Tc^2\int d\omt\, \omt^{(d-6)/2} \sc{C}^2(\omt)$;
compare\eq{CCd}. Inserting this into\eq{c_m2} gives
\bea
\fl\lefteqn {\Dc^{(2)}(t,\tw)=
\half \int\, dt'\,d\tw'\, M(t,t')M(\tw,\tw') \CCt(t',\tw')} 
\\
\fl &=&\half \int\, dt'\,d\tw'\, \frac{1}{m(t')m(\tw')}
\frac{1}{t}\sc{M}(t/t')\frac{1}{\tw}\sc{M}(\tw/\tw')
\CCt(t',t') \mathcal{G}(t',\tw')
\\
\fl &=&\frac{\tw \CCtd}{2\mu_d x} 
\int_0^1 d\yw\,\yw^{(d-2)/4}\sc{M}(1/\yw)
\int_0^{x}dy\, \sc{M}(x/y) y^{(6-d)/4} {\mathcal{G}}(y/\yw)
\\
\fl &=&\frac{\tw}{x} \int_0^1 d\yw\, \yw^2 \sc{M}(1/\yw)
\int_0^{x/\yw}du\, \sc{M}(x/u\yw)u^{(6-d)/4}
\frac{\CCtd}{2\mu_d}{\mathcal{G}}(u)
\label{general_Cm2}
\eea
In the second line we have used\eq{combination}
to write $M(t,t')=m(t)\delta(t-t')-t^{-1}m^{-1}(t')\sc{M}(t/t')$ up to
negligible corrections, and then 
immediately discarded the $\delta$-function contributions, which give
subleading corrections.

Let us denote by $U$ the value of the $u$-integral
in\eq{general_Cm2}. Since $\mathcal{G}(u)$ is defined separately for
$u>1$ and $u<1$ in\eq{CCt_scaling}, one splits the integral accordingly:
\bea 
\fl \frac{2\mu_d U}{\CCtd} &=&
\int_0^1\!\! du\, \sc{M}(x/u\yw)u^{(d-6)/4} {\mathcal{G}}(1/u)+
\int_1^{x/\yw}\!\! du\, \sc{M}(x/u\yw)u^{(6-d)/4} {\mathcal{G}}(u)
\\
\fl &=& \int_1^\infty\!\! du\, \sc{M}(xu/\yw)u^{-(d+2)/4} {\mathcal{G}}(u)+
\int_1^{x/\yw}\!\! du\, \sc{M}(x/u\yw)u^{(6-d)/4} {\mathcal{G}}(u)
\label{decomposition}
\eea
We now need ${\mathcal{G}}(u)$ for $u>1$. The denominator
in\eq{CCt_scaling} is $\CCtd/(\sigma_d \Tc^2)$,
and bearing in mind the
definition\eq{C_scaling_magn} of $\sc{C}$ with $\alpha=(d-2)/2$ gives
\bea
\fl \frac{\CCtd}{\sigma_d \Tc^2} u{\mathcal{G}}(u) &=&
\int d\omt\, \omt^{(d-6)/2}\sc{C}^2(\omt)e^{-2(u-1)\omt}
\\
\fl &=& 4\int d\omt\, \omt^{(d-2)/2}\int_0^1 dy\,\int_0^1 dy'\,
(yy')^{(d-2)/2}e^{-2\omt(1-y-y'+u)}
\\
\fl &=&2^{(4-d)/2}\Gamma\left({\ts\frac{d}{2}}\right)
\int_0^1 dy\,\int_0^1 dy'\, (yy')^{(d-2)/2}(1-y-y'+u)^{-d/2}
\label{3_dim_int}
\eea 
We now insert this into\eq{decomposition} and simplify the
numerical prefactors by using $\mu_d=\lambda_d^{-1}(d-2)(4-d)/4$ and
the explicit expression\eq{Ld} for $\lambda_d$ to get
\bea 
\fl
U&=& \frac{\Tc}{\Gamma(\frac{d-2}{2}) \Gamma(\frac{4-d}{2})}
\left[\int_1^{\infty}du\, \sc{M}(xu/\yw)u^{-(d+6)/4} \right. \nn
\fl & &\times \int_0^1 dy \int_0^1 dy'\,(yy')^{(d-2)/2}(1-y-y'+u)^{-d/2}
\nn
\fl &&{}+{}\left. \int_1^{x/\yw}du\, \sc{M}(x/u\yw)u^{(2-d)/4}
\int_0^1 dy \int_0^1 dy'\, \ldots\right]
\label{exact_I}
\eea
This is the $u$-integral from\eq{general_Cm2} and so overall we have
a four-dimensional integral over $\yw, u, y, y'$ for $\Dc^{(2)}$. In
general this cannot be evaluated in closed form; a series expansion is
given
in~\cite{Annibale_thesis}.
The large-$x$ behaviour, which will give
us the asymptotic FDR, is easier to extract. In the first $u$-integral
of\eq{exact_I} one can directly use the asymptotic form of
$\sc{M}(xu/\yw)$. One can show that for large $x$ the same replacement
can be made in the second integral, and the upper integration limit sent to
infinity thereafter.
This gives for the large-$x$ behaviour of $U$
\be
U= \frac{d-2}{2}\frac{\Tc V_d}{\Gamma(\frac{d-2}{2}) \Gamma(\frac{4-d}{2})} 
x^{(2-d)/4} \yw^{(d-2)/4} 
\label{U_large_x}
\ee
where $V_d$ is a $d$-dependent numerical constant given by
\be
\fl V_d=\int_1^{\infty}\!\!du\, (u^{-(d+2)/2}+1) \int_0^1\! dy \int_0^1\! dy'\,
(yy')^{(d-2)/2}(1-y-y'+u)^{-d/2}
\label{integral_sums}
\ee 
Inserting\eq{U_large_x} into\eq{general_Cm2}, the remaining
$\yw$-integral can be done explicitly to give
\bea
\fl \Dc^{(2)}(t,\tw)=\left(\frac{d-2}{2}\right)^2
\frac{\Gamma(\frac{d+4}{2}) V_d}{\Gamma(d+1)\Gamma(\frac{4-d}{2})}
\Tc\tw x^{-(d+2)/4}
\label{C_m2inf}
\eea 
As anticipated this has the same scaling as the first
contribution\eq{C_m1inf} to the correlation function, so that overall
for large $x$
\bea
\fl \Dc(t,\tw) = \Dc^{(1)}+\Dc^{(2)}&=&\frac{d-2}{2}\Tc\tw 
x^{-(d+2)/4}\left[\frac{4}{d+2} 
\left(1-\frac{d-2}{2}\frac{\Gamma(\frac{d+4}{2})\Gamma(\frac{d-2}{2})}
{\Gamma(d+1)}\right)\right. 
\nn 
& &{}+{}\left.\frac{d-2}{2}\frac{\Gamma(\frac{d+4}{2})V_d}
{\Gamma(d+1)\Gamma(\frac{4-d}{2})}\right]
\label{Cm_tot_infty}
\eea

The magnetization {\em response} function is rather easier to
find, by using\eq{R_simplification_magn} and\eq{combination}
in\eq{response_magn} 
and rescaling the integration variable to $y=t'/\tw$ as usual:
\be
\Q(t,\tw) = \int_1^x dy\,\left[\delta(y-x)-x^{-1}\sc{M}(x/y)\right]y^{-(d-2)/4}
\ee
The structure of this is rather similar to $\Dc^{(1)}$, and
by subtracting the vanishing term\eq{identity} one again gets a
significant cancellation,
\be
\Q(t,\tw) = \frac{1}{x}\int_0^1 dy\,\sc{M}(x/y)y^{(2-d)/4}
\ee
Inserting the explicit form\eq{M_scaling_final} of $\sc{M}$ one then
gets simply
\be
\Q(t,\tw) = x^{(2-d)/4}\left[1-\left(1-\frac{1}{x}\right)^{(d-2)/2}\right]
\label{R_finite}
\ee
which for $x\rightarrow\infty$ behaves as
\be
\Q(t,\tw) = \frac{d-2}{2}x^{-(2+d)/4}
\label{R_infty}
\ee

With the expressions $\eq{Cm_tot_infty}$ and $\eq{R_infty}$ for
correlation and response in the limit of long, well-separated times we
can now finally compute the asymptotic FDR defined by
\bea
\fl X_{\rm m}^\infty &=&
\lim_{t\gg\tw\gg 1}\frac{\Tc\Q(t,\tw)}{\Dc'(t,\tw)}
=\frac{4}{d+6}\left[\frac{4}{d+2}
\left(1-\frac{d-2}{2}\frac{\Gamma(\frac{d+4}{2})\Gamma(\frac{d-2}{2})}
{\Gamma(d+1)}\right)\right. 
\nn 
\fl & &{}+{}\left.\frac{d-2}{2}\frac{\Gamma(\frac{d+4}{2}) V_d}
{\Gamma(d+1)\Gamma(\frac{4-d}{2})}\right]^{-1}
\label{Xz}
\eea
where $V_d$ is given by\eq{integral_sums} and the prefactor $4/(d+6)$ accounts
for the fact that $\Dc\sim \tw x^{-(2+d)/4} \sim\tw^{(d+6)/4}$ and hence
$\Dc'(t,\tw)=[(d+6)/4\tw]\Dc(t,\tw)$.

Before exploring this result, let us comment briefly on the limit as
$d\to 4$, which should make contact with our results in the
previous subsection. The contribution from $\Dc^{(2)}$, which appears
in the second line of\eq{Cm_tot_infty} and\eq{Xz}, vanishes linearly
in $4-d$ in this limit because of the factor $\Gamma^{-1}((4-d)/2)$;
$V_d$ stays finite as we show below. This is consistent with the fact
that in dimension $d>4$ this term does not contribute.  In the limit
$d\to 4$ one has, from\eq{Cm_tot_infty}, $\Dc=\Tc\tw x^{-3/2}/2$ for
large $x$ which matches precisely\eq{Cm_dgt4} for 
$d>4$. Similarly, the large-$x$
magnetization response\eq{R_finite} for $d\to 4$ is $\Q=x^{-3/2}$ in
agreement with\eq{Rm_dgt4}.

We now look in more detail at the $d$-dependence of the asymptotic FDR\eq{Xz}
for the magnetization. Expanding in $\epsilon=4-d$ one has 
\bea
\fl X_{\rm m}^\infty
&=&\left(\frac{2}{5}+\frac{\epsilon}{25}+\order(\epsilon^2)\right)
\left[\left(\frac{2}{3}+\frac{\epsilon}{9}\right)
\left(\frac{3}{4}-\frac{\epsilon}{6}\right)+
\frac{\epsilon}{8} V_4+\order(\epsilon^2)\right]^{-1}
\nn
&=&\frac{4}{5}+\left(\frac{28}{225}-\frac{V_4}{5}\right)\epsilon
+\order(\epsilon^2)
\eea
where $V_4$ is the limiting value of $V_d$ for $d\to 4$, which can be
worked out explicitly as
\be
V_4=-\frac{2}{3}\ln 2+\frac{11}{12}+\frac{\pi^2}{24}
\ee
so that 
\be
\fl X_{\rm m}^\infty=\frac{4}{5}+\left(\frac{2\ln 2}{15}
-\frac{53}{900}-\frac{\pi^2}{120}\right)\epsilon+\order(\epsilon^2)
\label{X_m_epsilon_expansion}
\ee
It is remarkable that in a system as simple as the spherical model,
where standard critical exponents are rational functions of the
dimension $d$, the magnetization FDR for magnetized initial states is
very much more complicated, and irrational already {\em to first order} in
$\epsilon$. 

\begin{figure}
\begin{center}
\centerline{\includegraphics[width=10.0cm,clip=true]{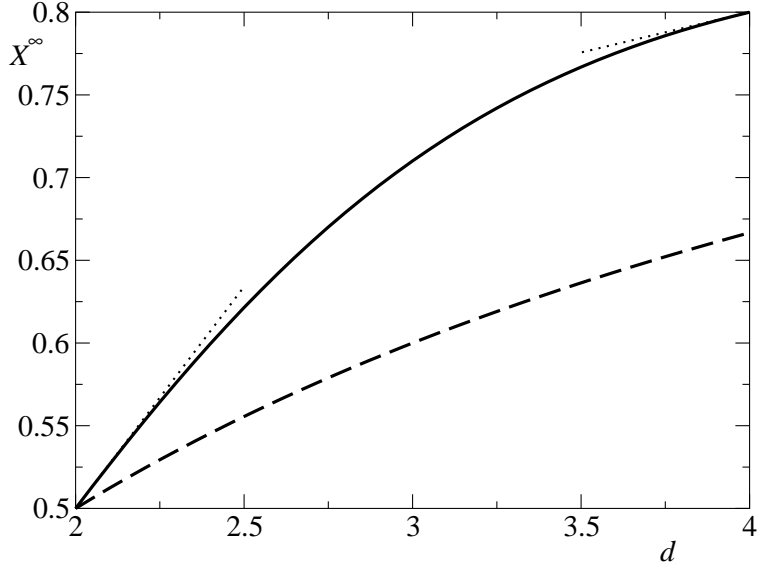}}
\caption{Asymptotic FDR $X^\infty_{\rm m}$ for the magnetization, for
critical coarsening with nonzero initial magnetization. Solid line:
Full theory~(\protect\ref{Xz}) including non-Gaussian corrections;
dotted lines indicate the first-order expansions near $d=2$ and
4. Dashed line: Gaussian theory~(\protect\ref{Xqneq}).
\label{fig:X_magnetized}
}
\end{center}
\end{figure}
In the opposite limit $d\to 2$, the $u$-integral in the
definition\eq{integral_sums} of $V_d$ diverges at the upper end;
dropping all non-divergent corrections gives the leading divergence as
\be
\fl V_d\approx\int_1^{\infty}du\,\int_0^1 dy \int_0^1 dy'\,(1-y-y'+u)^{-d/2}
\approx \int_1^{\infty}du\,u^{-d/2} = \frac{2}{d-2}
\label{limit_d2}
\ee
This divergence balances the vanishing prefactor
$(d-2)/2$ in\eq{Cm_tot_infty} and\eq{Xz} whereas the contribution in
the round brackets coming from $\Dc^{(1)}$ vanishes linearly with
$d-2$. Consequently the asymptotic behaviour of the correlation
function is, for $d$ close to $2$, $\Dc=[(d-2)/2]\Tc\tw x^{-1}$ and
the FDR becomes $X_{\rm m}^\infty=4/(d+6)=1/2$. One can also obtain
the leading order correction in $\epsilon'=d-2$, which is given by
\be
X_{\rm m}^\infty=
\frac{1}{2}+\left(\frac{1}{16}+\frac{\pi^2}{48}\right)\epsilon'
+\order(\epsilon'^2)
\label{X_m_epsilon_prime_expansion}
\ee
The only subtlety here is working out the subleading term of
$V_d$. Setting $V_d=2/\epsilon'+a_0+\ldots$, $a_0$ can be obtained as
the limit for $d\to 2$ of\eq{integral_sums} with $u^{-d/2}$ subtracted
from the integrand. The limit can be taken in the integrand itself for
finite $u$, giving a convergent integral with value
$3/2-\pi^2/12$. But one has to account separately for the large-$u$
tail $u^{-d/2}[-1+\int dy\,dy' (yy')^{(d-2)/2}]$ which integrates to
$(-1+4/d^2)[2/(d-2)]\to -2$ for $d\to 2$, giving $a_0=-1/2-\pi^2/12$
overall.

In summary, the asymptotic magnetization FDR $X_{\rm m}^\infty$ decays
from $4/5$ in $d=4$ to $1/2$ in $d=2$. Fig.~\ref{fig:X_magnetized} also
shows numerical values for intermediate $d$. $X_{\rm m}^\infty$ is
larger than the 
FDR\eq{Xqneq} from the Gaussian theory except in the limit $d\to2$; it also
remains larger throughout than the FDR\eq{X_baseline} for
unmagnetized initial states, shown in Fig.~\ref{fig:X_normal_and_E}.

\begin{figure}
\begin{center}
\centerline{\includegraphics[width=9.0cm,clip=true]{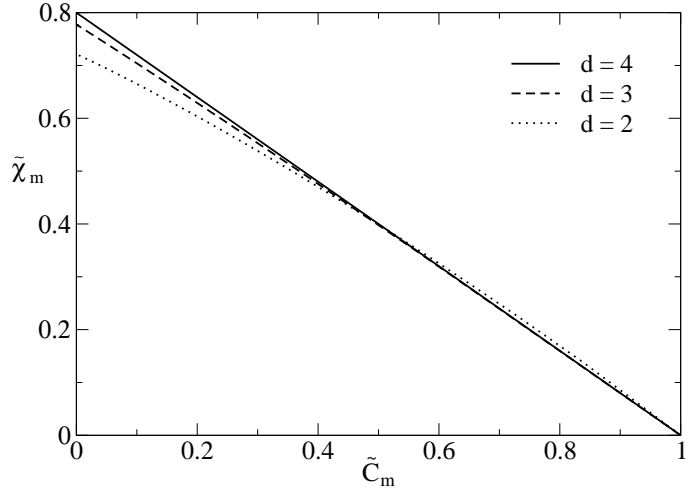}}
\caption{Normalized magnetization FD plot for $d=2, 3, 4$, showing
normalized susceptibility $\tilde{\chi}_{\rm m}$ versus normalized
correlation $\tilde\Dc$ in the limit of long times. For $d=4$ the plot
is a straight line with (negative) slope $4/5$ as expected. Increasing
deviations from this appear as $d$ decreases towards $2$.
\label{fig:mFD_plot}
}
\end{center}
\end{figure}
We next turn to the shape of the FD plot for the magnetization. As
explained in the introduction, this
is obtained by plotting the normalized susceptibility
$\tilde{\chi}_{\rm m}(t,\tw) = \Tc\chi_{\rm m}(t,\tw)/\Dc(t,t)$ versus
$\tilde\Dc(t,\tw)=\Dc(t,\tw)/\Dc(t,t)$.
In the limit $d\to 4$ the FD
plot must be a straight line with slope $X_{\rm m} = 4/5$, by
continuity with the results for $d>4$. Numerical evaluation
(see Fig.~\ref{fig:mFD_plot}%
) shows that, as the dimensionality decreases,
the FD plots deviate progressively from this straight lines. The most
extreme case is the limit $d\to 2$, where analytical forms can be
found.

To find the correlation function for $d\to 2$, it is useful to note that
the scaling function $\sc{M}(x)$ becomes equal to $\delta(x-1)$ in the
limit. Formally, one sees easily from\eq{M_scaling_final} that for
any smooth bounded function $f(x)$ and fixed $c>1$, 
$\int_1^c dx\, 
\sc{M}(x) f(x) \to f(1)$ because the divergence of $\sc{M}(x)$ at
$x=1$ becomes non-integrable in $d=2$. We now exploit this to
simplify $U$ from\eq{exact_I}.
In the first term in square brackets, the argument $xu/\yw$ of
$\sc{M}$ is always larger than $x/\yw$ and hence than $1$ (except at
the irrelevant boundary $\yw=x$ of the $\yw$-integral); in the limit
$d\to 2$, this contribution therefore vanishes. In the second term,
replacing $\sc{M}(x/u\yw)$ by $\delta(x/u\yw - 1)$ and expanding the
prefactor to leading order in $(d-2)/2$ gives
\be
\frac{U}{\Tc} = \frac{d-2}{2} \frac{x}{\yw} 
\int_0^1 dy \int_0^1 dy'\,(1-y-y'+x/\yw)^{-1} 
\label{d2_expansion}
\ee
Performing the integrals over $y$ and $y'$, one has
\be
\frac{U}{\Tc}=\frac{d-2}{2} \frac{x}{\yw} 
\left[\frac{x}{\yw}\ln\left(1-\frac{\yw^2}{x^2}\right)+
\ln\left(\frac{x+\yw}{x-\yw}\right)\right]
\ee
We can now use this limit form of $U$ to get the contribution
$\Dc^{(2)}$ to the correlation function; recall that $U$ was defined
as the $u$-integral in\eq{general_Cm2}. In the remaining
$\yw$-integral in this equation one can again replace $\sc{M}(1/\yw)$ by
$\delta(1/\yw -1)$ so that only $\yw=1$ contributes, giving
\be
\Dc = \Dc^{(2)}=\frac{d-2}{2}\Tc \tw \left[x\ln\left(1-\frac{1}{x^2}\right)+
\ln\left(\frac{x+1}{x-1}\right)\right]
\label{Cm_d2}
\ee
One can show that the other contribution $\Dc^{(1)}$ to the
magnetization correlator, given by\eq{C_m1}, vanishes as $\sim(d-2)^2$ for
$d\to 2$ so that to leading order $\Dc=\Dc^{(2)}$ as anticipated
in writing\eq{Cm_d2}. The corresponding response function
is found by expanding\eq{R_finite} to leading order in $d-2$:
\be
\Q = \frac{d-2}{2}\ln\left(\frac{x}{x-1}\right)
\ee
To get the FDR it only remains to work out the $\tw$-derivative of
$\Dc$:
\bea
\fl \Dc' \ \ \equiv \ \ \partial_{\tw}\Dc&=&
\frac{d-2}{2}\,\Tc\ln\left(\frac{x+1}{x-1}\right)
\eea
We therefore get, for the full dependence of the limiting FDR for
$d\to 2$ on scaled time $x=t/\tw$
\be
X_{\rm m}(x)=\frac{\Tc\Q}{\Dc'} = \ln\left(\frac{x}{x-1}\right)
\left[\ln\!\left(\frac{x+1}{x-1}\right)\right]^{-1}
\ee
For $x\to\infty$ this gives $X_{\rm m}^\infty=1/2$ consistent with the
discussion above. In the limit $x\to 1$ of comparable times, on the
other hand, $X_{\rm m}(x)$ approaches 1, logarithmically slowly; the FD
plot for $d\to 2$ therefore starts off with a pseudo-equilibrium
slope. Interestingly, this implies that the trends of the slope with
$d$ are different at the two ends of the plot: for well-separated
times ($x\to\infty$), the slope {\em decreases} from $4/5$ to $1/2$ as
$d$ decreases from 4 to 2; for comparable times ($x\to 1$) it {\em
increases} from 4/5 to 1.

To get the FD plot itself we need the susceptibility, which is found
by integration of the response\eq{R_finite} as
\bea
\chi_{\rm m}(t,\tw)&=&\int_{\tw}^t dt'\, \Q(t/t')=t\int_{1/x}^1dz\, \Q(1/z)
\\
&=& t\int_{1/x}^1 dz\,z^{(d-2)/4}
\left[1-\left(1-z\right)^{(d-2)/2}\right]
\label{chi_fin}
\eea
Expanding to linear order in $d-2$ and integrating gives
\be
\chi_{\rm m}(t,\tw)
=\frac{d-2}{2}\,t\,\frac{x-1}{x}\left[1-\ln\left(\frac{x-1}{x}\right)\right]
\ee
The normalized susceptibility $\tilde\chi_{\rm m}$ is obtained by
dividing by $\Dc(t,t)/\Tc$, which from\eq{Cm_d2} for $x\to 1$ equals
$2\ln 2[(d-2)/2]\,t$. 
The resulting FD plot for $d\to 2$ is shown in
Fig.~\ref{fig:mFD_plot}, together with the ones for $d=3$
(determined numerically%
) and $d=4$.  For $d\to
2$ the approach of the slope to the equilibrium value $X=1$ for $x\to
1$ is difficult to see because it is logarithmically slow. As
expected from the trends with $d$ in the initial and final slopes of
the FD plot, the curve for $d\to 2$ is the most strongly curved, while
in $d=4$ we have the anticipated straight line with slope $4/5$
that is required by continuity with the results for $d>4$.

To quantify the shape of the FD plot further one can also consider the
axis ratio $Y$, defined as the limiting value of $\tilde\chi_{\rm
m}(t,\tw)=\Tc\chi_{\rm m}(t,\tw)/\Dc(t,t)$ for well-separated times
$t\gg\tw\gg 1$. In equilibrium this would correspond to the FDT for
the static quantities, i.e. equal-time fluctuations and static
susceptibilities. Out of equilibrium, if the FD plot is straight then
$Y$ coincides with $X$; if it is not, then is has been argued that in
some circumstances $Y$ can be more relevant for characterizing
effective temperatures than $X$~\cite{OHeLiuNag04}. For the
magnetization FD plot in the current scenario of magnetized initial
states, we see from Fig.~\ref{fig:mFD_plot} that $Y$ decreases along
with $X_{\rm m}^\infty$ from $4/5$ as the dimension is lowered below
$d=4$. The two quantities begin to
differ more noticeably as $d$ decreases further, with $X_{\rm
m}^\infty=1/2$ and $Y=1/(2\ln 2)=0.7213\ldots$ in the limit $d\to 2$.

\section{Summary and discussion}
\label{sec:conclusion}

In this paper we have considered the non-equilibrium dynamics of the
spherical ferromagnet after a quench to the critical temperature
$\Tc$. Our focus has been the calculation of correlation and response
functions and the associated fluctuation-dissipation ratios (FDRs)
$X(t,\tw)$. The key quantity that can be extracted from the latter is
the asymptotic FDR $X^\infty$ for large and well-separated times
$t\gg\tw\gg 1$; it is independent of model-specific details within a
given dynamical universality
class. We were motivated by two questions: how does $X^\infty$
depend on the observable considered, both with regard to the
lengthscale and the type of observable (spin, bond, spin product)? And
what is the effect of initial conditions, in particular the presence
of a nonzero magnetization in the initial state? The first question
has implications for the interpretation of $T/X^\infty$ as an
effective temperature, which is plausible only if this quantity is
observable-independent. The second one allowed us to uncover whether
different initial conditions can lead to different universality
classes of critical coarsening.

A peculiarity of the spherical model is the weak infinite-range interaction
produced by the spherical constraint. This requires that one
distinguishes between long-range or ``block'' observables, which probe
lengthscales large compared to the (time-dependent) correlation length
but small compared to the system size, and global observables whose
behaviour depends on correlations across the entire
system. Technically, the first case is much easier to treat because
the standard theory where the spins have Gaussian statistics can be
used. Global correlation and response functions, on the other hand,
require non-Gaussian corrections arising from the fluctuations of the
effective Lagrange multiplier.

We dealt with the case of finite-range (i.e.\ either local or
long-range) observables in Sec.~\ref{sec:finite-range}. For spin
observables, we found in the long-range case\eq{X_baseline} the same
$X^\infty$ as for local spin correlations and
response~\cite{GodLuc00b}. This was as expected from the general
correspondence between local and long-range observables discussed in
the introduction. The FD plot for the long-range spin observable,
i.e.\ the magnetization, is a straight line in the long-time
limit. This is as in the Ising case in $d=1$~\cite{MayBerGarSol03},
but it is interesting to note that here it holds for all dimensions
$d>2$. On the other hand, in the Ising model with $d\geq 2$, RG
arguments have been adduced~\cite{CalGam05,MayBerGarSol04,CalGam02} to
suggest that the magnetization FD plot should not be straight, though
with deviations that are likely too small to be detectable
numerically~\cite{MayBerGarSol04}. It is likely that the Gaussian
statistics of the spherical model are responsible for producing a
simpler, straight-line magnetization FD plot in all dimensions,
although it would be interesting to know whether any other models have
this property.

We then looked at the effect of the type of observable on $X^\infty$,
considering both bond and spin product observables in either the local
or long-range versions. The results in equations~(\ref{X_bond_local}),
(\ref{X_bond_long-range}), (\ref{X_prod_local}), (\ref{Xbl_large_d})
show that, although the precise time-dependence of $X(t,\tw)$ varies,
the asymptotic FDR $X^\infty$ is the same in all cases. This is
consistent with general arguments~\cite{CalGam04} suggesting that for
a Gaussian theory all observables should yield the same $X^\infty$. In
contrast to the Ising case~\cite{MayBerGarSol03}, not all long-range
observables give nontrivial FD plots; in fact, only the block product
observable does so, and only for $d<4$, while all others produce
pseudo-equilibrium FD plots for long times.

The bulk of the paper was concerned with the more challenging analysis
of global correlation and response functions, focussing mostly on the
energy as a key observable. In Sec.~\ref{sec:setup} we constructed a
framework for calculating non-Gaussian corrections to the
spins, which are $\order(\rn)$ to leading order. This lead to the
general expression\eq{yt3} for these leading-order corrections. It
involves a two-time kernel $L(t,\tw)$ which from\eq{pinv_def} is the
functional inverse of $K(t,\tw)$ defined in\eq{p_def}. The basis of
all subsequent calculations is the determination of the long-time scaling of
these two functions, as summarized at the end of Sec.~\ref{sec:KandL}.

In Sec.~\ref{sec:energy_general} we obtained general expressions for
energy correlation and response functions, in terms of the kernel $L$
and other quantities known from the Gaussian theory; the results can
be found in eqs.~(\ref{C1},\ref{C2},\ref{C3},\ref{C4}) and\eq{RE}.
Evaluating these first for the equilibrium case, we found that the
energy correlation and susceptibility display a plateau for $T$ just
below $\Tc$ and $d>4$; this is caused by the $\qv=\zv$ wavevector,
i.e.\ by the slow relaxation of the global magnetization.  In
Sec.~\ref{sec:noneq_large_d}, we proceeded to the long-time analysis
of energy FD behaviour in the non-equilibrium case for $d>4$; the key
results are\eq{CE_d_gt_4_longtime} and\eq{RE_d_gt_4_longtime}.  The
associated FDR is given explicitly in\eq{XE} and has the {\em same}
asymptotic value $X^\infty=1/2$ as for all other (finite-range)
observables in $d>4$. The analysis of the case $d<4$ is more
difficult, and we were able to find closed-form
results\eqq{CE_prime}{RE_d_lt_4} only in the limit of well-separated 
times $t/\tw\gg 1$. This is, however, enough to determine $X^\infty$,
with the result\eq{Xinf_dlt4_explicit}. Evaluating this, both
numerically and by expansion in $4-d$ and $d-2$, the crucial
conclusion is that it does not coincide with the asymptotic FDR for
finite-range observables; see Fig.~\ref{fig:X_normal_and_E}. A naive
interpretation of $T/X^\infty$ as an
effective temperature for critical coarsening dynamics is therefore
ruled out, since such a temperaure ought to be observable-independent.
On the other hand, to first order in $4-d$ the result agrees with an
RG calculation~\cite{CalGam04} for the $O(n)$-model. We conclude that
non-Gaussian corrections to the FD behaviour of global observables in
the spherical model capture genuine physical effects that have close
counterparts in more realistic systems with only short-range
interactions.

Finally, in Sec.~\ref{sec:magnetized} we turned our attention to
critical coarsening starting from magnetized initial states;
physically this situation could be produced by an up-quench from an
equilibrated state at a starting temperature $T<\Tc$. We concentrated
on the simpler spin observables and found that already for them, the
presence of a nonzero magnetization makes global properties sensitive
to non-Gaussian corrections. As with the energy fluctuations, it is the
{\em global} correlation and response functions that make contact with
the results for short-range models, as obtained recently for the Ising
case~\cite{GarSolPagRit05}: we find $X^\infty_{\rm m}=4/5$ for $d>4$,
eq.\eq{X_m_d_gt_4}. This is distinct from the value $X^\infty=1/2$ for
the unmagnetized case, indicating that magnetized critical coarsening
is in a separate dynamical universality class. Surprisingly, even the
expressions 
for correlation and response functions themselves, which are not
expected to be universal, coincide with those for the Ising case.
It remains to be understood whether this is accidental or has more
profound origins. For the case $d<4$, we obtained new {\em exact} values for
the asymptotic FDR of magnetized critical coarsening. The
magnetization response\eq{R_finite} can be found 
explicitly for long times, while for the magnetization
correlator only the asymptotics for well-separated
times\eq{Cm_tot_infty} can be written in closed form. The resulting
$X^\infty_{\rm m}$, eq.\eq{Xz}, is surprisingly nontrivial: while it
matches continuously with $X^\infty_{\rm m}=4/5$ in $d>4$ and
approaches the simple value $X^\infty_{\rm m}=1/2$ for $d\to 2$ as
shown in Fig.~\ref{fig:X_magnetized}, it is irrational already to
first order in an expansion in $4-d$, eq.\eq{X_m_epsilon_expansion}, or
$d-2$, eq.\eq{X_m_epsilon_prime_expansion}.

While the conclusion of our calculation as regards the existence of a
well-defined effective temperature for critical coarsening is
negative, the issue of dynamic universality classes and new asymptotic
FDRs due to magnetized (and possibly other, different) initial
conditions clearly deserves further study. Results for systems with
short-range interactions, such as the $O(n)$ and $n$-vector models,
would be particularly welcome. After the present work was completed we
became aware that a first step in this direction has recently
been taken by the authors of Ref.~\cite{FedTri06}, who calculated the
FDR for the $n$-vector model with a magnetized initial state within an
$\epsilon$-expansion around $d=2$. Intriguingly, their result
$X^\infty=1/2$ for $d=2$ itself agrees with ours, but the first-order
correction in $d-2$ remains {\em rational} even for $n\to\infty$. It
therefore disagrees with our spherical model
result\eq{X_m_epsilon_prime_expansion}. This appears to be the first
example of genuine differences between the spherical and $n$-vector
(with $n\to\infty$) models, which are known to have identical
properties in equilibrium~\cite{Stanley68} and within a Gaussian
theory of the dynamics.

As regards future work, we note first that a complete classification
of dynamical universality classes within critical coarsening remains
to be achieved. An earlier study of the spherical model considered
initial conditions with long-range correlations but no overall
magnetization; this yields no new (non-zero) values of the asymptotic
FDR $X^\infty$~\cite{PicHen02}. The presence of a non-zero
magnetization thus appears to be important for observing new
phenomena, and is reflected in our calculation by the fact that
non-Gaussian fluctuations become important. Whether there are yet
other initial conditions that could give rise to distinct values of
$X^\infty$ is an open problem.

Our general framework for treating non-Gaussian
corrections to the dynamics can also be applied in other contexts.
For example, it can be used to analyse the {\em
fluctuations} across thermal histories of correlation and response
functions that have been coarse-grained across a finite-sized
system. The properties of these fluctuations should be useful for
understanding dynamical heterogeneities in coarsening
dynamics~\cite{CasChaCugIguKen03}, and we will report on the
results of such a study shortly. We have also extended our approach to
non-Gaussian corrections for the dynamics of e.g.\ the $O(n)$-model
with large but finite $n$, opening up the attractive prospect of
obtaining exact results analogous to the ones in this paper for models
with exclusively short-range interactions.

{\bf Acknowledgements}: We thank P Calabrese and A Gambassi for
sharing their field theoretic results with us before publication, and
for helpful discussions. PS acknowledges the hospitality of the KITP,
Santa Barbara, where this research was supported in part by the NSF
under grant no.\ PHY99-07949.

\appendix

\section{Evaluation of $X_E^\infty$}
\label{sec:X_E_infty}

In this appendix we evaluate the various numerical factors in the
asymptotic FDR\eq{Xinf_dlt4} for the energy in $d<4$. For $\Bd$ we
already have an expression\eq{CE_factor}. By definition, $\CCd$ is the
large-$t$ limit of $\CC(t,t)/t^{(4-d)/2}$. Its value can be deduced
from\eq{CC_gen_dlt4}: for small $q$, $q=\omega^2$ and hence $(dq) =
\sigma_d d\omega\, \omega^{(d-2)/2}$. (The value of the
proportionality constant, $\sigma_d=(4\pi)^{-d/2}\Gamma^{-1}(d/2)$, is
not actually needed explicitly because it cancels from the overall
result.) Thus from\eq{CC_gen_dlt4}
\be
\CCd = \sigma_d \Tc^2 \int\!d\omt\, \omt^{(d-6)/2} \sc{C}^2(\omt)
\label{CCd}
\ee
Next, $\Ld$ is defined by $\Ltwo\eql(t)=\Ld t^{(d-6)/2}$ for large
$t$. The Laplace transform ot $\Ltwo\eql$ therefore has a singular
term $\hat\Ltwo\eql(s) = \Ld \Gamma((d-4)/2) s^{(4-d)/2}$ for $s\to
0$. From\eq{ptwo_eql}, this must match the corresponding singularity
in $-1/\hat\p\eql(s)$. The latter follows from\eq{p_eql_LT} as
$\hat\p\eql(s) = s^{(d-4)/2} \Tc\sigma_d \int\!d\omt\,
\omt^{(d-4)/2}(1+2\omt)^{-1} = s^{(d-4)/2} \Tc\sigma_d
2^{(2-d)/2}\Gamma((4-d)/2)\Gamma((d-2)/2)$. This tells us that
\be
\Ld^{-1} = -\Tc\sigma_d
2^{(2-d)/2}\Gamma\left({\ts\frac{4-d}{2}}\right)
\Gamma\left({\ts\frac{d-4}{2}}\right)
\Gamma\left({\ts\frac{d-2}{2}}\right)
\label{Ld}
\ee

Finally, we need $\Ad$ from\eq{Ad}. Inserting\eq{C_scaling}
into\eq{CC_d_lt_4} and using\eq{CCd}, the scaling function $\G(x)$ can
be written for $x\geq 1$ as
\bea
\fl \G(x) &=&
\frac{4\sigma_d \Tc^2}{\CCd}
\int d\omt\, \omt^{(d-6)/2} \omt^2 \int_0^1 dy \int_0^1 dy' 
(yy')^{(d-4)/2} e^{-2(1+x-y-y')\omt}
\\
\fl &=& \frac{\Gamma(\frac{d}{2})2^{(4-d)/2}\sigma_d \Tc^2}{\CCd}
\int_0^1 dy\, y^{(d-4)/2} \int_0^1 \frac{dy'}{y'^2} [(1+x-y)/y'-1]^{-d/2}
\\
\fl &=& \frac{2}{d-2}\frac{\Gamma(\frac{d}{2})2^{(4-d)/2}\sigma_d \Tc^2}{\CCd}
\int_0^1 dy\, y^{(d-4)/2} \frac{(x-y)^{(2-d)/2}}{1+x-y}
\label{Gt_def}
\eea
We define the integral as $\Gt(x)$. It is an unnormalized version
of $\G(x)$; because of $\G(1)=1$ one then has
$\G(x)=\Gt(x)/\Gt(1)$. The corresponding unnormalized value of $\Ad$,
$\Atd=\Ad\Gt(1)$, is
\bea
\fl \Atd &=& \int_0^\infty dx\,x\left[\Gtd x^{-d/2}-\Gt(x)\right]
\\
\fl &=& \frac{2\Gtd}{4-d} - \int_0^1 dx\,x\Gt(x) + 
\int_1^\infty dx\,x\left[\Gtd x^{-d/2}-\Gt(x)\right]
\\
\fl &=& \frac{4}{(4-d)(d-2)}-\int_1^\infty dx\,x^{-(d+2)/2}\Gt(x) + 
\int_1^\infty dx\,x\left[\frac{2 x^{-d/2}}{d-2}-\Gt(x)\right]
\label{Atd_aux}
\eea
Here we have used\eq{CC_d_lt_4} to express the values of $\Gt(x)$ for
$x<1$ in terms of those for $x>1$; we also inserted
$\Gtd=\Gd\Gt(1)=\lim_{x\to\infty}\Gt(x)x^{d/2} = 2/(d-2)$ which
follows from\eq{Gt_def}. The first integral has no divergences.  The
second one could in principle be left as it is for numerical
evaluation, but it is useful to rewrite it using a dimensional
regularization trick, as follows. If we extend the definition of
$\Gt(x)$ to $d>4$ using\eq{Gt_def}, then
both parts of the second integral
in\eq{Atd_aux} are separately finite for $4<d<6$, and we can
evaluate them first in that range of $d$ and then analytically
continue to $d<4$. (The region $x\approx 1$ causes no difficulty since
$\Gt(x) \sim (x-1)^{(4-d)/2}$ for $x\to 1$, which remains integrable
for $d<6$.) The first part gives $[2/(d-2)][2/(d-4)]$ and just
cancels the first term in\eq{Atd_aux}, so that
\bea
\fl\lefteqn{\int_1^\infty dx\,x\left[\frac{2 x^{-d/2}}{d-2}-\Gt(x)\right] +
  \frac{4}{(4-d)(d-2)}= }\nonumber\\
\fl &=&
 - \int_1^\infty dx \int_0^1 dy\, y^{(d-4)/2}
(x-y)^{(2-d)/2} \frac{x}{1+x-y}
\\
\fl &=&
-\frac{\pi}{\sin[\pi(d-4)/2]}
+ \int_1^\infty dx \int_0^1 dy\, y^{(d-4)/2}
\frac{(x-y)^{(2-d)/2}(1-y)}{1+x-y}
\eea
The cancellation of the pure power-law term proportional to $\Gtd$
from the integral\eq{Ad} is a feature well-known from dimensional
regularization in field theory. Inserting the last expression into\eq{Atd_aux}
then gives\eq{Atd} in the main text.

Collecting the above results, the asymptotic FDR\eq{Xinf_dlt4} for the
energy in $d<4$ is
\bea
X_E^\infty &=& \frac{4\Tc}{d\Atd}\Biggl[
\frac{2}{d-2}\frac{\Gamma\left(\frac{d}{2}\right)
2^{(4-d)/2}\sigma_d \Tc^2}{\CCd}
\left(-
\frac{\Gamma\left(\frac{d-4}{2}\right)\Gamma\left(\frac{d+4}{2}\right)}
{\Gamma(d)}\right) \CCd \Biggr]^{-1}
\nonumber\\
& & \times \left[-\Tc\sigma_d 
2^{(2-d)/2}\Gamma{\ts\left(\frac{4-d}{2}\right)}
\Gamma{\ts\left(\frac{d-4}{2}\right)}
\Gamma{\ts\left(\frac{d-2}{2}\right)}\right]
\\
&=&\frac{2}{d\Atd}
\frac{\Gamma(d)\Gamma{\ts\left(\frac{4-d}{2}\right)}}
{\Gamma\left(\frac{d+4}{2}\right)}
\eea
which is the result stated in\eq{Xinf_dlt4_explicit}.

\section{Solution for $\sc{L}(x)$ for magnetized case in $d<4$}
\label{sec:L_solution}

In this appendix we prove that the solution
$\sc{L}(x)=2/(4-d)x^{(2-d)/2}+(2-d)/(4-d)x^{-d/2}$ given
in\eq{solution_L} does indeed satisfy the integral
equation\eq{pinv_cond}
\be
\fl
\int_1^x dy\,
(x-y)^{(2-d)/2}(y-1)^{(d-6)/2}\left[\sc{\p}\left({x}/{y}\right)
\sc{L}\left(y\right)-\sc{\p}(x)\right] = 0
\label{change_var}
\ee
with $d<4$ and $\sc{K}(x)$ given by\eq{Ktot_final} for the magnetized
case considered here.

The difference in the square 
brackets can be written as the integral of a partial derivative 
\be
\sc{\p}\left({x}/{y}\right)
\sc{L}\left(y\right)-\sc{\p}(x)=\int_1^ydz\,
\partial_z\left[\sc{\p}\left({x}/{z}\right)
\sc{L}\left(z\right)\right]
\ee
so that, after exchanging the order of the integrals, the left-hand side
of\eq{change_var} becomes
\bea
\fl
l&=& \int_1^x dz\, \partial_z\left[\sc{\p}(x/z)
\sc{L}(z)\right]\int_z^x dy\, (x-y)^{(2-d)/2}(y-1)^{(d-6)/2}
\\
\fl&=&\frac{1}{x-1}\frac{2}{4-d}
\int_1^x dz\,
(x-z)^{(4-d)/2}(z-1)^{(d-4)/2}\partial_z
\left[\sc{\p}(x/z)\sc{L}\left(z\right)\right]
\label{start_int_eq}
\eea
where in the second step the $y$-integral has been performed.
Eq.\eq{start_int_eq} can equivalently be obtained from\eq{change_var}
by integrating by parts.
From\eq{Ktot_final} for $\sc{K}(x)$ and the form given above for
$\sc{L}(x)$, we can find the $z$-derivative of
$\sc{\p}\left({x}/{z}\right)\sc{L}(z)$ explicitly. After simplifying
the latter, 
equation\eq{start_int_eq} becomes
\bea
\fl l&=&\eta \int_1^x dz\,(x-z)^{(4-d)/2}(z-1)^{(d-4)/2}\left\{-\frac{1}{z}+
\frac{d}{2z^2}\right.+\frac{d-2}{2}\left[\frac{d}{2}x-z(x+1)\right] \nn
\fl &&\times\left.
z^{-(d+2)/2}(x-z)^{(d-4)/2}\int_{z/x}^1dy\,(1-y)^{(2-d)/2}y^{(d-4)/2}\right\}
\label{after_diff}
\eea
where 
\be
\eta=\frac{x^{(2-d)/2}}{x-1}\frac{2(d-2)}{(4-d)^2}
\ee
collects all the prefactors. By rescaling $y$ by a factor $x$, the
last integral in\eq{after_diff} can be transformed to $\int_{z}^x
dy\,(x-y)^{(2-d)/2}y^{(d-4)/2}$. Interchanging the $y$ and
$z$-integrals then gives
\bea
\fl \frac{l}{\eta}&=&\int_1^x dz\,(x-z)^{(4-d)/2}(z-1)^{(d-4)/2}
\left(-\frac{1}{z}+\frac{d}{2z^2}\right)
\nn
\fl &&+\int_{1}^x dy\,(x-y)^{(2-d)/2}y^{(d-4)/2}
\nn
\fl &&\times\int_1^{y} dz\,z^{-(d+2)/2}(z-1)^{(d-4)/2}
\,\frac{d-2}{2}\left[\frac{d}{2}x-z(x+1)\right]
\label{middle}
\eea 
The integral on the last line can now be calculated explicitly to give
\bea
g(y)
&=&\left(\frac{y-1}{y}\right)^{(d-2)/2}\left(\frac{d-2}{2}\frac{x}{y}-1\right)
\label{g_y2}
\eea
Relabelling $y$ to $z$, eq.\eq{middle} can thus be written as a single
integral:
\bea
\fl \frac{l}{\eta}
&=&\int_1^x dz\, (x-z)^{(2-d)/2}
\left[(z-1)^{(d-4)/2}(x-z)\left(-\frac{1}{z}+\frac{d}{2z^2}\right)
+z^{(d-4)/2}g(z)\right]
\\
\fl &=&\int_1^x dz\, (x-z)^{(2-d)/2} (z-1)^{(d-4)/2}
\left[
\frac{x}{z^2}-\left(\frac{d-2}{2}+\frac{4-d}{2}x\right)\frac{1}{z}
\right]
\label{rescaled}
\eea
Finally, the variable change $v=(x/z-1)/(x-1)$ transforms the
integration range to $0\ldots 1$ and leads after a few simplifications
to straightforward Beta-function integrals:
\bea
\fl \frac{l}{\eta}
&=& x^{(2-d)/2}(x-1)\int_0^1
dv\,v^{(2-d)/2}\left(-\frac{4-d}{2}+v\right)(1-v)^{(d-4)/2}
\\
\fl &=& x^{(2-d)/2}(x-1)\left[
-\frac{4-d}{2}\frac{\Gamma\left(\ts\frac{4-d}{2}\right)
\Gamma\left(\ts\frac{d-2}{2}\right)}{\Gamma(1)}
+\frac{\Gamma\left(\ts\frac{6-d}{2}\right)
\Gamma\left(\ts\frac{d-2}{2}\right)}{\Gamma(2)}
\right]
= 0
\eea
This proves that\eq{change_var} is indeed satisfied, as required.

\section*{References}

\bibliography{/home/psollich/references/references}
\bibliographystyle{unsrt}

\end{document}